\documentclass[prb,aps,amssymb,twocolumn,superscriptaddress,notitlepage]{revtex4-2}
\usepackage{amsmath}
\usepackage{tikz}
\usepackage{graphicx}
\usepackage[normalem]{ulem}
\usepackage{hyperref}
\usepackage{braket}
\usepackage{bbold}
\usepackage{bm}
\usepackage{siunitx}
\newcommand{\br}{{\bf r}}
\newcommand{\bk}{{\bf k}}
\newcommand{\bQ}{{\bf Q}}
\DeclareMathOperator{\tr}{tr}

\begin{document}

\title{A moir\'e superlattice  on the surface of a  topological insulator}

 \author{Jennifer Cano}
 \affiliation{Department of Physics and Astronomy, Stony Brook University, Stony Brook, New York 11974, USA}
 \affiliation{Center for Computational Quantum Physics, Flatiron Institute, New York, New York 10010, USA}
\author{Shiang Fang}
\affiliation{Department of Physics and Astronomy, Center for Materials Theory, Rutgers University, Piscataway, NJ 08854 USA}
\author{J. H. Pixley}
\affiliation{Department of Physics and Astronomy, Center for Materials Theory, Rutgers University, Piscataway, NJ 08854 USA}
\author{Justin H. Wilson}
\affiliation{Department of Physics and Astronomy, Center for Materials Theory, Rutgers University, Piscataway, NJ 08854 USA}
\date{\today}

\begin{abstract}
Twisting van der Waals heterostructures to 
induce correlated many body states  
provides a novel tuning mechanism in solid state physics. In this work, we theoretically investigate the fate of the surface Dirac cone of a three-dimensional topological insulator subject to a superlattice potential.
Using a combination of diagrammatic perturbation theory, lattice model simulations, and ab initio calculations we elucidate the unique aspects of twisting a single Dirac cone with an induced moir\'e potential and the role of the bulk topology on the reconstructed surface band structure. We report a dramatic renormalization of the surface Dirac cone velocity as well as demonstrate a topological obstruction to the formation of isolated minibands. Due to the topological nature of the bulk, surface band gaps cannot open; instead additional satellite Dirac cones emerge, which can be highly anisotropic and made quite flat.
 We discuss the implications of our findings for future experiments.
 \end{abstract}

\maketitle

\section{Introduction}
\label{sec:introduction}

Observing and controlling  (pseudo)relativistic quasiparticle excitations has become a central aspect of modern condensed matter physics. 
Massless Dirac fermions appear in the electronic band structure of a number of materials that host linear touching points, i.e. Dirac cones, such as in graphene \cite{geim2007}, Weyl semimetals \cite{Wan11,Weng15,Huang15,Xu15,Lv15,Xu15a,Lv15a,Xiong2015} and their symmetry-protected generalizations \cite{Young12,Wang12,Liu14,Liu14a,Steinberg14,Bradlyn2016,Chang2017,Schroter2018,rao2019observation,sanchez2019topological,cano2019multifold,klemenz2020systematic}, and 
on the surface of a three-dimensional topological insulator (3D TI) \cite{fu2007topological,moore2007topological,roy2009topological,qi2008topological,schnyder2008classification}. 
On the surface of a 3D TI, the Dirac cone is  anomalous, with a corresponding partner of equal and opposite helicity on the opposing surface.
Such anomalous Dirac cones have been observed experimentally on the surfaces of topological insulators such as Bi$_2$Se$_3$, Bi$_2$Te$_3$ and Sb$_2$Te$_3$ \cite{xia2009observation,zhang2009topological,Chen2009experimental,hsieh2009observation,hasan2010colloquium}. 

Recent  experiments have demonstrated an unprecedented amount of control over the nature of two-dimensional (2D) Dirac excitations by twisting van der Waals heterostructures, i.e., orienting adjacent layers with a relative twist angle \cite{andrei2020graphene}. 
The twist induces a moir\'e superlattice that dramatically renormalizes the velocity of the low energy Dirac excitations. In some cases, the velocity can even vanish at precise ``magic'' values of the twist angle \cite{bistritzer2011moire,Santos-2012}. This quenches the kinetic energy, thus enhancing the effective interaction strength and promoting the formation of exotic many-body states.
The velocity renormalization has been demonstrated experimentally in twisted bilayer graphene (TBG), where the flat bands and enhanced correlation strength lead to exotic superconducting, correlated insulating, and quantum anomalous Hall phases \cite{cao2018unconventional,cao2018correlated,yankowitz2019tuning,lu2019superconductors,sharpe2019emergent,serlin2019intrinsic,kerelsky2019maximized,xie2019spectroscopic,jiang2019charge}. Moir\'e superlattices have also been used to induce correlated insulating states in transition metal dichalcogenides \cite{Tang2020simulation,regan20mott,shimazaki2020strongly,wang2019magic} and various multilayer graphene heterostructures such as trilayer graphene \cite{tsai2019correlated}, double bilayer graphene \cite{cao2019electric,burg2019correlated,shen2020correlated,liu2020tunable}, and graphene layers twisted relative to hexagonal boron nitride \cite{chen2019evidence,chen2019signatures,chen2020tunable,Chen2020}.
Furthermore,  superlattices can also produce magic-angle conditions in ultracold atom experimental set ups~\cite{fuMagicangleSemimetals2020,PhysRevA.100.053604,PhysRevLett.125.030504,luo2020spin}. In the limit of an incommensurate superlattice  an Anderson-like single-particle phase transition can occur at the magic-angle~\cite{Pixley-2018,fuMagicangleSemimetals2020,chouMagicangleSemimetalsChiral2020,fuFlatTopologicalBands2020,gonccalves2020incommensurability}.

Motivated by the success of twisted van der Waals heterostructures, we ask, can the enhanced and tunable interaction strength observed in Dirac cones in twisted graphene heterostructures can be applied to the Dirac cone on the surface of a 3D TI? 
If the answer is in the affirmative, it provides a new route to engineer interacting instabilities on the surface of 3D TIs, such as surface magnetism \cite{baum2012magnetic,baum2012density,marchand2012lattice,schmidt2012strong,sitte2013interaction,mendler2015magnetic} or topological superconductivity \cite{Fu2008superconducting,Santos2010}.
Vortices in the latter phase host the long sought-after Majorana fermions \cite{read2000paired,kitaev2001unpaired,volovik1999fermion}.
As experimentally observing these strongly correlated phases on the surface of a 3D TI remains elusive, a novel approach is necessary. 

\begin{figure}
    \centering
    \includegraphics[width=\columnwidth]{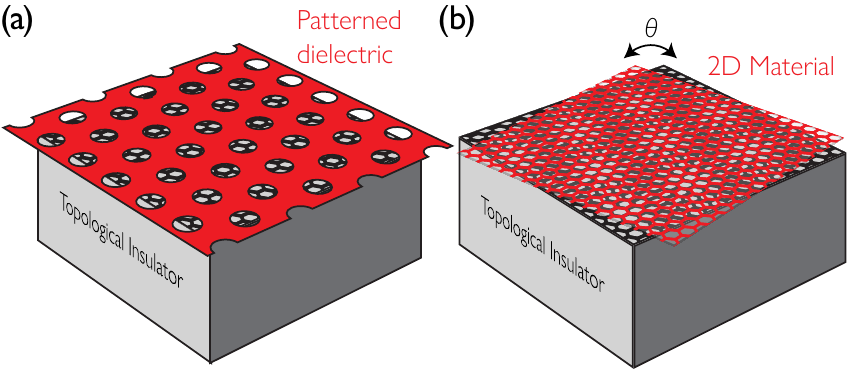}
    \caption{{\bf Schematic depictions of experimental realizations of a surface-moir\'e potential.} (a) Surface potential induced by gating a periodic patterned dielectric. (b) Twisting a gapped 2D material on the surface of the 3D TI. }
    \label{fig:schematic}
\end{figure}

On the other hand, the single-particle band theory of TBG does not directly apply to a twisted heterostructure on the surface of a 3D TI.
In TBG, the superlattice potential introduces single-particle band gaps, creating a low-energy miniband that describes excitations on the moir\'e lattice 
\cite{lopes2007graphene,Shallcross2010electronic,morell2010flat,bistritzer2011moire,Lopes2012continuum}. 
In the vicinity of the magic angle, the hard gaps separating the half-filled miniband from the filled and empty bands allow the miniband to become increasingly flat, which enhances the relative strength of correlations.
The crucial difference between the Dirac cones in TBG and the Dirac cone on a 3D TI surface is that because the latter is anomalous,
the existence of a gap to the superlattice miniband is forbidden.
Specifically, a moir\'e induced gapped miniband would violate the requirement that extended surface states exist at all energies in the bulk band gap \cite{fu2007topological,moore2007topological,schnyder2008classification,qi2008topological,roy2009topological}.
Thus, the Dirac cone on the surface of a topological insulator must exhibit fundamentally different behavior in a moir\'e potential than the Dirac cones in TBG.

The main goal of this manuscript is to reconcile which features of TBG 
can be utilized to create strongly correlated topological phases on TI surfaces.
To this end, we develop the theory of a 3D TI subject to a superlattice potential.
The superlattice potential could be imposed by a patterned dielectric superlattice, which has been achieved in graphene \cite{dielectric_patterning} and is expected to be more widely applicable \cite{shi2019gate,li2020anisotropic}.
A second possibility is to build a moir\'e heterostructure by layering a lattice-matched 2D material on top of the 3D TI surface with a relative twist angle or lattice mismatch.
This set-up is also within current experimental reach, as graphene-3D TI heterostructures have already been 
realized \cite{jin2013proximity,zhang2014proximity,cao2016heavy,steinberg2015tunneling,bian2016experimental,tian2016electrical,chong2018topological,jafarpisheh2018proximity,Khokhriakov2018}.
We will consider both of these possibilities, which are depicted in Fig.~\ref{fig:schematic}. 

We demonstrate that, distinct from TBG, a moir\'e potential induces additional gapless satellite Dirac cones (SDCs) in the renormalized band structure instead of forming a moir\'e superlattice miniband gap.
The SDCs emerge at energies nearby the original Dirac cone and are protected by time-reversal symmetry. 
They can have either isotropic or strongly anisotropic Dirac cone velocities, depending on their symmetry.
If the SDCs are isolated in energy, they will appear as a \emph{pseudogap} in the surface density of states. 
However, if they coincide in energy with other metallic bands, they may not be directly visible to spectroscopic probes. 
Using a diagrammatic perturbative approach, we develop a theory for the SDCs that we compare, in detail, with exact numerical simulations of a lattice model and find excellent agreement in the regime of applicability. We then extend these considerations to a patterned gate potential on the surface of Bi$_2$Se$_3$ to demonstrate the experimental feasibility of our theory.

Both our perturbative and numerically exact results demonstrate that a moir\'e potential on a TI surface will lead to a renormalization of the original surface Dirac cone, but does not yield a magic-angle condition for a vanishing velocity. 
However, the same is not true of SDCs: we demonstrate that the velocities of SDCs can vanish at magic twist angles perturbatively (and in exact numerics, they become quite small).
Moreover, we numerically demonstrate that the SDCs can be made very flat, inducing a significant enhancement of the surface density of states of the TI.
Thus, despite the absence of superlattice miniband gaps, our results suggest that twisting can promote Hartree-Fock like instabilities on the surface of a 3D TI.

The paper is organized as follows.
In Sec.~\ref{sec:model} we derive the superlattice potential for the patterned dielectric and twisted heterostructure depicted in Fig.~\ref{fig:schematic}.
In Sec.~\ref{sec:continuum} we obtain expressions for the energy and renormalized velocity of the original Dirac cone and SDCs to leading order in perturbation theory using a continuum model.
We then validate the perturbative results by studying the two models numerically in Secs.~\ref{sec:toymodel} and \ref{sec:abinitio}.
In Sec.~\ref{sec:toymodel} we develop a minimal model consisting of a bulk 3D TI tunnel-coupled on its surface to a gapped 2D material.
We match the renormalized velocity and SDCs to the perturbative results.
In Sec.~\ref{sec:abinitio} we apply the conceptually simpler gate potential to an ab initio model of Bi$_2$Se$_3$; this more realistic model demonstrates the experimental feasibility of our results.
We conclude with a discussion in Sec.~\ref{sec:discussion}. Further details of our calculations can be found in the Appendices. 

\section{Models}
\label{sec:model}

In the following manuscript, we study the effect of a moir\'e potential on the surface Dirac cone of a 3D TI.
The models that we study take the form
\begin{equation}
    H= H_\mathrm{TI} + H_\mathrm{potential},
    \label{eq:Hgeneral}
\end{equation}
where $H_\mathrm{TI}$ describes the 3D TI and
\begin{equation}
   H_\mathrm{potential} = \sum_{\bk,z,\alpha,\mathbf{Q}}  c^\dagger_{\mathbf{k} + \mathbf{Q},z,\alpha} V_{\mathbf{Q}} c_{\mathbf{k},z,\alpha} \delta_{z,L},
    \label{eq:Hpot}
\end{equation}
where $V_\mathbf{Q}$ is a Fourier component of the superlattice potential $V(\mathbf{r})$ on the TI surface (at $z=L$), and $c_{\bk,z,\alpha}$ denotes the fermionic annihilation operators of the TI, where $\bk$ indicates momentum perpendicular to the surface, $z$ indicates position in the direction of the open surface, and $\alpha$ indicates spin and sublattice degrees of freedom.
The delta function specifies that the potential only acts on the top layer of the TI (or the top quintuple layer of Bi$_2$Se$_3$ as defined in Sec.~\ref{sec:abinitio}).
More generally, $V_\mathbf{Q}$ could have spin or sublattice indices  as well as $z$-dependence (i.e. replacing $\delta_{z,L}$ with $\sim e^{-z/\xi}$), but in this work, we only consider $V_\mathbf{Q}$ proportional to the identity matrix and acting on the top layer of the TI.

Fig.~\ref{fig:schematic} shows the two scenarios we have in mind that produce such a potential: a patterned dielectric (a) and a twisted van der Waals heterostructure (b).
In Secs.~\ref{sec:gateintro} and \ref{sec:twistintro}, we will derive the potential and explain approximations in each case.

\subsection{Patterned Dielectric Superlattice}
\label{sec:gateintro}
The first situation we consider is a 3D TI with a surface potential applied through a patterned dielectric superlattice depicted in Fig.~\ref{fig:schematic}(a). 
This is described by the Hamiltonian
\begin{equation}
    H = H_\mathrm{TI}+H_\mathrm{G},
    \label{eq:generalmodelgate}
\end{equation}
where $H_\mathrm{TI}$ describes the topological insulator and $H_\mathrm{G}$ is produced by the gate. The gate potential is milled to produce a periodic potential on a much larger length scale than the original lattice spacing and can be described by
\begin{equation}
    H_\mathrm{G}=\sum_{{\bf r}\in S,\alpha} c_{\br,\alpha}^{\dag}V_G({\bf r})c_{\br,\alpha},
    \label{eq:HG}
\end{equation}
where, throughout, fermionic annihilation operators in the TI are denoted as $c_{\br,\alpha}$, where $\alpha$ denotes sublattice and spin, and ${\bf r}\in S$ denotes sites on the surface of the TI.
Ignoring higher harmonics that could be produced due to the patterned structure, we approximate the periodic gate potential as  
\begin{equation}
    V_G({\bf r})= \frac{W}{2} \sum_{j=1}^N \exp(i\mathbf Q_j \cdot (\mathbf r - \mathbf r_0) ), 
    \label{eq:defVG}
\end{equation}
where the ${\bf Q}_j$ are a set of minimal-length reciprocal lattice vectors of the milled lattice structure in the gate, $N$ is determined by its symmetry, and $\mathbf r_0$ allows the origin of the patterned structure to be shifted relative to the 3D TI surface. 
We will return to this model in Sec.~\ref{sec:abinitio} when we consider Bi$_2$Se$_3$.

\subsection{Twisted Surface}
\label{sec:twistintro}

We now demonstrate that the twisted van der Waals heterostructure depicted in Fig.~\ref{fig:schematic}(b), made from an electrically gapped 2D material layered on the surface of a 3D TI, such as hexagonal boron nitride on Bi$_2$Se$_3$, will also induce a superlattice potential.
The Hamiltonian takes the form
\begin{equation}
    H = H_\mathrm{TI}+H_\mathrm{2D}+H_{\mathrm{TI}-\mathrm{2D}},
    \label{eq:generalmodel}
\end{equation}
where $H_\mathrm{TI}$ describes the topological insulator, $H_\mathrm{2D}$ captures the 2D layer, and $H_{\mathrm{TI}-\mathrm{2D}}$ is the tunnel coupling between the TI and the 2D material. We will denote fermionic annihilation operators in the TI  as $c_{\br,\alpha}$ and in the 2D layer as $d_{\br,\alpha}$, where $\br$ denotes position and $\alpha$ denotes sub-lattice and spin.

The tunnel coupling in real space takes the form:
\begin{equation}
    H_{\mathrm{TI}-\mathrm{2D}} = \sum_{\br,\br',\alpha,\beta} T_{\alpha\beta}(\br,\br')c_{\br\alpha}^{\dag}d_{\br' \beta}+\mathrm{h.c.} ,
    \label{eq:HTI2D}
\end{equation}
where $\br,\br'$ denote real space positions on the 3D TI surface and 2D layers, respectively, $\alpha$ and $\beta$ denote sub-lattice and spin, and $T_{\alpha \beta}(\br, \br')$ is the tunnel coupling matrix. 
We assume that the tunneling from the 2D layer is only into the surface of the TI and not the layers below and that $T(\br, \br')$ is a function only of $\br - \br'$.
This is a good approximation  as the topological surface states are exponentially bound to the surface.
The Fourier transform of Eq.~(\ref{eq:HTI2D}) (derived in Appendix~\ref{sec:derivecoupling}) is then
\begin{equation}
    H_{\mathrm{TI}-\mathrm{2D}} \approx \sum_{\mathbf k, j} c_{\mathbf k,\alpha}^\dagger T_{\alpha\beta}({\mathbf Q}_{j}) d_{\mathbf k+{\mathbf Q}_{j},\beta} + \mathrm{h.c.},
    \label{eqn:HTI2D-approx}
\end{equation}
where the approximation comes from limiting ourselves to a finite number of moir\'e reciprocal lattice vectors ${\mathbf Q}_{j}$, $j=1,..,m$ with 
$m$ sub-extensive, such that the set of $\mathbf{Q}_j$ is closed under the symmetry group of the interface and such that Eq.~(\ref{eqn:HTI2D-approx}) becomes more precise as more values of $j$ are included.
For a small twist angle and no lattice mismatch, the $\mathbf{Q}_j$ are given by
\begin{equation}
    \mathbf{Q}_j = \theta \hat{\mathbf{z}} \times \mathbf{G}_j,
    \label{eq:defQ}
\end{equation}
where $\mathbf{G}_j$ is a reciprocal lattice vector in a single layer.

Since the 2D layer is gapped, we assume that its dispersion varies slowly on the scale of the Dirac cone and, for simplicity, treat it as a completely flat band with spin degeneracy, offset by energy $\Delta$ from the 3D TI surface Dirac crossing:
\begin{equation}
    H_\mathrm{2D} =- \sum_{\bk,s,s'} \left( \Delta \sigma_0\right)_{ss'} d_{\bk s}^{\dag}d_{\bk s'},
    \label{eq:H2Dsurf}
\end{equation}
where $\sigma_0$ is the $2\times 2$ identity matrix acting in spin space.

Time reversal symmetry is implemented by:
\begin{equation}
T = i\sigma_y K,
\label{eq:defTR}
\end{equation}
where $K$ is the complex conjugation operator.
Time reversal imposes the following constraint on the coupling terms:
\begin{equation}
T_{-\mathbf{Q}} = \sigma_y T_\mathbf{Q}^* \sigma_y,
\label{eq:TRconstraint}
\end{equation}
where $T_\mathbf{Q}$ is a shorthand for $T(\mathbf{Q})$.

\subsubsection{Induced potential on the surface of the TI}
\label{sec:qpderivation}
Because the 2D material is gapped, the $d_{\bk,\uparrow}$ and $d_{\bk,\downarrow}$ degrees of freedom can be integrated out, yielding an effective potential on the surface of the 3D TI that we will now derive.
We will apply this effective potential to a continuum model in Sec.~\ref{sec:continuum}, where we perturbatively compute the renormalized velocity and SDCs.
Later, in Sec.~\ref{sec:toymodel}, we introduce a 3D lattice model of a TI and numerically compute the density of states resulting from the effective potential computed here.

Using standard techniques, we write the model in Eqs.~(\ref{eqn:HTI2D-approx}) and (\ref{eq:H2Dsurf}) in terms of a fermionic path integral over Grassman fields.
Integrating out the gapped 2D layer yields the effective action:
\begin{equation}
S_{\rm c}^{\rm eff} = -\int d\omega %d^2\mathbf{k}
\sum_{\bk,\mathbf{Q}_1,\mathbf{Q}_2} c_{\omega,\mathbf{k}+\mathbf{Q}_1}^\dagger  \frac{T_{\mathbf{Q}_1} T_{\mathbf{Q}_2}^\dagger}{\Delta+\omega}  c_{\omega,\mathbf{k}+\mathbf{Q}_2},
\end{equation}
where $c_{\omega,\mathbf{k}} = \left( c_{\omega,\mathbf{k},\uparrow}, c_{\omega,\mathbf{k},\downarrow} \right)$ is a two-component spinor.
In the low-energy limit, $\omega \ll \Delta $, $S^{\rm eff}_c$ becomes:
\begin{equation}
    S_{\rm c}^{\rm eff} 
   \approx\int d\omega \sum_{\br} c^{\dag}_{\omega}(\br) \mathcal{T}(\br)
    \left( \frac{\omega}{\Delta^2} -\frac{1}{\Delta} \right)
    \mathcal{T}(\br)^{\dag}c_{\omega}(\br),
    \label{eq:Sceff}
\end{equation}
where $\mathcal{T}(\br) = \sum_{\bQ}e^{i \bQ\cdot\br} T_{\bQ}$.
To interpret this result as a renormalized Hamiltonian, we need the coefficient of the $\omega$ term in the action be unity. To achieve this we rescale the TI creation operators operators by the quasiparticle weight $Z(\br)$, i.e.
$
 c_\omega(\br)\rightarrow   c_\omega(\br) \sqrt{Z(\br)}
 $
where the quasiparticle weight  is given by
 \begin{equation}
 Z(\br)=
  \frac{1}
  {
  1+
    \frac{\mathcal{T}(\br)\mathcal{T}^{\dag}(\br)}
    {
     \Delta^2
     }
  }.
\end{equation}
Under this definition, the induced potential on the surface of the TI is given by
\begin{equation}
    V_{\mathrm{2D}}(\br) = \frac{Z(\br)}{\Delta}\sum_{\bQ_1,\bQ_2}e^{i\br \cdot (\bQ_1 - \bQ_2)} T_{\bQ_1}T_{\bQ_2}^{\dag}.
    \label{eqn:subpot}
\end{equation}
Rescaling the $c$ operators in $H_\mathrm{TI}$ (whose form we have not yet specified) will produce spatial dependence in the hopping coefficients from $Z(\mathbf{r})$. 
This amounts to a renormalization of the hopping that is suppressed by one order of $\Delta$ relative to the induced potential in Eq.~(\ref{eqn:subpot}).
Thus, we neglect this additional renormalization to all of the model parameters due to $Z(\br)$ 
and only consider the induced potential with $Z(\br)$ set to unity in Eq.~(\ref{eqn:subpot}). 

After integrating out the gapped degrees of freedom in the 2D material and setting $Z(\mathbf{r}) \rightarrow 1$, 
the resulting effective Hamiltonian consists of $H_\mathrm{TI} + H_\mathrm{2D}^{\rm eff}$, where $H_\mathrm{2D}^{\rm eff}$ describes the effective superlattice potential:
\begin{equation}
    H_\mathrm{2D}^{\rm eff} = 
    \sum_{\bk,z,\alpha,\mathbf{Q}}  c^\dagger_{\mathbf{k} + \mathbf{Q},z,\alpha} V_{\mathbf{Q}} c_{\mathbf{k},z,\alpha}\delta_{z,L},
    \label{eq:defH2D}
\end{equation}
with
\begin{equation}
    V_{\mathbf{Q}} = \frac{1}{\Delta} \sum_{\mathbf{Q}_1} T_{\mathbf{Q} + \mathbf{Q}_1} T_{\mathbf{Q}_1} ^\dagger.
    \label{eq:defVQ}
\end{equation}
Time reversal symmetry (\ref{eq:TRconstraint}) and hermiticity require $V_\mathbf{Q} = \sigma_y  V_\mathbf{Q}^T \sigma_y$ (see Appendix~\ref{sec:VQidentity}).
Thus, in the low-energy model of a Dirac cone, where $V_\mathbf{Q}$ is a $2\times 2$ matrix, $V_\mathbf{Q}$ is proportional to the identity matrix, i.e., it has no spin structure.
Except where otherwise indicated, we will always take $V_\mathbf{Q}$ to be proportional to the identity, and, therefore we will use $V_\mathbf{Q}$ to indicate a number, not a matrix.
With this assumption, the twisted heterostructure is also described by the Hamiltonian in Eq.~(\ref{eq:Hgeneral}).

\section{Perturbation theory}
\label{sec:continuum}
To begin, we derive the continuum theory of a time-reversal invariant single Dirac cone subject to a superlattice potential perturbatively. 
We focus on the renormalization of the Dirac cone velocity and the development of satellite Dirac cones at finite energies due to scattering between degenerate states. 
As explained in Sec.~\ref{sec:introduction}, the satellite Dirac cones cannot gap due to the nontrivial topology of the bulk; at best they form a pseudogap density of states at the satellite peak energy. This topological obstruction to forming true minibands separated from other states by a hard electronic gap is a manifestation of the bulk-boundary correspondence and makes this problem fundamentally distinct from  twisted graphene multi-layers. We will verify these predictions beyond perturbation theory using exact numerical calculations in a lattice model of a 3D TI in Sec.~\ref{sec:toymodel} and a Wannier-ized model obtained from ab initio calculations of Bi$_2$Se$_3$ in Sec.~\ref{sec:abinitio}.

\subsection{Continuum surface Hamiltonian}

We start by considering a low-energy effective Hamiltonian for the surface Dirac cone of a 3D TI.
Following the notation in Eq.~(\ref{eq:Hgeneral}), we take $H_\mathrm{TI}$ to be the continuum model for the surface Dirac cone:
\begin{equation}
    H_\mathrm{TI} = \int \frac{d^2\bk}{(2\pi)^2} \sum_{s,s'} \left( v\mathbf{k} \cdot \sigma \right)_{ss'} c_{\bk s }^{\dag}c_{\bk s' },
    \label{eq:HTIsurf}
\end{equation}
where $s \! = \uparrow, \downarrow$ indicates spin.
In this section, to obtain analytical results, we consider only the linear dispersion of the Dirac cone, therefore ignoring warping that generically occurs at higher orders in $\mathbf{k}$ \cite{fu2009hexagonal}.
However, in subsequent sections, we will numerically study Dirac cones on the surface of bulk 3D TI Hamiltonians, which naturally contain all orders in $\mathbf{k}$.
We find that higher orders in $\mathbf{k}$ are necessary to quantitatively match the results, but that the linear Dirac cone (\ref{eq:HTIsurf}) correctly captures the qualitative effects of the twisted interface.

\subsection{Renormalized velocity}
\label{sec:renormalizedvelocity}
To determine the effect of the surface potential (\ref{eq:Hpot}) on the surface Dirac cone, we perturbatively compute the surface electron Green's function $G({\bf k},\omega)^{-1}=G_0({\bf k},\omega)^{-1}-\Sigma({\bf k},\omega)$, where the free surface Green functions is $G_0({\bf k},\omega)^{-1}=\omega - v {\bf k}\cdot \bm {\sigma}$ with a self energy $\Sigma({\bf k},\omega)$.
To leading order, the effective potential gives rise to a self-energy:
\begin{equation}
    \Sigma(\mathbf{k},\omega) = -  \omega \gamma ,
    \label{eq:sigma}
\end{equation}
where
\begin{equation}
    \gamma = \sum_\mathbf{Q\neq 0} \frac{ |V_\mathbf{Q}|^2}{v^2|\mathbf{Q}|^2}.
    \label{eq:defalpha}
\end{equation}
Note, we have not included $V_{\mathbf{Q}=0}$ since it can be taken into account in the free Hamiltonian by a chemical potential shift: $G_0({\bf k},\omega)^{-1}=\omega - v {\bf k}\cdot \bm {\sigma} - V_{\mathbf{Q}=0} + \mu$, with $\mu = V_{\mathbf{Q}=0}$.
Eqs.~(\ref{eq:sigma}) and (\ref{eq:defalpha}) are derived to second order in perturbation theory in Appendix~\ref{sec:derivesigma} and apply when the TI surface (with the potential) has an $n$-fold rotation symmetry with $n>2$. The rotation symmetry prevents an anisotropic velocity renormalization.
As the surfaces of most known 3D TIs have a three-fold rotation symmetry, we do not discuss the anisotropic case further, although the additional terms are found in Appendix~\ref{sec:derivesigma}.

The self-energy in Eq.~(\ref{eq:sigma}) results in a renormalization of the Dirac cone velocity:
\begin{equation}
    v \rightarrow v_* = \frac{v}{1+\gamma}
    \label{eq:vstar}
\end{equation}
From the definition of $\gamma$ in Eq.~(\ref{eq:defalpha}), the renormalized velocity decreases with a stronger potential strength, $V_\mathbf{Q}$, or smaller Fourier wavevectors, $|\mathbf{Q}|$.
The latter occurs by increasing the size of the superlattice in the case of a gated potential or by decreasing the twist angle in the twisted heterostructure.

Importantly, unlike in graphene, the self-energy in Eq.~(\ref{eq:sigma}) does not contain a term proportional to $\mathbf{k} \cdot \sigma$ to this order.
Consequently, there is not a ``magic angle'' where the velocity vanishes. Instead, the velocity decreases while remaining positive to this order in perturbation theory.

\subsection{Satellite Dirac cones at finite energy}
\label{sec:satellite}

Graphene in a superlattice potential exhibits a family of ``satellite'' Dirac cones (SDCs) \cite{park2008anisotropic,yankowitz2012emergence,dean2013hofstadter,hunt2013massive,ponomarenko2013cloning,dielectric_patterning}.
The SDCs occur when the superlattice potential couples degenerate eigenstates at distinct momenta, resulting in an effective Dirac Hamiltonian near the degenerate points.
The energies of the SDCs are determined by the superlattice wavelength and Dirac cone velocity.
In TBG, these SDCs gap, resulting in a superlattice miniband.

In this section, we will show that analogous SDCs appear on the surface of a topological insulator in a superlattice potential.
Specifically, we will derive the energy and dispersion of the lowest-energy SDCs that arise from the surface Hamiltonian in Eq.~(\ref{eq:HTIsurf}) to linear order in perturbation theory in the superlattice potential in Eq.~(\ref{eqn:subpot}). 
Like the renormalization of the original Dirac cone, we find that SDCs invariant under time-reversal and an $n$-fold rotational symmetry with $n>2$ are isotropic (see proof in Appendix~\ref{sec:isotropic}), while those at other momenta are anisotropic.
Importantly, and in contrast to graphene, we also show that the SDCs cannot be gapped because they are protected by time-reversal symmetry.
This protection is what prevents a gapped miniband from forming on the surface of the 3D TI.

To derive the energy and momenta of the SDCs, we denote the eigenstates of the Dirac Hamiltonian $\mathbf{k} \cdot \sigma$ with positive/negative energy using first-quantized notation:
\begin{equation}
    \ket{\mathbf k,\pm} = \ket{\mathbf k} \otimes \tfrac1{\sqrt2} \begin{pmatrix}1 & \pm e^{i\varphi_{\mathbf k}} \end{pmatrix}^T ,
    \label{eq:kspinor}
\end{equation}
where $\ket{\mathbf k}$ is an eigenstate of the momentum operator and $\varphi_\mathbf{k}$ is the angle between $\mathbf{k}$ and the $x$-axis.
For simplicity, we restrict our analysis of SDCs to positive energies, i.e., the $\ket{\mathbf{k},+} = \ket{\mathbf{k}}\otimes \ket{e^{i\varphi_\mathbf{k}}}$ states in Eq.~(\ref{eq:kspinor}), where we have introduced the shorthand,
\begin{equation}
    \ket{e^{i\varphi}} = \tfrac1{\sqrt2} \begin{pmatrix}1 & e^{i\varphi} \end{pmatrix}^T.
    \label{eq:spinor}
\end{equation}
Time-reversal is implemented by $i\sigma_y K$, so that:
\begin{equation}
    {\rm TR}\ket{\mathbf k,\pm} = e^{-i\varphi_\mathbf{k}}\ket{-\mathbf{k},\pm}
    \label{eq:TRspinor}
\end{equation}

\begin{figure*}[t]
   \includegraphics[]{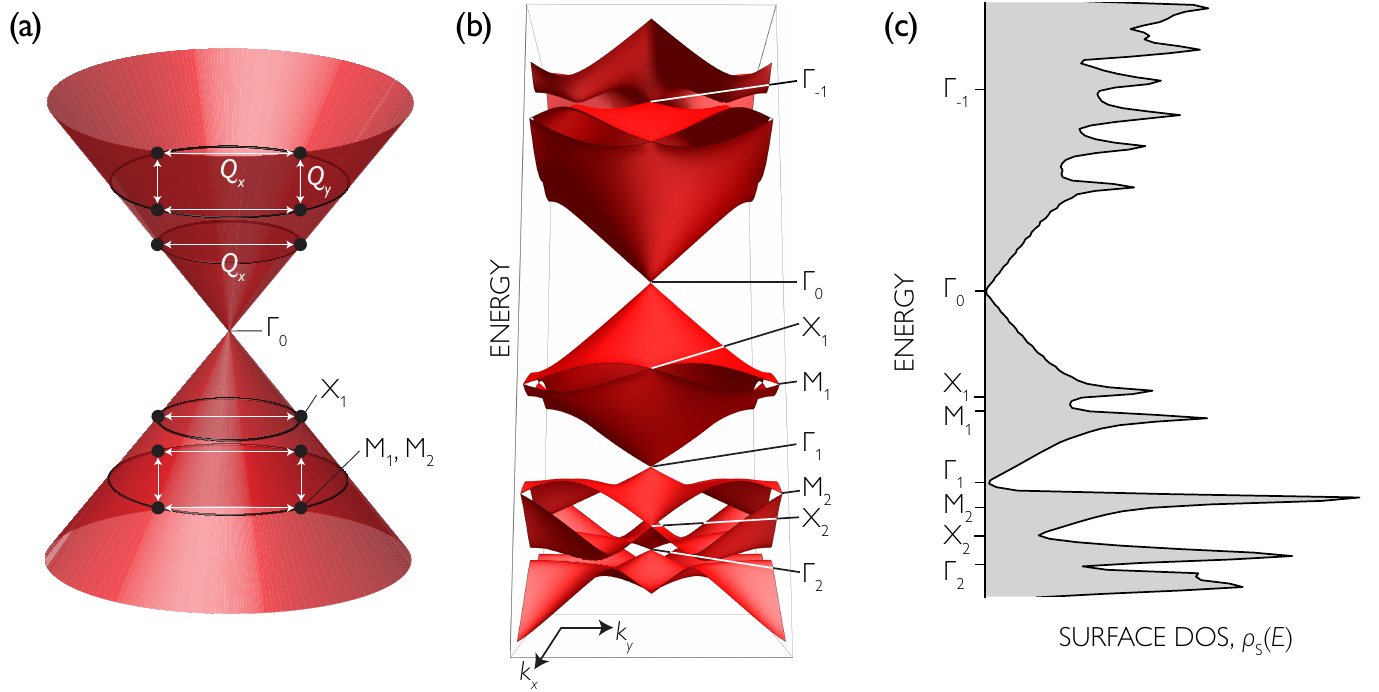}
\caption{{\bf Momentum exchanges on the surface Dirac cone, satellite Dirac cone labels and density of states.} (a) The lowest SDC ($X_1$) is a result of the superlattice potential $V_{\mathbf{Q}_x}$ coupling the states at momenta $\pm \mathbf{Q}_x/2$, shown by the white line connecting two black dots. (A SDC at the same energy occurs from coupling the states at $\pm \mathbf{Q}_y/2$, not shown.) On the square lattice, the next lowest energy SDC ($M_1$) results from the superlattice potential coupling the four degenerate states at $\pm \mathbf{Q}_x/2 \pm \mathbf{Q}_y/2$, indicated by the white square connecting four black dots. (b) The resulting downfolded cone and satellites computed with Eq.~\eqref{eqn:square-subpot} at $W=0.2t$. (c) The resulting density of states in arbitrary units. The cone $\Gamma_1$ is visible and isolated from other features in the spectrum.}
\label{fig:satelliteQ}
\end{figure*}

We first derive the lowest energy SDC: let $\mathbf{Q}$ be the shortest reciprocal lattice vector of the superlattice such that $V_{\mathbf{Q}}\neq 0$ (e.g.\ this corresponds to the $X_1$ SDC in Fig.~\ref{fig:satelliteQ}). 
Then the lowest positive energy SDC occurs at the smallest momenta that differ by $\mathbf{Q}$, i.e., $\pm \mathbf Q/2$, as shown in Fig.~\ref{fig:satelliteQ}.
Since the states at $\pm \mathbf{Q}/2$ are time-reversed partners, they are degenerate and will remain degenerate after being coupled by the potential.
Therefore, we must use degenerate perturbation theory to determine the effect of the superlattice potential.

Without loss of generality, let $\mathbf{Q}$ be oriented in the $x$-direction (there may be symmetry-related reciprocal lattice vectors of the same length in other directions).
The effective Hamiltonian to first order in degenerate perturbation theory and linear order in $\mathbf{k}$ near $\pm \mathbf{Q}/2$ is
\begin{equation}
    H_{\mathbf Q/2} = \begin{pmatrix} \tfrac{v}2 Q + v k_x &  V_{\mathbf Q} \braket{e^{i\varphi_{\mathbf k + \mathbf Q/2}} | e^{i\varphi_{\mathbf k - \mathbf Q/2}}} \\  V_{\mathbf Q}^* \braket{e^{i\varphi_{\mathbf k -\mathbf Q/2}} | e^{i\varphi_{\mathbf k + \mathbf Q/2}}} & \tfrac{v}2 Q - v k_x \end{pmatrix},
\end{equation}
given in the basis $\ket{\mathbf{Q}/2+\mathbf{k},+}, \ket{-\mathbf{Q}/2+\mathbf{k},+}$.
The diagonal terms come from expanding $v\mathbf{\mathbf{k}} \cdot\sigma$ to linear order in $\mathbf{k}$ and the off-diagonal terms come from the effective potential in Eq.~(\ref{eq:defH2D}).
We are measuring the energy relative to any potential constant energy shift $V_{\mathbf{Q} = \mathbf{0}}$.
We evaluate the off-diagonal matrix elements to linear order in $\mathbf k$ using the definition of $\ket{e^{i\phi}}$ in Eq.~(\ref{eq:spinor}):
\begin{equation}
    \braket{e^{i\varphi_{\mathbf k + \mathbf Q/2}} | e^{i\varphi_{\mathbf k - \mathbf Q/2}}} = 2 i \frac{k_y}{Q},
\end{equation}
yielding the simplified Dirac Hamiltonian
\begin{equation}
    H_{\mathbf Q/2} = \begin{pmatrix} \tfrac{v}2 Q + vk_x &  2 i V_{\mathbf Q} \tfrac{k_y}{Q} \\  -2 i V_{\mathbf Q}^*\tfrac{k_y}{Q}  & \tfrac{v}2 Q - vk_x \end{pmatrix},
    \label{eq:HQ2}
\end{equation}
which describes an anisotropic SDC at energy:
\begin{equation}
    E^{\mathrm{sat}} = \tfrac{v}{2} Q + O(|V_{\mathbf{Q}}|^2) ,
    \label{eq:Esat2}
\end{equation}
with anisotropic velocities:
\begin{equation}
    v_{\parallel} = v + O(|V_{\mathbf{Q}}|^2), \quad v_{\perp} = \tfrac{2|V_{\mathbf Q}|}{Q} +  O(|V_{\mathbf{Q}}|^2).
    \label{eq:vQ2}
\end{equation}
Eqs.~(\ref{eq:Esat2}) and (\ref{eq:vQ2}) describe the energy and dispersion of the SDCs closest in energy to the original Dirac cone.
Eq.~(\ref{eq:vQ2}) applies to $\mathbf{Q}$ in any direction, where the subscripts denote the velocities parallel and perpendicular to $\mathbf{Q}$.
It shows that when exactly two momenta are coupled by the moir\'e potential, the resulting SDC is anisotropic.
The anisotropy results because $\mathbf{Q}$ defines a preferred axis.

We now turn to the next lowest energy SDC (e.g. on the square lattice this corresponds to $M_1$ in Fig.~\ref{fig:satelliteQ}).
If the lattice has an $N$-fold rotational symmetry, the next lowest energy SDC occurs from coupling $N$ degenerate states at momenta symmetrically positioned around the original Dirac cone, such that neighboring points are connected at first order in perturbation theory by the superlattice potential.
This is shown for $N=4$ (corresponding to a square lattice) in Fig.~\ref{fig:satelliteQ}; only the cases $N=2,3,4$ or $6$ can occur in crystals.
Together, the $N$ momenta make an $N$-regular polygon, with side $Q$ and distance from center $k_0$ (in the simplest case, $k_0 = Q/(2\sin(\pi/N))$, but this is not the only possibility).
We label the $N$ states at these momenta as 
$\ket{n} \equiv \ket{k_0(\cos\frac{2\pi n}{N}, \sin \frac{2\pi n}{N}),+}$,
using the notation of Eq.~(\ref{eq:kspinor}), where $n$ is defined mod $N$.
This yields the Hamiltonian to first order in degenerate perturbation theory in $V$:
\begin{equation}
    H_N = \sum_{n=0}^{N-1} V_{\mathbf Q_n} \braket{ e^{2\pi i (n+1)/N} | e^{2\pi i n/N}} \ket{n+1}\bra{n} + \mathrm{H.c.}. \label{eq:HN-degenpert}
\end{equation}
Since the set of $\mathbf{Q}_n$ are related by symmetry, all $V_{\mathbf{Q}_n}$ have the same magnitude, $W \equiv |V_{\mathbf Q_n}|$.
Further, we proved in Sec.~\ref{sec:qpderivation} that $V_\mathbf{Q} = V_{-\mathbf{Q}}^*$; thus, if $N$ is even, $\prod_n V_{\mathbf Q_n} = W^N$.
Consequently, the phases of $V_{\mathbf{Q}_n}$ can be eliminated by a gauge transformation.
The matrix element is evaluated by using Eq.~(\ref{eq:spinor}):
\begin{equation}
    \braket{ e^{2\pi i (n+1)/N} | e^{2\pi i n/N}} = e^{-i\pi/N} \cos(\pi/N).
    \label{eq:N-overlap}
\end{equation}
Restricting ourselves to $N$ even ($N$ odd is discussed in Appendix~\ref{sec:Nodd}), the Hamiltonian simplifies to
\begin{equation}
    \tilde H_N = W\cos(\pi/N) \sum_{n=0}^{N-1} e^{-i\pi/N} \ket{n+1}\bra{n} + \mathrm{H.c.}.
    \label{eq:HN}
\end{equation}
This is a tight-binding model enclosing a $\pi$-flux from the central Dirac node.
It is invariant under the $N$-fold rotation symmetry $C_N: \ket{n} \mapsto \ket{n+1}$.
Therefore, its eigenstates are
\begin{equation}
    \ket{j} = \frac1{\sqrt{N}}\sum_{n=0}^{N-1} e^{2\pi i j n/ N} \ket{n},
    \label{eq:defj}
\end{equation}
where $j$ is defined mod $N$.
The state $\ket{j}$ is an eigenvector of $C_N$ with eigenvalue $e^{-2\pi i j/N}$ and has energy:
\begin{equation}
    E^{\mathrm{sat}}_j  = v k_0 + 2W \cos(\pi/N) \cos(\tfrac{2\pi(j + 1/2)}{N}).
    \label{eq:EsatN}
\end{equation}
Eq.~(\ref{eq:EsatN}) shows that the states $\ket{j}$ and 
$\ket{-1-j}$ are degenerate.
According to Eq.~(\ref{eq:TRspinor}), these states are time-reversed partners.
Thus, the degeneracy is protected to all orders in perturbation theory.
Each time-reversed pair forms the degenerate point of a gapless SDC.

To determine the velocity of the SDCs, we include a small perturbation $\mathbf k$ to the states $\ket{n}$.
The derivation can be found in Appendix~\ref{sec:SDCvelocity}.
To linear order in $W$, the velocities of the Dirac cones formed by the pairs $\ket{0},\ket{N-1}$ and $\ket{N/2-1},\ket{N/2}$ are given by:
\begin{equation}
    v_{\pm}^\mathrm{sat} \equiv 
    \begin{cases}
        \tfrac v2 + \tfrac{W}{k_0} \sin^2(\pi/N) & \text{ for } \ket{0},\ket{N-1} \\
        \tfrac v2 - \tfrac{W}{k_0} \sin^2(\pi/N) & \text{ for } \ket{N/2-1},\ket{N/2}\\
        %0 & \text{ else.}
    \end{cases}
        \label{eq:vsatGamma}
\end{equation}
Eq.~(\ref{eq:vsatGamma}) shows that unlike the lowest-energy Dirac cones derived in Eqs.~(\ref{eq:Esat2}) and (\ref{eq:vQ2}), the Dirac cones that form from a set of $N>2$ degenerate $\mathbf{k}$ points are isotropic.
This isotropy is required to all orders in $W$ due to the combination of time reversal and $N$-fold rotational symmetry (proof in Appendix~\ref{sec:isotropic}).

Eq.~(\ref{eq:vsatGamma}) only shows the velocity of two Dirac cones.
However, for $N>4$, there are $N/2-2$ degenerate pairs whose velocity is not shown in Eq.~(\ref{eq:vsatGamma}): these remaining degenerate cones have zero velocity to linear order in $W$ (proof in Appendix~\ref{sec:SDCvelocity}).
We expect these cones develop a non-zero dispersion to higher order in $\mathbf{k}$.

In Appendix~\ref{sec:higherpert} we use Green's functions to extend the degenerate perturbation theory to arbitrarily high order. 
We use the results at higher orders in perturbation theory to compare the theory we have developed in this section with lattice model simulations in the following section.

\section{Twisting the surface of a 3D TI}
\label{sec:toymodel}

We now consider a bulk tight-binding model of a 3D TI and study the effect of an induced moir\'e superlattice potential [derived in Eq.~\eqref{eqn:subpot}] on its surface in a numerically exact fashion. 
The numerical results are well described by our perturbative theory when high enough orders are considered. 
Our results demonstrate that once SDCs are produced, they can be made quite flat, which in some cases yields a corresponding magic-angle condition perturbatively. 
The flat SDCs produce a large enhancement of the surface density of states, which raises the exciting possibility of twist induced weak coupling instabilities on the surface.

\subsection{Model}

We consider the 3D generalization of the Bernevig-Hughes-Zhang model \cite{Bernevig2006quantum} on a simple cubic lattice, given in real space by:
\begin{align}
    H_{TI} & = \sum_{\mathbf{r},\mu=x,y,z} \left( \frac{i}{2} t_\mu \psi_\mathbf{r}^\dagger \alpha_\mu \psi_{\mathbf{r} + \hat{\mu}} - \frac{1}{2} m_2 \psi_\mathbf{r}^\dagger\beta\psi_{\mathbf{r}+\hat{\mu}} + {\rm h.c.} \right)
    \nonumber
    \\
    & + \sum_\mathbf{r} \psi_\mathbf{r}^\dagger \left[ (m_0 + 3m_2) \beta  \right] \psi_\mathbf{r},
    \label{eq:HTI}
\end{align}
where $\psi_\mathbf{r}$ is a four component spinor made of electron annihilation operators $c_{\mathbf{r},\tau,s}$ at site $\mathbf{r}$ with parity $\tau=\pm$, spin $s=\uparrow/\downarrow$, and the matrices $\alpha_\mu$ and $\beta$ are given by:
\begin{equation}
\alpha_\mu = \tau_x \otimes \sigma_\mu , \quad \beta = \tau_z \otimes \sigma_0.
\end{equation}
Time-reversal symmetry is implemented by ${\rm TR} = \tau_0 \otimes i\sigma_y K$.
We consider the parameters $t_\mu =t= 1, m_0 = -1, m_2=1$, for which Eq.~(\ref{eq:HTI}) describes a 3D TI.

\begin{figure}
\includegraphics[width=\columnwidth]{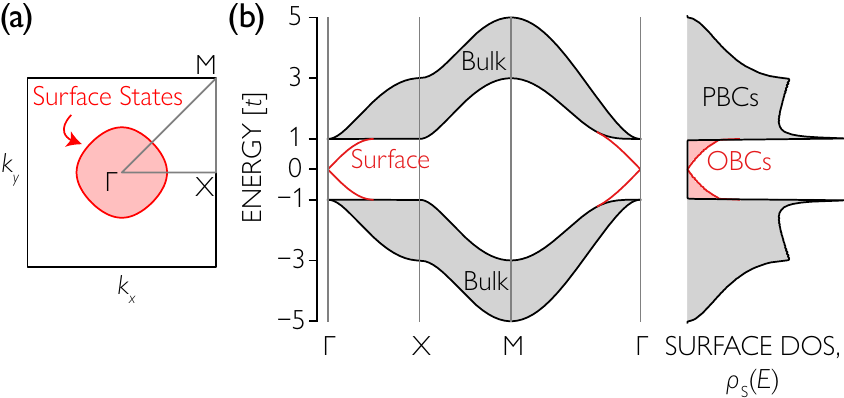}
\caption{{\bf Surface state properties of the 3D TI lattice model in Eq.~\eqref{eq:HTI}}. (a) A depiction of the full surface Brillouin zone and the region where surface state solutions exist is marked in red for the model parameters considered here $t=1, m_0=-1,m_2=1$. (b, left) The band-structure depicting bulk states (grey) and the surface Dirac cones (red) along the high-symmetry cut shown in (a). (b, right) The corresponding surface density of states in the layer $z=L$ for periodic and open boundary conditions (PBCs and OBCs, respectively) along the $z$-direction.}
    \label{fig:cleanDOS}
\end{figure}

With open boundary conditions in the $z$-direction, this band structure gives rise to a surface Dirac cone at the Gamma point in the surface Brillouin zone on the two open faces of the slab.
Following Ref.~\cite{mong2011edge}, the surface state solutions only exist in the regime $|m_0+3m_2-m_2(\cos k_x + \cos k_y)|<|m_2|$. In this reduced part of the surface Brillouin zone, the surface states have a dispersion $E_S(k_x,k_y)=\pm t \sqrt{\sin^2 k_x+ \sin^2 k_y}$ and are exponentially bound to the surface. In the low energy regime near the $\Gamma$ point the surface dispersion is a Dirac cone $E_S(k_x,k_y)\approx \pm v_0\sqrt{k_x^2+k_y^2}$ with velocity $v_0=t$. These  surface states are demonstrated in Fig.~\ref{fig:cleanDOS}.

\subsection{Approach}
The topological surface states can be clearly seen by computing the surface density of states in  the top layer $z=L$ [denoted as the surface layer $S(z=L)$], which is defined as
\begin{align}
    \rho_S(E) &= \frac{1}{L^2}\sum_{{\bf r}\in S(z=L)}\rho_{{\bf r}}(E),
    \\
    \rho_{{\bf r}}(E) &= \sum_{n,\tau,s} |\langle n | {\bf r},\tau,s \rangle |^2 \delta(E-E_n),
\end{align}
where $E_n$ is an energy eigenvalue, $| n\rangle$ is the corresponding eigenstate, $\rho_{{\bf r}}(E)$ is the local density of states at site ${\bf r}$ summed over the internal states of sub-lattice ($\tau$) and spin ($s$), and $ | {\bf r},\tau,s \rangle=\psi_{{\bf r},\tau,s}^{\dag}|0\rangle$. 
We compute the layer-resolved density of states using the kernel polynomial method (KPM)~\cite{RevModPhys.78.275} and take advantage of the stochastic trace method projected onto the layer $S(z)$~\cite{Wilson-2018}.
Here we focus on the top surface with $z=L$.

To track the original surface Dirac cone at the $\Gamma$ point [labeled $\Gamma_0$ in Fig.~\ref{fig:satelliteQ}(b)] we use the scaling of the surface density of states for an isotropic Dirac cone
\begin{equation}
    \rho_S(E\approx E_{\Gamma_0})=\frac{1}{2\pi (v_{\Gamma_0})^2}|E-E_{\Gamma_0}| + O(|E-E_{\Gamma_0}|^2).
    \label{eqn:rhoSED}
\end{equation}
As shown in Fig.~\ref{fig:cleanDOS},  we find the low energy density of states scales like $\rho_S(E) \sim v_{\Gamma_0}^{-2}|E|$, characteristic of a two-dimensional Dirac semimetal with the expected velocity  $v_{\Gamma_0}=v_0=t$ and energy $E_{\Gamma_0}=0$. 
To show the low energy states are surface states, we compare this calculation with one that has periodic boundary conditions in layer $z=L$. 
As shown in Fig.~\ref{fig:cleanDOS}(b), we find a clear insulating gap in the case of periodic boundary conditions.
We deduce that the gapless states correspond to topological surface states.

As we will discuss in detail below, we are also able to track the SDC at $\Gamma_{1}$ [see Fig.~\ref{fig:satelliteQ}(b)] through a similar scaling of the surface density of states in Eq.~\eqref{eqn:rhoSED} and extract its energy $E^{\mathrm{sat}}_{\Gamma_1}$ and its velocity $v^{\mathrm{sat}}_{\Gamma_1}$. 
These expressions were derived more generally, perturbatively, in Eqs.~\eqref{eq:EsatN} and \eqref{eq:vsatGamma} (the ``$-$'' case).

On the other hand, for the anisotropic SDCs [e.g.\ in Eq.~\eqref{eq:vQ2}] the expression in Eq.~\eqref{eqn:rhoSED} does not apply because anisotropic Dirac cones produce a contribution to the density of states at low energy that goes like $(2\pi v^{\mathrm{sat}}_{\perp} v^{\mathrm{sat}}_{\parallel})^{-1}|E-E^{\mathrm{sat}}| + O(|E-E^{\mathrm{sat}}|^2)$.
Further, for the model parameters we have investigated, we always find that the anisotropic Dirac cones coexist with metallic bands at a similar energy and thus do not appear as a true psedugogap; i.e.\ there is not a vanishing surface density of states as $E \rightarrow E^{\mathrm{sat}}_{X_1,M_1}$ [see Fig.~\ref{fig:satelliteQ}(b) for SDC labels]. Therefore their velocity cannot be reliably extracted from the surface DOS.

\begin{figure}
\includegraphics[width=0.8\columnwidth]{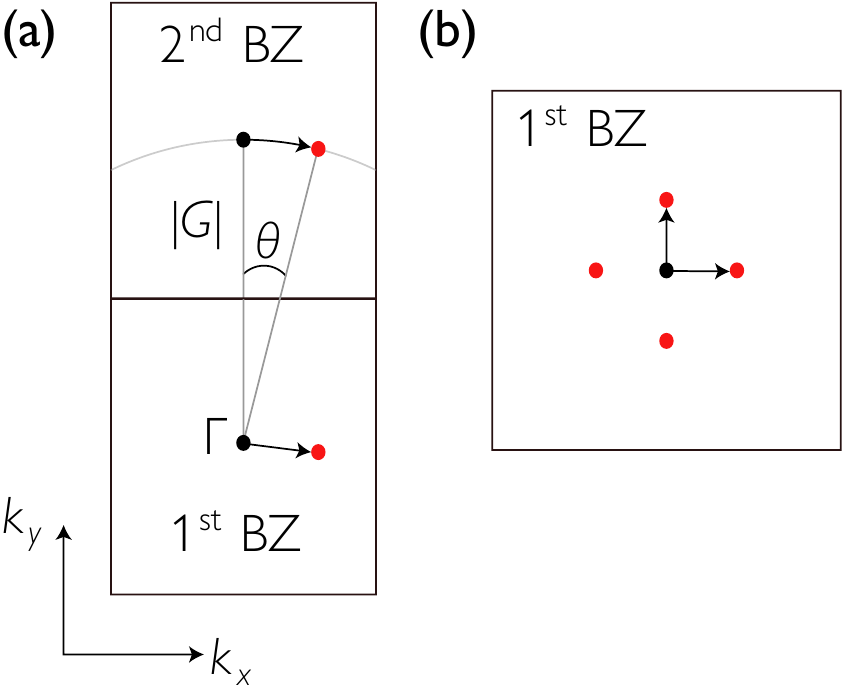}
\caption{{\bf Momentum exchange of the surface Dirac cone in the first and second BZ.} (a) The Gamma point (black circle) in the first and second Brillouin zone (BZ) is scattered by a momentum exchange ${\bf Q}$ to the red circle in the second BZ. (b) The scattered momentum point (red) is shifted back into the first BZ and its angular rotation about the $C_4$ axis is neglected, as described in the main text.}
    \label{fig:2BZ}
\end{figure}

Thus, for a full numerical treatment of the SDCs $X_1, M_1,$ and $\Gamma_1$ [labeled in Fig.~\ref{fig:satelliteQ} (b)]  we perform a calculation of the band structure treating the entire system as a supercell, twisting the boundary conditions in the $x$ and $y$ directions, and obtaining the surface energy eigenstates through Lanczos.
We filter the eigenstates by weight on the surface with the potential to avoid contamination of states from the other surface.
The velocities are then extracted through numerical twist derivatives of the energy eigenstates.

\subsection{Surface moir\'e potential}

We now consider a heterostructure consisting of a 2D material layered on top of a lattice-matched 3D TI with a relative twist angle.
As derived in Sec.~\ref{sec:qpderivation}, the twisted heterostructure induces a potential that acts only on the top layer of the 3D TI,
\begin{equation}
H_{\mathrm{2D}} = \sum_\mathbf{r} \psi_\mathbf{r}^\dagger V_{\mathrm{2D}}(\mathbf{r}) \psi_\mathbf{r} \delta_{z,L},
\end{equation}
where the potential $V_{\mathrm{2D}}(\mathbf{r})$ is given in Eq.~\eqref{eqn:subpot} without specifying any details about the form of $\mathcal{T}(\br)$.
We now derive an effective 2D potential specific to the case of a square lattice with a small twist angle.

Focusing on the C$_4$ symmetry of the lattice,  a rotation of the surface Dirac cone at the surface Gamma point will perturbatively induce scattering with itself at some order in the potential. Each scattering process can be considered as a hop in momentum space; for enough hops  the Dirac cone at Gamma will mix with a state in the second (rotated) Brillouin zone (BZ), which defines the wavectors $\bQ_i$ in Eq.~\eqref{eqn:subpot}. As shown in Fig.~\ref{fig:2BZ}(a), for a twist $\theta$, this gives rise to a wavevector  with magnitude
\begin{equation}
    |\bQ_i | \equiv Q= 2|{\bf G}|\sin(\theta/2),
    \label{eq:qtheta}
\end{equation}
where $|{\bf G}|=2\pi$ is the magnitude of a reciprocal lattice vector. We ignore the slight rotation in the wavector (which introduces a correction of order $\sim\theta^2$) and take
\begin{equation}
   {\bf Q}_i/|{\bf Q}| = 0, \pm \hat x, \pm \hat y
\end{equation}
as the  momentum transfer due to the 2D layer, as shown in Fig.~\ref{fig:2BZ}(b). We expect that the effect of neglecting this additional angular dependence for small $\theta$ is negligible.
For simplicity, we assume the scattering matrices $T_\mathbf{Q}$ are diagonal in spin and sublattice space.
Taking $T_{\mathbf{Q}=0} = W_0\mathbb{I}$ and $T_{\mathbf{Q} = \pm |\mathbf{Q}|\hat{x},\pm |\mathbf{Q}|\hat{y}} = W_1\mathbb{I}$, Eq.~(\ref{eqn:subpot}) yields the potential:
\begin{widetext}
\begin{align}
    V_{\mathrm{2D}}(\br) & = \mu_V + \frac{1}{\Delta}\left(4W_0 W_1[\cos(Qx) + \cos(Qy)]
    + 2W_1^2(\cos(2 Qx)+\cos(2 Qy)
    +2[\cos(Q(x-y))+\cos(Q(x+y))] ) \right),
    \label{eqn:square-subpot}
\end{align}
\end{widetext}
where $\mu_V = (W_0^2 + 4W_1^2)/\Delta$ is a constant chemical potential shift on the surface. To account for the fact that the origin of rotation is random, we add two random phases $\phi_x$ ($\phi_y$) to each term $Q x$ ($Qy$) such that $Qx \rightarrow Qx + \phi_x$ ($Qy \rightarrow Qy + \phi_y$). Thus, the induced potential contains a surface chemical potential contribution and an incommensurate potential modulating with a wavector $Q$ (determined by the twist angle) that is composed of three harmonics.
In the following, for simplicity, we fix the interlayer tunneling to  $W_0=W_1\equiv W$. In the subsequent numerical calculations we average over 100 realizations of different phases $\phi_x$ and $\phi_y$ sampled independently from $[0,2\pi]$. 
To reduce finite size effects we also average over twisted boundary conditions in the $x$- and $y$-direction.

\begin{figure*}
\centering
\includegraphics[]{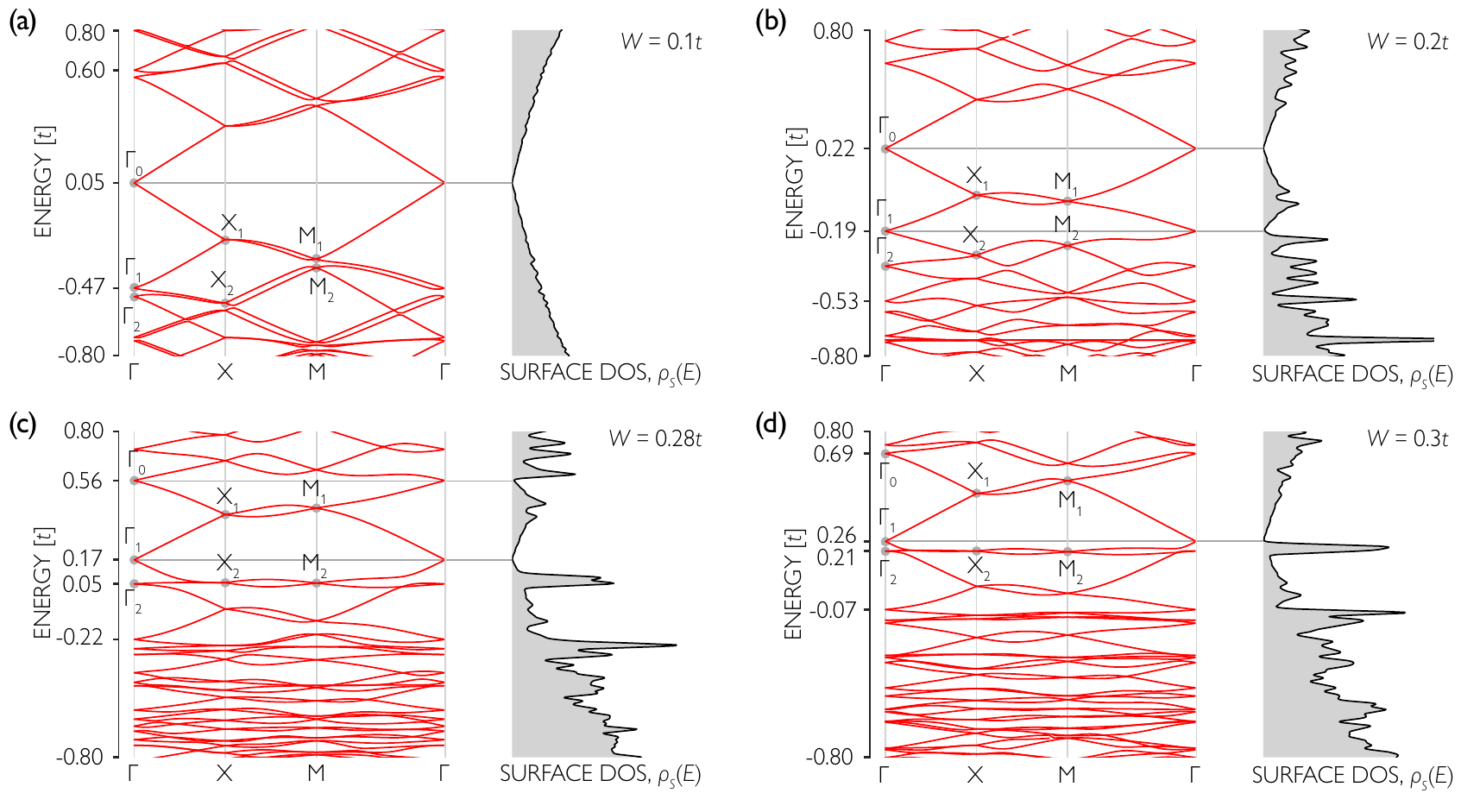}
\caption{{\bf Surface dispersion and density of states due to the surface potential on the 3D TI lattice model in Eq.~\eqref{eq:HTI}.} This data is obtained on a cubic system of linear size $L=89$ with the moir\'e surface potential in Eq.~\eqref{eqn:square-subpot} with a twist parameterized by the wave vector $Q=2\pi 8/89$ [see Eq.~\eqref{eq:qtheta}]. We use Lanczos with twisted boundary conditions in the $x-y$ plane to compute the surface dispersions whereas the surface DOS is computed using KPM with an expansion order $N_C=2^{11}$. A selection of relevant SDCs are labeled by $\Gamma_{0,1,2}$, $X_{1,2}$ and $M_{1,2}$ depending on where in the mini-Brillouin zone they appear. (a) At $W=0.1t$, the Dirac cone is only slightly perturbed. The dispersion reveals the formation of SDCs but the surface DOS is relatively unaffected. (b) At $W=0.2t$, the SDC $M_2$ moves below $\Gamma_1$ in energy, revealing it in the surface DOS. (c) At $W=0.28t$, both $\Gamma_0$ and $\Gamma_1$ are visible in the surface DOS. (d) at $W=0.3t$, $\Gamma_0$ becomes obscured by higher bands and the band associated with $\Gamma_2$ has become flat, causing a large increase in the density of states.}
 \label{fig:disanddos}
\end{figure*}

\subsection{Perturbative results}
\label{sec:ptlattice}
Applying the surface perturbation theory of Sec.~\ref{sec:continuum} (generalized in Appendix~\ref{sec:higherpert} to higher orders) we obtain the description for the renormalized Dirac cone at $\Gamma_0$ and the SDC at $\Gamma_1$ [see the labeling for $\Gamma_0$ and  $\Gamma_1$ in Fig.~\ref{fig:satelliteQ}(b)]. 
In Appendix~\ref{sec:ptlatticeA} we list our full results, including those for $\Gamma_{-1}$ and $\Gamma_2$, to fifth order in the parameter $\alpha \equiv W^2/(\Delta v_0 Q)$. 
Here, we restrict ourselves to  expressions up to order $\alpha^3$, for brevity. 
For $\Gamma_0$, we obtain the renormalized velocity
\begin{equation}
    v_{\Gamma_0} = \frac{v_0}{1+25\alpha^2},
    \label{eq:vmodel}
\end{equation}
which is exactly Eq.~(\ref{eq:vstar}) applied to this model at the shifted Dirac node energy
\begin{equation}
    E_{\Gamma_0}  = \mu_V + \frac{80\alpha^3}{1+25\alpha^2},
    \label{eq:modelshift}
\end{equation}
where for our parameter choice $v_0 = t$ and $\mu_V=5W^2/\Delta$.
Eq.~(\ref{eq:vmodel}) shows that the original ($\Gamma_0$) Dirac cone velocity is decreased by increasing the tunneling strength, $W$, or by decreasing the superlattice reciprocal lattice vectors, $\mathbf{Q}$, since $\alpha \propto W^2/Q$. 
However, Eq.~(\ref{eq:modelshift}) shows that $\Gamma_0$ shifts up in energy at the same time as its velocity is decreasing.
Eventually, as we show numerically in Fig.~\ref{fig:disanddos}, the renormalization of $\Gamma_0$ becomes obscured as it shifts into higher energy bands.

However, Fig.~\ref{fig:disanddos} also shows that as $\Gamma_0$ becomes hidden in higher energy bands, the SDC $\Gamma_1$ (see labelling in Fig~\ref{fig:satelliteQ}(b)) moves through the Fermi level (from negative to positive energy). Ultimately, it is $\Gamma_1$ that 
possesses a true pseudogap and semimetallic behaviour, away from other bands near the Fermi level. At large enough $W$ the SDCs directly below $\Gamma_1$, labelled as $\Gamma_2$, $X_2$, and $M_2$, merge into a very flat band that contributes to a large peak in the DOS.

Applying perturbation theory to $\Gamma_1$ yields its energy 
\begin{multline}
    E_{\Gamma_{1}}^{\mathrm{sat}}=-v_0\sin(Q) + \mu_V 
    \\
    +
    Qv_0Z_{\Gamma_1}\Big(2\alpha + 31\alpha^2 + \frac{497}{2}\alpha^3 \Big),
\end{multline}
and velocity
\begin{equation}
     v_{\Gamma_{1}}^{\mathrm{sat}}/v_0= \frac{Z_{\Gamma_1}}{4} \left( 2\cos(Q) - 6\alpha + 195\alpha^2 + 2175\alpha^3 \right),
     \label{eqn:vsatlat}
\end{equation}
where the quasiparticle residue is given by
\begin{equation}
    Z_{\Gamma_{1}}^{-1} = 1 + 183\alpha^2/2 + 1977\alpha^3/2.
\end{equation}
In the above, we have included Dirac cone curvature corrections only in the $O(1)$ terms; the rest of the terms assume a perfectly linear cone.
While $\Gamma_0$ does not have an accessible magic-angle condition (i.e.\ where its velocity vanishes) at leading order,
the perturbative expression for the velocity of $\Gamma_1$ can vanish, although our exact results in the next section (Fig.~\ref{fig:SatelliteEnergies}) show that we do not probe this parameter regime.

It is important to note that the diagrammatic perturbation theory neglects surface-bulk scattering processes, which are certainly present in the model. 
Therefore, we now compare these perturbative results with numerically exact results where we extract the energy and velocity of SDCs from the surface density of states.

\subsection{Tuning the interlayer tunneling}
\label{sec:tuneW}

We first consider the effect of varying the interlayer tunneling strength at a fixed $Q$ in Eq.~\eqref{eqn:square-subpot}. 
This is practically more straightforward than varying $Q$, which requires careful consideration of the boundary conditions (and which we consider in the next section).
As illustrated by the perturbative calculations in the previous section, decreasing $Q$ or increasing $W$ alters the surface spectrum in a similar way because the physics is largely determined by the parameter $\alpha = W^2/(\Delta v_0 Q)$.

In the following we take a commensurate approximate for $Q=2\pi F_{n-5}/F_n$ where $L=F_n$ is the $n$th Fibonacci number and consider cubic system sizes of $L=89=F_{11}$ for density of states and $L=13=F_7$ for dispersions. 
We set the energy difference $\Delta=t$.
Despite the simplicity of our model, it displays many similarities with the ab initio calculation in Sec.~\ref{sec:abinitio}.

As demonstrated in Fig.~\ref{fig:disanddos}, as we increase the tunneling strength $W$, the surface dispersion and the surface density of states become strongly renormalized. 
The original Dirac cone ($\Gamma_0$) shifts and its velocity decreases. 
We are able to clearly track it in the density of states until $W=0.3t$ where the density of states becomes finite at the renormalized Dirac node energy. The SDC at $\Gamma_1$ remains visible in the surface DOS throughout. 
At $W=0.1t$ [Fig.~\ref{fig:disanddos} (a)] we can also see the formation of SDCs that are not visible in the surface DOS appearing at $X_{1,2},M_{1,2}$ and $\Gamma_{1,2}$ as labelled in Fig.~\ref{fig:satelliteQ}(b).

\begin{figure*}
    \centering
    \includegraphics{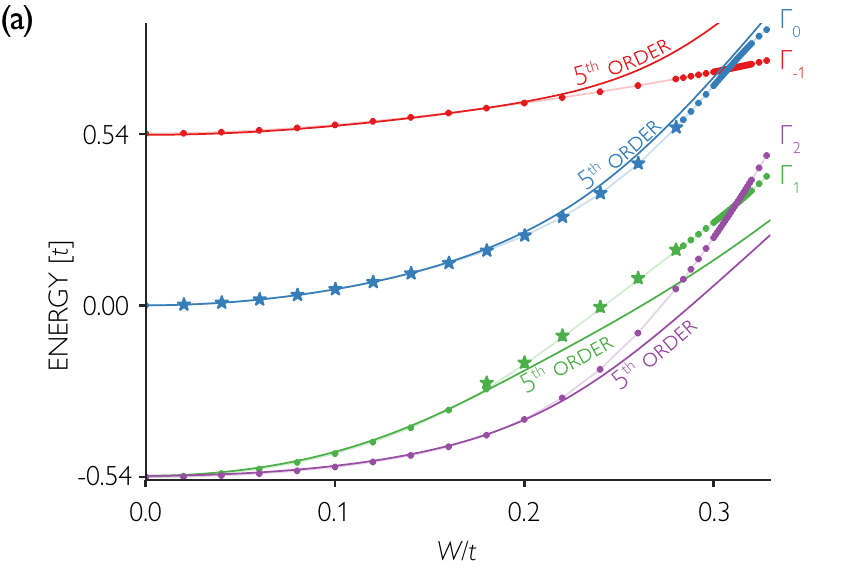}
        \includegraphics{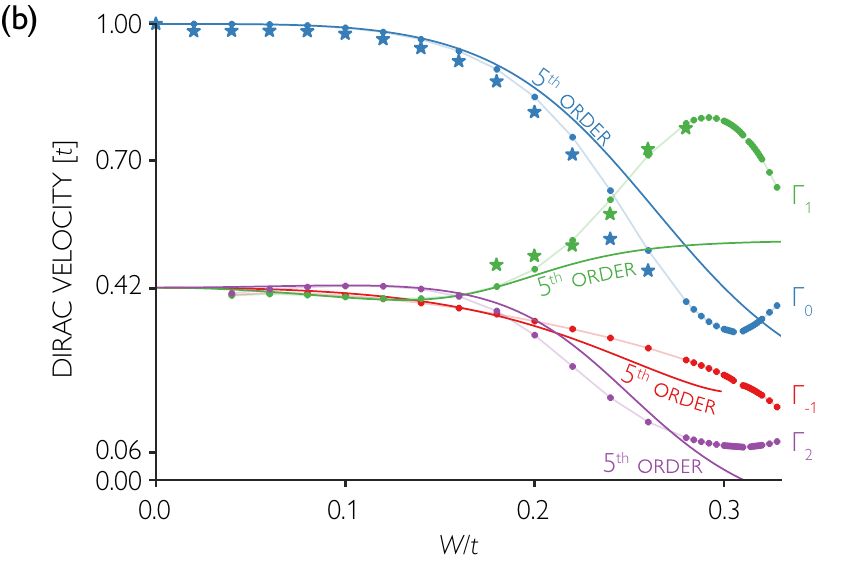}
    \caption{{\bf Comparison of surface perturbation theory and numerically exact results for the model in Eq.~\eqref{eq:HTI} with a surface moir\'e potential from Eq.~\eqref{eqn:square-subpot} characterized by the wavevector $Q\approx 4\pi /(11+5\sqrt{5})$ for various SDCs at the $\Gamma$ point.}  Dirac point location (a) and corresponding velocity (b).
    Dots indicates values extracted from dispersion curves on the surface of a TI as a function of the interlayer tunneling $W$. 
    The darker lines labeled ``5th order'' indicate perturbative results for that Dirac point computed to fifth order (see Sec.~\ref{sec:ptlattice} and Appendix~\ref{sec:ptlatticeA}), which are in good agreement with our numerically exact results. Perturbatively, $\Gamma_2$ has a magic angle condition; however this feature is rounded out and the velocity remains finite in the exact calculation. Notice that Dirac points cross without any level repulsion, as explained in the text.
    When the energies cross in (a) the velocity calculation becomes unreliable (taken as finite differences on the dispersion curve), so we have omitted those points from the curves.
    Results are obtained for a linear system size $L=11$ ($Q=2\pi/11$) using Lanczos.
    The points indicated with $\star$ are extracted from the surface DOS from fitting the data to Eq.~\eqref{eqn:rhoSED} with a linear system size $L=89$ ($Q=2\pi 8/89$) and agree well with the results obtained from the dispersion.}
    \label{fig:SatelliteEnergies}
\end{figure*}

As we increase $W$ in Fig.~\ref{fig:disanddos}(b) the SDC at $\Gamma_1$ opens a pseudogap in the surface DOS at negative energy,  for $W\ge 0.2 t$. 
In contrast, we always find that the SDCs at $X_{1,2}$ and $M_{1,2}$ are subleading to nearby metallic bands; while they do not display a pseudogap, they are responsible for some of the non-trivial structure of peaks and dips in the surface DOS. 
The locations of the SDCs move monotonically in energy for increasing $W$ and the bands are renormalized in a non-trivial fashion. 
We find that in the vicinity of $W\approx 0.3 t$ [Fig.~\ref{fig:disanddos}(c) and (d)] the SDCs at $X_2,M_2,\Gamma_2$ become essentially flat, which induces a large enhancement of the surface DOS. 
Importantly, to reach the regime with flat SDC's does not require fine tuning as we show by tuning $Q$ in Sec.~\ref{sec:tuneQ} as well as by demonstrating that a similar phenomena occurs on the surface of Bi$_2$Se$_3$ in Sec.~\ref{sec:abinitio}.

The original Dirac cone at the $\Gamma$ point ($\Gamma_0$) moves towards positive energy upon increasing $W$. 
Applying the scaling at low energies near the Dirac point in Eq.~(\ref{eqn:rhoSED}), we extract $E_{\Gamma_0}$ and $v_{\Gamma_0}$; we similarly use the scaling near the pseudogap induced by the SDC to find $E_{\Gamma_1}$ and $v_{\Gamma_1}$.
We compare these to the perturbative results in Fig.~\ref{fig:SatelliteEnergies} (DOS-extracted values indicated by $\star$).
To access the SDCs that are not visible in the surface DOS we use twist dispersions (e.g.\ as shown in Fig.~\ref{fig:disanddos}) to compute the locations and velocities of each Dirac point that are shown in Fig.~\ref{fig:SatelliteEnergies}. 
We find good qualitative agreement between the numerical results and the perturbation theory at fifth order, demonstrating the success of our theory.
Our results show that the surface DOS is not controlled by $\Gamma_0$; instead, as $W$ is increased, a complex rearrangement of the other bands creates a Fermi surface and finite density of states on top of the original surface Dirac cone.

We now focus on the SDC at $\Gamma_2$ in Fig.~\ref{fig:SatelliteEnergies}.
The fifth order perturbative result yields a magic angle condition with a vanishing velocity near $W\approx 0.3 t$. Our exact numerical results indicate a small but non-vanishing velocity. Nonetheless, this produces a large enhancement of the surface density of states as demonstrated in Fig.~\ref{fig:disanddos} (c) and (d). 
This finding is one of our main results: the essentially flat SDC and corresponding large enhancement in the surface density of states represents an ideal starting point to search for weak coupling instabilities on the surface of a TI.

Lastly, Fig.~\ref{fig:SatelliteEnergies}(a) shows \emph{non-avoided} crossings of $\Gamma_0$ with $\Gamma_{-1}$ and $\Gamma_1$ with $\Gamma_2$.
To explain the latter case, we have shown that to arbitrarily high order in perturbation theory (see Appendix~\ref{sec:higherpert}), $\Gamma_1$ and $\Gamma_2$ are orthogonal for $N=4$ because they have different rotational eigenvalues, as defined in Eq.~\eqref{eq:defj}.
Thus, they have no level repulsion.
While this is always the case for SDCs originating from states degenerate at $W=0$, it does not explain the un-avoided crossing between $\Gamma_0$ and $\Gamma_{-1}$.

To understand this band crossing, we show that the potential $V$ does not mix these two Dirac cones.
Specifically, using Eqs.~\eqref{eq:kspinor} and \eqref{eq:defj}, we compute the overlap between $V\ket{j}$ and the states at $\mathbf{k}=0$:
\begin{equation}
\ket{j} \rightarrow \frac1{\sqrt N} \sum_n\begin{pmatrix} e^{2\pi i j n/N} \\ e^{2\pi i(j+1) n/N}  \end{pmatrix} \propto \begin{pmatrix} \delta_{j,0} \\ \delta_{j,N-1}  \end{pmatrix},
\end{equation}
where the arrow indicates $\braket{\mathbf k=0| V |j }/W$.
Therefore, the only vectors that have overlap with the original Dirac cone are $\ket{j}$ with $j=0,N-1$, which is precisely the ``$+$'' satellite cone in Eq.~\eqref{eq:vsatGamma}.
The other satellite cones do not have matrix elements with $\Gamma_0$,
which explains the  crossing between  $\Gamma_{-1}$ and $\Gamma_0$ in Fig.~\ref{fig:SatelliteEnergies}(a).

\subsection{Varying the twist}
\label{sec:tuneQ}
 In the following we take $\Delta=t$ (recall $\Delta$ is the energy difference between the top of the 2D valence band and the charge neutrality point of the Dirac cone), fix the interlayer tunneling to $W_0=W_1\equiv W=0.3 t$, and vary $Q$ through the twist $\theta$ in Eq.~\eqref{eq:qtheta}. We note that $Q$ and $W$ enter through the ratio $\alpha=W^2/(Qv_0\Delta)$ and thus the behavior we see when varying $Q$ is similar to varying $W$.

\begin{figure}
\centering
\includegraphics[]{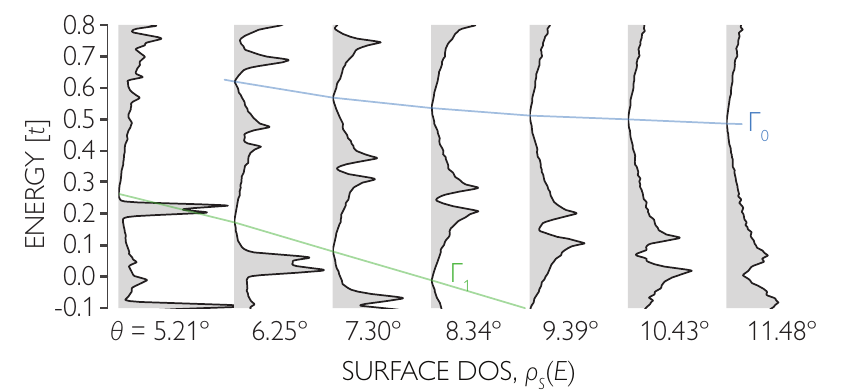}
\caption{{\bf Effect of the induced potential as a function of the twist angle on the surface DOS $\rho_S(E)$ computed from the lattice model in Eq.~\eqref{eq:HTI}.} Results for an interlayer tunneling $W_0=W_1=0.3t$ across various values of the twist value $\theta$ parameterized by $Q=2\pi m/L=4\pi \sin(\theta/2)$ for the integers $m=1,\dots,L$  with a KPM expansion order $N_C=2^{12}$ and a  system size $L=55$ averaged over 100 realizations of the potential.
The density of states as a function of energy  for various twist angles demonstrates the original Dirac cone $\Gamma_0$ is shifted to the close proximity of $\mu_V=0.45t$ and  experiences its most dramatic velocity renormalization at small angles.
The figures also reveal the satellite Dirac cone at $\Gamma_1$ where an additional ``V-shape'' appears in the DOS (i.e. a Dirac semimetal scaling $\rho(E)\sim |E-E_{\Gamma_1}|$) at lower energy. The location of the Dirac cones are marked with blue ($\Gamma_0)$ and green ($\Gamma_1$) lines. }
  \label{fig:square-subpot}
\end{figure}

In Fig.~\ref{fig:square-subpot} we show the effect of the twist on the energy dependence of the surface density of states. We parameterize $Q$ as a rational number $Q=2\pi n/L$ for $0<n\leq L$ via the possible commensurate momenta of the system, which allows us to access angles in the range $0 \leq \theta \leq 60^{\circ}$. We find that small angles (and $Q$ close to $\pi$) have the most dramatic effect on the surface Dirac cone. 

\begin{figure}
\centering
\includegraphics[]{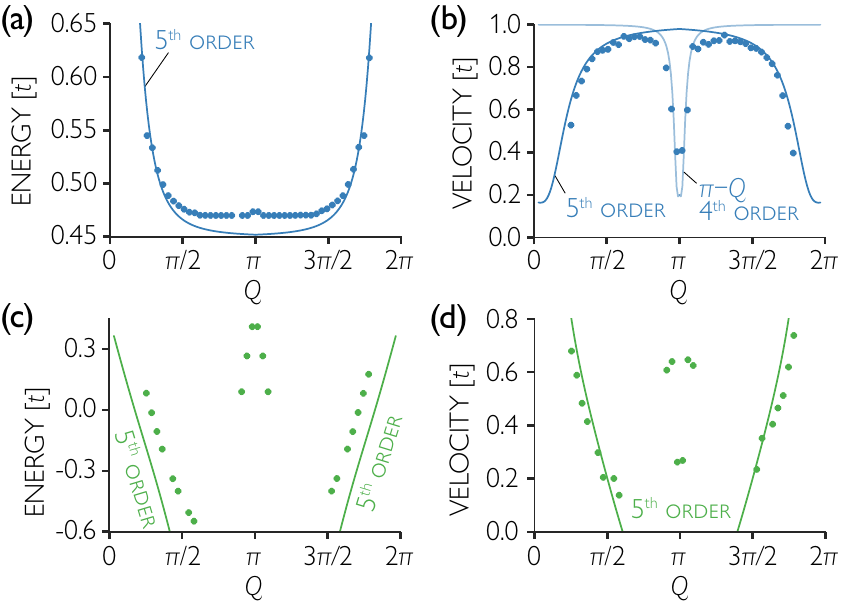}
\caption{{\bf The evolution of the original surface Dirac point at $\Gamma$ (top) and the  SDC at $\Gamma_1$ (bottom) in the the lattice model in Eq.~\eqref{eq:HTI} as a function of the twist via $Q=4\pi \sin(\theta/2)$.} This data is obtained  from the low energy scaling of the surface density of states in Eq.~\eqref{eqn:rhoSED}  on a cubic system of size $L=55$ and a KPM expansion order $N_C=2^{12}$. (a) The location in energy and (b) the velocity of the Dirac point at $\Gamma$.  (c) The location in energy and (d) the velocity of the SDC at $\Gamma_1$. Near $Q=\pi$ distinct behavior occurs due to a change in the most relevant momentum exchange processes in perturbation theory, which needs to be reformulated about $\pi-Q$. We performed this reformulation only for the velocity in (b), which shows excellent agreement. In each case, the perturbation theory qualitatively describes the numerical results.}
  \label{fig:EsatvsQ03}
\end{figure}

Our first observation is that the Dirac cone moves from zero energy to sit close to (but not at) $\mu_V (= 0.45 t)$ and persists for a large range of twist angles [see Fig.~\ref{fig:EsatvsQ03}(a)].  
As we increase the twist angle we find a moderate renormalization of the Dirac cone velocity $v_{\Gamma_0}$, shown in more detail in Fig.~\ref{fig:EsatvsQ03}(b), in good agreement with our perturbative theory. 
However, at the smallest twist angles considered, $\theta \lesssim 6^{\circ} $, we find that the surface Dirac cone scaling at $\Gamma_0$ is not clearly visible. 

The appearance of satellite Dirac cones are visible at each angle presented. 
Through a similar comparison of the twist dispersions we performed in Fig.~\ref{fig:disanddos} (not shown here), we find that the large enhancement of the DOS at small twists is due to the essentially flat bands at the SDCs $X_2$, $M_2$, and $\Gamma_2$. 
This occurs close to the pseudogap induced by the SDC at $\Gamma_1$. 
Upon increasing the twist angle, the SDC at $\Gamma_1$ renormalizes and moves down in energy, and as a result an almost pseudogap appears in between $\Gamma_0$ and $\Gamma_1$. 
We have checked that this feature is due to the SDCs at $X_1$ and $M_1$.

Here, instead of computing the twist dispersion we use the low-energy scaling of the surface density of states in Eq.~\eqref{eqn:rhoSED}
to estimate the velocity renormalization of the original Dirac cone $\Gamma_0$ and the SDC $\Gamma_1$ and compare with our perturbative results at fifth order (see Sec.~\ref{sec:ptlattice} and Appendix~\ref{sec:ptlatticeA}). 
As shown in Fig.~\ref{fig:EsatvsQ03}, we find good qualitative agreement between our theoretical predictions and the exact numerical results for both the locations $E_{\Gamma_0},E_{\Gamma_1}$ and velocities $v_{\Gamma_0},v_{\Gamma_1}$.
In addition to the fifth order perturbation theory, Fig.~\ref{fig:EsatvsQ03}(b) shows a line labeled ``$\pi-Q$ 4$^{\mathrm{th}}$ order.'' 
This expression comes from terms in Eq.~\eqref{eqn:square-subpot} that go as $2Q$ and thus when $Q\approx \pi$, they connect nearby momenta by wrapping around the Brillouin zone.
This effect can be included in the continuum calculations, and matches what is observed in the lattice model.

The lattice topological insulator model has complicated features not captured by the continuum model in Sec.~\ref{sec:continuum} (such as surface-to-bulk scattering and Dirac cone warping). Nonetheless, the continuum model not only qualitatively captures the resulting physics, but agrees quantitatively when the perturbation theory is extended to high-enough order.
To conclude, in this section we have demonstrated the clear success of our continuum theory in Sec.~\ref{sec:continuum} in describing a twisted 3D TI surface.
We now apply it to  a patterned gate potential on the surface of Bi$_2$Se$_3$.

\section{A patterned dielecric superlattice on the surface of Bi$_2$Se$_3$}
\label{sec:abinitio}
In this section, we consider realistic numerical simulations of a thin slab of the 3D TI Bi$_2$Se$_3$ on top of a patterned dielectric substrate. The Bi$_2$Se$_3$ slab displays a single Dirac cone at a surface termination. 
The modulated potential from a patterned dielectric substrate \cite{dielectric_patterning}, whose potential strength is tunable by an electric backgate, induces scattering of the surface states. 
From the effective theory point of view, this scenario is similar to the twisted topological insulator heterostructure studied in the previous section. 
However, finding material candidates to realize that heterostructure may be challenging.
The patterned dielectric substrate overcomes this challenge because it can be engineered to a custom superlattice.
Therefore, it provides a promising platform to realize tunable Dirac cone and satellite Dirac cone renormalization.
Ultimately, this may be favorable for realizing interacting states on a topological insulator surface.
In the following, we discuss how our model is derived based on the first principle calculations and then present numerical calculations of the density of states upon varying the potential strength.

\begin{figure}
    \centering
    \includegraphics{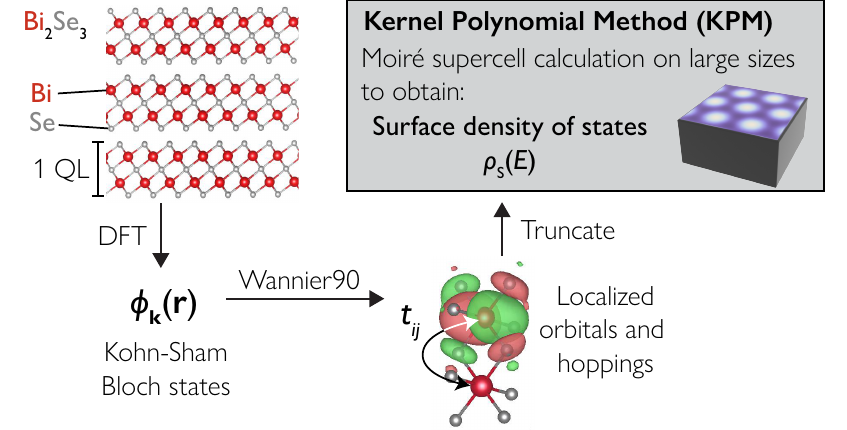}
    \caption{{\bf DFT-to-KPM pipeline for moir\'e calculations.} 
    Top left: Side view of a 3 quintuple layer (QL) Bi$_2$Se$_3$ crystal. 
    The larger, red atoms represent Bismuth atoms while the smaller, gray atoms are Selenium, and one QL is labelled.
    We use DFT to compute the Kohn-Sham Bloch states and then find the Wannier functions and hopping matrix elements (pictured in bottom right).
    Truncating this matrix (see Fig.~\ref{fig:Bi2Se3-fulldispersion}), we perform simulations on the full supercell with the KPM. 
    }
    \label{fig:Bi2Se3-schematic}
\end{figure}

\begin{figure*}
    \centering
    \includegraphics{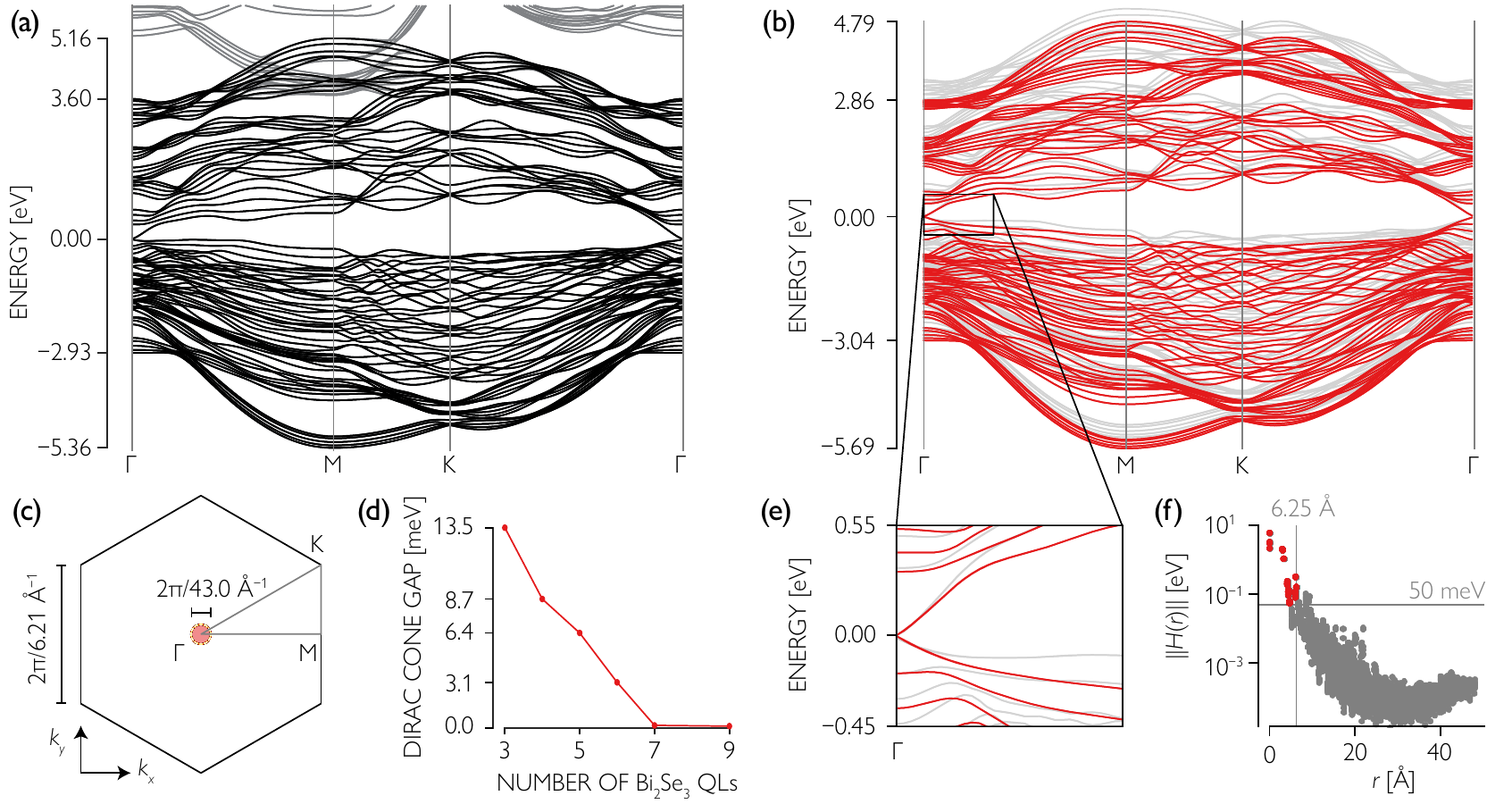}
    \caption{{\bf Truncating the Wannierized model to efficiently model Bi$_2$Se$_3$.} 
    (a) Comparing the electronic structure from the full DFT calculation (gray) to the reconstructed Wannier tight-binding Hamiltonian with spin-orbit coupling (black) for a slab with 5 QLs. 
    A basis with Bi and Se $p$ orbitals effectively captures the low-energy DFT band structure. 
    (b) The bands from the truncated tight-binding Hamiltonian (red) compared to the full Wannierization (gray), with the zoomed in view of the Dirac cone in (e). 
    (c) The 2D surface Brillouin zone of the Bi$_2$Se$_3$ slab with the cut pictured in (a,b) drawn as the gray line. The small red zone indicates crystal momenta that support surface states (as calculated by a cut $\SI{250}{meV}$ above the Dirac cone energy).
    (d) The dependence of the surface Dirac cone gap (from hybridization between the two surfaces) on the number of QLs. 
    (f) The scatter plot for hopping strength versus the bond pair distance in the tight-binding Hamiltonian, as well as the cutoffs used to construct the truncated Hamiltonian (the Hamiltonian terms retained are colored in red).
    }
    \label{fig:Bi2Se3-fulldispersion}
\end{figure*}

\subsection{Density Functional Theory and Wannier Function Analysis}
We employed density functional theory (DFT) calculations to perform electronic structure simulations for finite Bi$_2$Se$_3$ slabs (see Fig.~\ref{fig:Bi2Se3-schematic} for the crystal structure and a quintuple layer of Bi and Se as well as an overview of our numerical pipeline). 
The first principle computations are implemented in the Vienna Ab initio Simulation Package (VASP) code \cite{vasp1,vasp2}, using Projector Augmented-Wave (PAW) formalism \cite{PAW} for the pseudopotential and Perdew-Burke-Ernzerhof (PBE) parametrized exchange-correlation energy functional \cite{VASP_pbe}. 
The electronic ground state was converged with an energy cutoff of 300 eV and an $11 \times 11 \times 1$ Brillouin Zone Monkhorts-Pack sampling grid \cite{MP_grid}. 
We computed the slabs with $3 \leq L_z \leq 9$ quintuple layers (QL).
The surface cones take up a small amount of the full Brillouin zone [around $0.6 \%$, see Fig.~\ref{fig:Bi2Se3-fulldispersion}(c)] and display a small gap from the top/bottom surface hybridization, as shown in Fig. \ref{fig:Bi2Se3-fulldispersion}(d). 
In the following, we have used the 5 QL Bi$_2$Se$_3$ crystal for the simulations; its hybridization gap of 6 meV is small compared to the semi-conducting bulk gap ($\sim 300$ meV) \cite{zhang2009topological}.

We performed the Wannier transformation \cite{wannier_review} on top of the converged DFT calculations of a 5-QL slab, as implemented in the Wannier90 code \cite{mlwf,mlwf_new} (this is the second stage of our pipeline in Fig.~\ref{fig:Bi2Se3-schematic}). 
This transformation gives a real space description of the band structure in terms of localized Wannier functions derived from periodic Bloch wavefunctions. 
The Wannier transformation not only interpolates the DFT electronic structure efficiently, but also gives an atomic interpretation of the electronic properties in terms of on-site potentials and hopping terms between neighbors. 
Based on the converged DFT calculations of Bi$_2$Se$_3$ slabs, we derived a Wannier model in a basis of all $p$ orbitals on Bi and Se atoms. 
In Fig.~\ref{fig:Bi2Se3-fulldispersion}(a), we compare the electronic structure with spin-orbit coupling from the projected Wannier model (black) and full DFT calculations (gray), which show good agreement.

\subsection{Modeling the Superlattice Potential}

We now discuss how to model the potential generated from the patterned dielectric substrate \cite{dielectric_patterning} in close contact with the surface Bi$_2$Se$_3$ QL. 
The experimental technique utilizes lithography to etch the substrate material, leaving holes with a controllable pattern and size. 
When an electric bias is applied to the backgate under the patterned dielectric substrate, the electric potential at the substrate surface is modulated as well. 
The COMSOL simulations in Ref.~\cite{dielectric_patterning} show a rather smooth electric potential is generated that can be controlled by the superlattice patterning and the gating. 
Typically, the characteristic length scale is around 100 nm, with potential energy variations around 50 meV (SI in \cite{dielectric_patterning}). 
For the theoretical modeling, such a smooth electrostatic potential from the patterned dielectric substrate can be expanded with the lowest dominant harmonic components. 
Therefore, in the modeling below, we consider a substrate potential of the form
\begin{equation}
    V(\mathbf{r})= W \sum_{j=1}^3 \cos(\mathbf q_j \cdot \mathbf r + \phi_j), \quad \sum_j \phi_j = 0, \label{eq:patterned_dielectric}
\end{equation} applied to the surface Bi$_2$Se$_3$ QL.
For simplicity, we assume it is applied uniformly to all the states in that surface QL. 
We take the length scale set by the pattern to be $2\pi/|\mathbf q_j|\approx \SI9{nm}$ for most calculations and vary $W$ up to $\SI{250}{meV}$. 
Recall that much of the physics is driven by $W/|\mathbf q_j|$ so one can simultaneously increase the pattern's length scale and lower the potential size to see similar results.
This could make the experimental verification more feasible, allowing larger patterns to be machined.

Before diving into the numerical results with a patterned dielectric on the Bi$_2$Se$_3$ slab, we discuss an alternative scenario, where the superlattice potential is generated by stacking layers together in a two-dimensional van der Waals structure. 
Hexagonal boron nitride (hBN) stacks are often used as encapsulating layers to protect graphene devices. Depending on the stacks geometry, these hBN layers can alter the electronic structure of the device, as exemplified by the massive Dirac gap \cite{GhBN_gap} and miniband structure in encapsulated graphene sheets \cite{GhBN_miniband}. 
In these examples, the hBN substrate layers introduce an electrostatic potential generated from the charged ionic boron and nitrogen atoms. 
In the graphene-hBN interface, such an hBN substrate potential was studied based on first principle calculations \cite{Jung2014}. 
Although the length scale of the potential generated by hBN itself is determined by its lattice constant ($a \sim 2.5$ \AA), the length scale for the potential landscape in the graphene-hBN interface can be greatly enhanced due to the moir{\'e} pattern formed by the small lattice constant difference at a small twist angle. 
Other more complicated configurations could also lead to proximity effects that modify the electronic structure, such as spin-orbit coupling \cite{Li2019SOCproximity}. 

It is certainly interesting to consider inducing a surface potential via layer stacking on top of the Bi$_2$Se$_3$ crystal. 
However, we do not expect the hBN layers to achieve the desired effect due to the substantial difference between the lattice constants of Bi$_2$Se$_3$ and hBN. 
In Bi$_2$Se$_3$, the topological surface Dirac cone occupies only a small portion of the 2D Brillouin zone near the $\Gamma$ point.
The momentum scattering at the interface would have a length scale much larger than the Dirac cone size [see Fig.~\ref{fig:Bi2Se3-fulldispersion}(c)], preventing the surface states within the topological Dirac cone from effectively coupling to themselves by the hBN atomic potential.
Instead, to avoid scattering into the bulk, the gapped 2D material must have a
lattice constant within $10\%$ of the lattice constant of Bi$_2$Se$_3$.
This requirement comes directly from computing where the surface states exist in the Brillouin zone (Fig.~\ref{fig:Bi2Se3-fulldispersion}): any moir\'e pattern, originating from either lattice mismatch or twist, needs to have a length scale much greater than $\SI{4.3}{nm}$.

\begin{figure}
    \centering
    \includegraphics{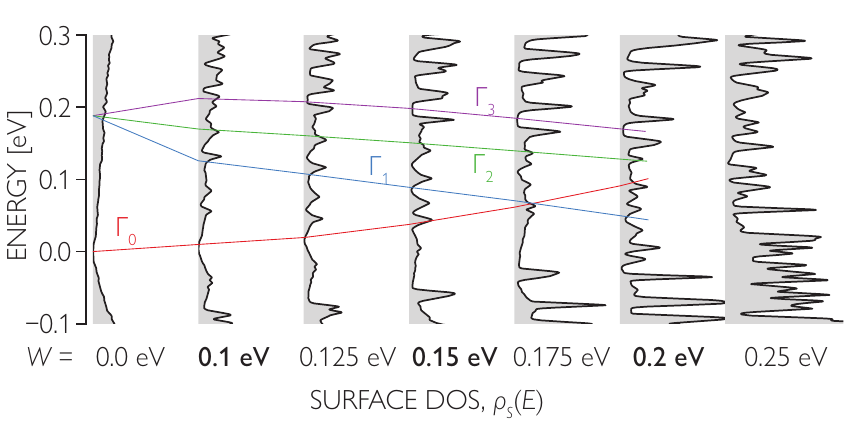}
    \caption{{\bf Effect of gating with the  patterned dielectric on the surface density of states $\rho_S(E)$ of  Bi$_2$Se$_3$.} 
    Within the bulk gap, the density of states indicates a Dirac cone (far left, $W=0$). As we tune the gating on the patterned dielectric (with length scale $\approx \SI{9}{nm}$), features consistent with SDCs emerge (red, blue, green, purple lines track SDCs at the Gamma point). For energies less than $\Gamma_0$, there is an enhancement of the DOS for moderate values of $W$.
    }
    \label{fig:DOS_Bi2Se3}
\end{figure}

\begin{figure}
    \centering
    \includegraphics{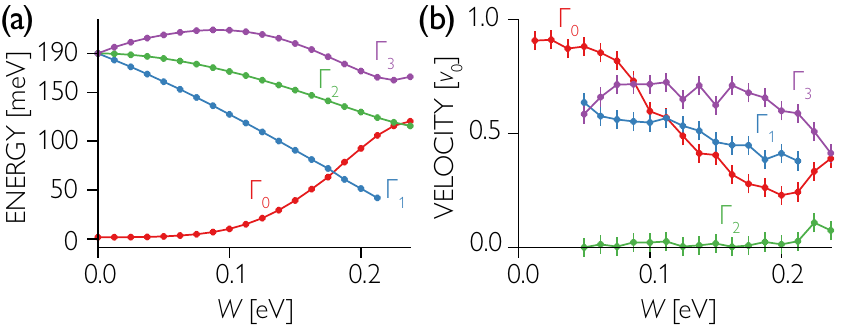}
    \caption{{\bf Tracking SDCs energies and velocities for Bi$_2$Se$_3$.} 
    The original Dirac cone $\Gamma_1$ and the three SDCs originating from the Gamma point of the moir\'e Brillouin zone ($\Gamma_{2,3,4}$).
    (a) Shows the energies of the Dirac cones and (b) shows the velocities of the Dirac cones.
    Notice that $\Gamma_0$ has unavoided crossings with $\Gamma_{1,2}$. 
    All cones have renormalized velocities except $\Gamma_2$ which, within error, has zero velocity. 
    Error bars are dominated by the energy grid of the density of states from which the energies of the states are extracted.}
    \label{fig:energy_velocity_Bi2Se3}
\end{figure}

\subsection{Tuning the patterned dielectric}

We now apply the potential in Eq.~\eqref{eq:patterned_dielectric}, fixing $|\mathbf q_j| \approx (2\pi/9) \mathrm{nm}^{-1}$ and varying $W$.
This leads to physics very similar to the previous sections: satellite Dirac cones appear, the surface density of states is enhanced, and Dirac cone velocities are renormalized.

In order to perform these simulations efficiently, we truncate terms in the \emph{ab initio} Hamiltonian while retaining the relevant physics (this represents the last step in our numerical pipeline shown in Fig.~\ref{fig:Bi2Se3-schematic}). 
The \emph{ab initio} and truncated Hamiltonians are compared in Fig.~\ref{fig:Bi2Se3-fulldispersion}(b), where the red curves represent the dispersion of the truncated Hamiltonian and the light gray represents the untruncated spectrum [the black curves shown in Fig.~\ref{fig:Bi2Se3-fulldispersion}(a)]. 
This truncation preserves the Dirac cone, as shown in Fig.~\ref{fig:Bi2Se3-fulldispersion}(e).
We now describe specifically how we truncate the Hamiltonian: we define hoppings between atoms at positions $\mathbf r_i$ and $\mathbf r_j$ by a matrix $H(\mathbf r_i - \mathbf r_j)$, where the columns are orbitals of the atom hopped from and rows are orbitals of the atom hopped to.
We discard this entire matrix if (1) the norm defined by $|| H(\mathbf r) || \equiv \sqrt{\tr[ H(\mathbf r)^\dagger H(\mathbf r) ]}$ falls below $\SI{50}{meV}$ or (2) the distance between atoms $r$ exceeds $\SI{6.25}{\AA}$ [see Fig.~\ref{fig:Bi2Se3-fulldispersion}(f)].
This truncation scheme naturally preserves all symmetries.

We now describe our results.
First, consider Fig.~\ref{fig:DOS_Bi2Se3}, the density of states on the top QL (see Fig.~\ref{fig:Bi2Se3-schematic} to visualize a QL).
At $W=0$, we see the usual Dirac cone density of states. As we increase the potential strength, we begin seeing structure in the DOS consistent with satellite Dirac cones.
Here, we do not track all satellite Dirac cones, only the cones at the Gamma point of the moir\'e Brillouin zone.
Notice that $\Gamma_0$ becomes obscured by other states near $W=\SI{150}{meV}$ while $\Gamma_{1,3}$ become visible for increased potential strength.
Importantly, for energies less than $\Gamma_0$, we observe an enhancement of the DOS in a similar manner to the toy model described in the previous section.

Contrariwise, $\Gamma_2$ never appears to have a Dirac-cone like structure in the density of states: in fact, our theory in Eq.~\eqref{eq:vsatGamma} predicts a vanishing dispersion to linear order in $\mathbf k$, giving formally zero Dirac-cone velocity for this SDC.
The simulations here are non-perturbative and suggest that this zero velocity persists; we expect $E_{\Gamma_2}\sim  k^2$.

In order to access all the energies and velocities of the satellite points at the Gamma point, we first simulate the Hamiltonian at the Gamma point (zero crystal momentum in the moir\'e BZ), and obtain a density of states using KPM that can resolve individual eigenstates (technically, the expansion order $N_C$ is made very large).
With this density of states, we then vary $W$ and subtract off the parts of the density that are independent of $W$.
This method allows us to track the centers of the resulting peaks in the density of states and extract the approximate energies.
To obtain the velocities, we perform the same calculation at the crystal momentum $k_1\approx 0.03\times 2\pi/\SI{9}{nm}$ (close to the Gamma point) and compute $v_{\Gamma_i} \approx [E_{i}(\mathbf k_1) - E_{i}(\Gamma)]/k_1$; we check that this gives consistent results as we change the direction of $\mathbf k_1$.
The results are shown in Fig.~\ref{fig:energy_velocity_Bi2Se3}.
We can very accurately keep track of the energy of the satellite peaks. Our results show $\Gamma_0$ has an unavoided crossing with $\Gamma_{1,2}$ as discussed at the end of Sec.~\ref{sec:tuneW}. (In addition, $\Gamma_0$ appears to strongly avoid $\Gamma_3$ at large $W$).
Furthermore, we see that the velocities of the Dirac cones also get renormalized (Fig.~\ref{fig:energy_velocity_Bi2Se3}), particularly the velocity of $\Gamma_0$.
This calculation confirms that $\Gamma_2$ indeed has vanishing Dirac cone velocity to within numerical accuracy.

This verifies our perturbative theory of the SDCs as well as our hypothesis that an enhancement of the density of states generically appears on the surface of a 3D TI subject to a moir\'e potential.

\section{Discussion}
\label{sec:discussion}

In this manuscript, we presented a comprehensive study of a moir\'e superlattice potential on the surface of a topological insulator.
We derived the potential induced by a gapped 2D material coupled to the TI surface with a small twist angle and showed analytically and numerically that the potential both shifts and flattens the Dirac cone.
SDCs, protected by time reversal symmetry, emerge at higher/lower energies; we have characterized their energies and dispersion perturbatively, yielding excellent agreement with numerics.
We independently verified these results and their experimental relevance by an ab initio calculation of a superlattice potential on Bi$_2$Se$_3$, which also displayed flat bands embedded within the surface Dirac spectrum.

Our work lays a framework for future studies of topological twistronics.
We have established that a twisted heterostructure or a patterned dielectric superlattice can lead to flattened topological Dirac cones and flat bands with a corresponding enhanced density of states within the surface Dirac cone.
There is mean field evidence that this increased density of states will promote symmetry-breaking instabilities that can gap the surface Dirac cone, leading to an anomalous interaction-driven Hall effect or topological superconductivity \cite{baum2012magnetic,marchand2012lattice,schmidt2012strong,sitte2013interaction,Santos2010}.
Further, in the case of Bi$_2$Se$_3$, surface phonons have been theoretically argued to be able to mediate surface superconductivity, with a predicted transition temperature of on the order $\sim 1$K~\cite{PhysRevB.88.081404}. Upon applying a moir\'e superlattice potential, our results demonstrate that the greatly enhanced surface density of states can appreciably raise this transition temperature in an exponential fashion [assuming a mean field transition temperature $T_c\sim \exp(-1/g\rho_S(E_F))$ for an electron-phonon coupling $g$], making surface superconductivity in Bi$_2$Se$_3$ more viable, despite the absence of a gapped moir\'e miniband.

An important direction will be ab initio studies to optimize the material parameters: realistic parameters should be computed for various combinations of 2D layers and 3D topological insulators to determine the strength of the interlayer coupling.
While there is some previous ab initio studies of moir\'e patterns on 3D TI surfaces \cite{zhang2014proximity,Schouteden2016}, future work should study the dependence on twist angle.
Extending the theory to include magnetic, superconducting, and gapless layers will yield heterostructures with new properties that are predisposed to different instabilities, giving rise to a large phase diagram to be investigated in future work.

\emph{Note Added}: After this work was completed, we became aware of a related and independent work \cite{LiangFupaper}.

\acknowledgements
We acknowledge useful conversations with Shafique Adam, Sankar Das Sarma, Lia Krusin-Elbaum, David Vanderbilt, and Weida Wu.
This work was partially supported by the Air Force Office of Scientific Research under Grant No.~FA9550-20-1-0136 (J.H.P.) and Grant No.~FA9550-20-1-0260 (J.C.).
This work was performed in part (by J.C., J.H.P., J.H.W.) at the Aspen Center for Physics, which is supported by National Science Foundation grant PHY-1607611.
J.C. is also grateful for the hospitality of the Kavli Institute for Theoretical Physics, supported by the National Science Foundation under Grant No. NSF PHY-1748958, and support from the Flatiron Institute, a division of the Simons Foundation.
S. F. is supported by a Rutgers Center for Material Theory Distinguished Postdoctoral Fellowship.
The authors acknowledge the Beowulf cluster at the Department of Physics and Astronomy of Rutgers University, and the Office of Advanced Research Computing (OARC) at Rutgers, The State University of New Jersey (http://oarc.rutgers.edu), for providing access to the Amarel cluster and associated research computing resources that have contributed to the results reported here.

\appendix

\section{Derivation of momentum-space coupling between two layers with different unit cells}
\label{sec:derivecoupling}

In the following, we derive the tunnel coupling in momentum space between two layers with different unit cells.
The derivation is quite general; in Sec.~\ref{sec:twistintro}, we apply the results by taking layer 1 to be the 3D TI surface and layer 2 to be the 2D material.

To derive the tunneling between layers, we consider the electron creation operator in layer 1 to be $c_{\mathbf r}^\dagger$ and an electron in layer 2 to be created by $d_{\mathbf r}^\dagger$.
The tunneling from an atom at position $\mathbf r'$ in layer 2 to an atom at position $\mathbf r$ in layer 1 is then given by the function $T(\mathbf r - \mathbf r')$.
This form assumes that only the relative position of the two atoms is important.
We further assume that $T(\mathbf r - \mathbf r')$ is largest when $\mathbf r = \mathbf r'$ and dies off exponentially with $|\mathbf r - \mathbf r'|$.

The tunneling Hamiltonian then takes the form
\begin{equation}
    H_T = \sum_{\mathbf r, \mathbf r'}c_{\mathbf r}^\dagger  T(\mathbf r - \mathbf r') d_{\mathbf r'} + \mathrm{h.c.},
\end{equation}
where we allow for $T$ to be a matrix and $c$ and $d$ to be multi-component spinors.
Let $\mathbf a_{1,2}$ and $\mathbf a_{1,2}'$ be the lattice vectors in layers 1 and 2, respectively.
The position operators follow
\begin{equation}
    \begin{split}
        \mathbf r & = n_1 \mathbf a_1 + n_2 \mathbf a_2, \\
        \mathbf r' & = n_1' \mathbf a_1' + n_2' \mathbf a_2' + \mathbf r_0.
    \end{split}
\end{equation}
While $c_{\mathbf r}$ and $d_{\mathbf r'}$ can be Fourier transformed into their respective crystal momentum, $T(\mathbf r)$ is not periodic (it is the tunneling between an atom position $0$ in the top layer and position $\mathbf r$ in the bottom layer; as an example $T(\mathbf r) \sim e^{-r/\xi}$ would be a reasonable approximation).
We can, nonetheless, Fourier transform $T(\mathbf r)$.
Using $\mathbf k$ and $\mathbf k'$ as crystal momentum in the Brillouin zone for layer 1 and 2, respectively, and $\mathbf p$ for momentum in real space, 
\begin{equation}
    H_T = \sum_{\mathbf r, \mathbf r'} \int_{\mathbf k}
     \int_{\mathbf k'} \int_{\mathbf p}
      c_{\mathbf k}^\dagger  T(\mathbf p) d_{\mathbf k'} e^{i\mathbf p \cdot(\mathbf r - \mathbf r')} e^{-i\mathbf k \cdot \mathbf r} e^{i\mathbf k'\cdot \mathbf r'} + \mathrm{h.c.},
\end{equation}
where the sum over $\mathbf r$ and $\mathbf r'$ is over the integers $n_{1,2}$ and $n_{1,2}'$ respectively, $\int_{\mathbf k} \equiv \int \frac{d^2 \mathbf k}{(2\pi)^2}$ with integral domain over the entire Brillouin zone, and $\int_{\mathbf p} \equiv \frac{d^2p}{(2\pi)^2}$ with integral domain over $\mathbb R^2$.
We expect that while $T(\mathbf r)$ is short-ranged, $T(\mathbf p)$ will be peaked about $\mathbf p = 0$ and die off as $|\mathbf p| \rightarrow \infty$.

The sum over $\mathbf r$ and $\mathbf r'$ can be completed.
For example,
\begin{align}
    \sum_{\mathbf r} e^{i (\mathbf p - \mathbf k) \cdot \mathbf r} & = \prod_{j}\sum_{n_j} e^{i n_j (\mathbf p - \mathbf k)\cdot \mathbf a_j} \\
    & = \prod_j \sum_{m_j} \delta[(\mathbf p - \mathbf k) \mathbf \cdot \mathbf a_j + 2\pi m_j] \\
    & = \frac1{|\det{A}|}\sum_{\mathbf m} \delta(\mathbf p - \mathbf k + m_1 \mathbf G_1 + m_2 \mathbf G_2),
\end{align}
where $\mathbf G_{1,2}$ are reciprocal lattice vectors in layer 1, $A$ is the matrix whose columns are $\mathbf a_j$, and $\mathbf m = (m_1,m_2)$ run over all pairs of integers.
This allows us to do the integrals over $\mathbf p$ and $\mathbf k'$ in addition to the sums over $\mathbf r$ and $\mathbf r'$ to obtain
\begin{widetext}
\begin{equation}
    H_T = \sum_{\mathbf m, \mathbf m'} \int_{\mathbf k} c_{\mathbf k}^\dagger  T(\mathbf k - m_1 \mathbf G_1 - m_2 \mathbf G_2) d_{\mathbf k + m_1' \mathbf G_1' + m_2' \mathbf G_2'-  m_1 \mathbf G_1 - m_2 \mathbf G_2} \Theta(\mathbf k + m_1' \mathbf G_1' + m_2' \mathbf G_2'-  m_1 \mathbf G_1 - m_2 \mathbf G_2 \in \mathrm{BZ}') + \mathrm{h.c.},
\end{equation}
\end{widetext}
where $\Theta(\mathbf k' \in \mathrm{BZ}')$ is 1 when $\mathbf k'$ is in the Brillouin zone of layer 2, and zero otherwise.
At this point, we have neglected the offset $\mathbf r_0$ which can be absorbed into phases in $T$.

We can specify unique points in the BZ to determine the tunneling.
First, for $\mathbf k$ near the $\Gamma$ point, $m_1 = m_2 = 0$ will dominate and we will thus have the leading term
\begin{equation}
    H_T \approx \sum_{\mathbf k\text{ near }\Gamma} c_{\mathbf k}^\dagger T(0) d_{\mathbf k} + \mathrm{h.c.}.
\end{equation}
To compute the next leading term, we assume, without loss of generality, that the layer 1 BZ is smaller than or equal to the layer 2 BZ.
Then $T(\mathbf G_1)$, for instance, represents part of the next leading term.
Labeling the vectors that are equidistant as $\mathbf G_j$ (e.g., for triangular lattice $\mathbf G_1 = \mathbf G_1$, $\mathbf G_2 = \mathbf G_2$, $\mathbf G_3 = -\mathbf G_1 - \mathbf G_2$, etc.), we define $T_{\mathbf Q_j} \equiv T(\mathbf G_j)$.

To proceed, we need some geometric information regarding the reciprocal lattice vectors. We assume first that $\mathbf G_{1,2}' \approx \mathbf G_{1,2} + \epsilon \mathbf G_{1,2} + \theta \hat{\mathbf z} \times \mathbf G_{1,2}$, i.e., the two layers are arranged with a small twist angle $\theta$ and have a small mismatch $\epsilon$.
Since this holds for all directions, we can enumerate $\mathbf Q_j \equiv \mathbf G_j' - \mathbf G_j \approx \epsilon \mathbf G_j + \theta \hat{\mathbf z} \times \mathbf G_j$.

The result for $\mathbf k$ near the $\Gamma$ point is then
\begin{equation}
    H_T \approx \sum_{\mathbf k\text{ near }\Gamma} \left[c_{\mathbf k}^\dagger T(0) d_{\mathbf k} + \sum_j  c_{\mathbf k}^\dagger T_{\mathbf Q_j} d_{\mathbf k - \mathbf Q_j} + \mathrm{h.c.} \right].
\end{equation}

This math underlies our illustration in Fig.~\ref{fig:2BZ}.

%%%%%%%%%%%%%%%%%%%%%%
%%%%%%%%%%%%%%%%%%%%%%
%%%%%%%%%%%%%%%%%%%%%%
%%%%%%%%%%%%%%%%%%%%%%

\section{The superlattice potential on a TI surface is spin-independent}
\label{sec:VQidentity}

In this appendix, we derive the constraint of time-reversal symmetry on the superlattice potential $V_\mathbf{Q}$.
In the twisted heterostructure, we show that $V_\mathbf{Q}$ must satisfy $ V_\mathbf{Q} = \sigma_y  V_\mathbf{Q}^T \sigma_y$.
Therefore, if $V_\mathbf{Q}$ is a $2\times 2$ matrix, it must be proportional to the identity matrix in spin space.
This holds for any time-reversal preserving model of a superlattice potential on a 3D TI Dirac cone where $V_\mathbf{Q}$ is a $2\times 2$ matrix.

We derived in Sec.~\ref{sec:twistintro} that in the twisted heterostructure, $V_{2D}(\mathbf{r})$ gives rise to an effective Hamiltonian (\ref{eq:defH2D}), which we repeat here for convenience:
\begin{equation}
    H_{2D}^{\rm eff} = \int \frac{d^2\mathbf{k}}{(2\pi)^2} \sum_{\mathbf{Q}}  c^\dagger_{\mathbf{k} + \mathbf{Q}} V_{\mathbf{Q}} c_{\mathbf{k}},
    \label{eq:defH2D-2}
\end{equation}
where
\begin{equation}
    V_{\mathbf{Q}} = \frac{1}{\Delta} \sum_{\mathbf{Q}_1} T_{\mathbf{Q} + \mathbf{Q}_1} T_{\mathbf{Q}_1} ^\dagger.
    \label{eq:defVQ-2}
\end{equation}
Hermiticity requires
\begin{equation}
    V_\mathbf{Q} = V^\dagger_{-\mathbf{Q}},
    \label{eq:VQhermitian}
\end{equation}
while time reversal symmetry (Eq.~(\ref{eq:TRconstraint})) constrains $T_{-\mathbf{Q}} = \sigma_y T_\mathbf{Q}^* \sigma_y$, which enforces
\begin{equation}
    V_{-\mathbf{Q}} = \sigma_y  V_\mathbf{Q}^* \sigma_y
    \label{eq:VQTR}
\end{equation}
Combining Eqs.~(\ref{eq:VQhermitian}) and (\ref{eq:VQTR}) yields:
\begin{equation}
    V_\mathbf{Q} = \sigma_y  V_\mathbf{Q}^T \sigma_y.
\end{equation}
Thus, if $V_\mathbf{Q}$ is a $2\times 2$ matrix -- as in the low-energy theory of a Dirac cone -- then $V_\mathbf{Q}$ is proportional to the identity matrix (although $T_\mathbf{Q}$ need not be), which completes the proof that $V_\mathbf{Q}$ is spin-independent.
In this case, 
\begin{equation}
    V_\mathbf{Q} = V_{-\mathbf{Q}}^*
    \label{eq:VQconstraint}
\end{equation}
Notice that Eqs.~(\ref{eq:VQhermitian}) and (\ref{eq:VQTR}) do not require $V_\mathbf{Q}$ to take the form of Eq.~(\ref{eq:defVQ-2}), but apply more generally to any potential applied to the surface of a topological insulator that takes the form of Eq.~(\ref{eq:defH2D-2}) and satisfies time reversal symmetry.

%%%%%%%%%%%%%%%%%%%%
%%%%%%%%%%%%%%%%%%%%
%%%%%%%%%%%%%%%%%%%%
%%%%%%%%%%%%%%%%%%%%

\section{Perturbative correction to the velocity from the effective potential}
\label{sec:derivesigma}

The effective potential generates a self-energy in the Green's function, which, to leading order in $V_\mathbf{Q}$, takes the form:
\begin{equation}
    \Sigma(\mathbf{k},\omega) = V_{\mathbf{Q}=0}+ \sum_\mathbf{Q}  V_\mathbf{-Q} G_0(\mathbf{k}+\mathbf{Q},\omega) V_{\mathbf{Q}},
    \label{eq:defSE}
\end{equation}
where $V_\mathbf{Q}$ is determined by the Fourier transform of the superlattice potential and $G_0(\mathbf{k},\omega) = \left( \omega - v\mathbf{k} \cdot \sigma \right)^{-1}$ is the single-particle Green's function describing the surface Dirac cone of the 3D TI.
Using the results of Appendix~\ref{sec:VQidentity}, $V_\mathbf{Q} = V_{-\mathbf{Q}}^*$ is proportional to the identity matrix; therefore, the self-energy in Eq.~(\ref{eq:defSE}) can be written as:
\begin{equation}
    \Sigma(\mathbf{k},\omega) = V_{\mathbf{Q}=0}+ \sum_\mathbf{Q}
    \frac{ |V_\mathbf{Q}|^2}{\omega - v(\mathbf{k} + \mathbf{Q})\cdot \sigma}
    \label{eq:SE}
\end{equation}

We now evaluate the self-energy in the low-energy limit where $|\mathbf{k}|,\omega \ll |\mathbf{Q}|$:
\begin{widetext}
\begin{align}
    \Sigma(\mathbf{k},\omega) &= V_{\mathbf{Q} = 0 } + \sum_\mathbf{Q} |V_\mathbf{Q}|^2 \frac{\omega + v(\mathbf{k} +\mathbf{Q}) \cdot\sigma }{\omega^2 - v^2|\mathbf{k} +\mathbf{Q}|^2} \nonumber\\
&=  V_{\mathbf{Q}=0} -\sum_\mathbf{Q} \frac{ |V_\mathbf{Q}|^2}{v^2|\mathbf{Q}|^2} \left(  \omega + v(\mathbf{k} +\mathbf{Q}) \cdot\sigma  \right) \left( 1 - \frac{2\mathbf{k} \cdot \mathbf{Q} }{|\mathbf{Q}|^2} + \cdots \right) \nonumber\\
&=  V_{\mathbf{Q}=0} -\sum_\mathbf{Q} \frac{ |V_\mathbf{Q}|^2 }{v^2|\mathbf{Q}|^2} \left(  \omega + v\mathbf{Q} \cdot \sigma + v\mathbf{k} \cdot \sigma - \frac{2v (\mathbf{Q} \cdot \sigma) (\mathbf{k} \cdot \mathbf{Q})}{|\mathbf{Q}|^2}   + \cdots \right)  \nonumber\\
&=  V_{\mathbf{Q}=0} -\sum_\mathbf{Q} \frac{ |V_\mathbf{Q}|^2 }{v^2|\mathbf{Q}|^2} \left(  \omega + v\mathbf{k} \cdot \sigma - \frac{2v (\mathbf{Q} \cdot \sigma) (\mathbf{k} \cdot \mathbf{Q})}{|\mathbf{Q}|^2}   + \cdots \right).
\label{eq:SE2}
\end{align}
\end{widetext}
In the last line, we have used the constraint $V_\mathbf{Q} = V_{-\mathbf{Q}}^*$ (Eq.~(\ref{eq:VQconstraint})), which requires the term odd in $\mathbf{Q}$ to cancel.

We now make an assumption: assume that the TI surface (with the potential) has an $n$-fold rotational symmetry, where $n>2$.
Let $R_{\theta}$ denote the matrix that rotates $\mathbf{Q}$ by an angle $\theta$ about the $\hat{z}$ axis.
Then by symmetry, $|V_{\mathbf{Q}}|^2 = |V_{R_{2\pi/n}\mathbf{Q}}|^2$
and the last two terms in Eq.~(\ref{eq:SE2}) cancel, due to the following identity:
\begin{align}
    &\sum_{j=0}^{n-1} \frac{1}{| \mathbf{Q}|^2} \left( (R_{2\pi j/n}\mathbf{Q}) \cdot\sigma\right) \left( \mathbf{k} \cdot (R_{2\pi j/n} \mathbf{Q}) \right)\nonumber\\
    &= \sum_{j=0}^{n-1} (\cos\theta_j \sigma_x + \sin\theta_j \sigma_y) (k_x\cos\theta_j  + k_y\sin\theta_j) \nonumber\\
    &= \sum_{j=0}^{n-1} (k_x\sigma_x \cos^2\theta_j + k_y\sigma_y \sin^2\theta_j) + \nonumber\\
    & \quad\quad + \sum_{j=0}^{n-1} (k_y\sigma_x + k_x\sigma_y)\cos\theta_j\sin\theta_j \nonumber\\
    &= \frac{n}{2}\mathbf{k} \cdot \sigma \text{ for }n>2
\end{align}
where $\theta_j = \theta_0 + 2\pi j/n$ for some initial angle $\theta_0$.

Therefore, if the surface of the 4D TI has an $n$-fold symmetry with $n>2$ (absorbing $V_{\mathbf{Q}=0}$ into a shift in the chemical potential),
\begin{equation}
    \Sigma(\mathbf{k},\omega) = -\omega \gamma,
\end{equation}
to second order in $V_\mathbf{Q}$,
where $\gamma$ is defined in Eq.~(\ref{eq:defalpha}) and repeated here for convenience:
\begin{equation}
    \gamma  = \sum_\mathbf{Q} \frac{ |V_\mathbf{Q}|^2}{v^2|\mathbf{Q}|^2}
    \label{eq:defalpha2}
\end{equation}

We obtain the velocity renormalization by computing the Green's function to second order in $V_\mathbf{Q}$:
\begin{align}
    G(\mathbf{k},\omega) &= \left( G_0^{-1}(\mathbf{k},\omega) - \Sigma(\mathbf{k},\omega) \right)^{-1} \nonumber\\
    &= \left( \omega - v\mathbf{k} \cdot \sigma +\omega\gamma \right)^{-1} \nonumber\\
    &= \frac{1}{1+\gamma} \left( \omega - v_* \mathbf{k} \cdot \sigma  \right)^{-1},
    \label{eq:Greensfunc}
\end{align}
where the renormalized velocity is
\begin{equation}
    v_* = \frac{v}{1+\gamma},
    \label{eq:vstar2}
\end{equation}
as stated in the main text in Eq.~(\ref{eq:vstar}). In addition,
 taking into account the chemical potential shift from the $V_{\mathbf{Q}=0}$ term, the Dirac cone is shifted in energy to
\begin{equation}
    E_D = V_{\mathbf{Q}=0},
    \label{eq:Estar2}
\end{equation}
with a quasiparticle residue $Z$ given by
\begin{equation}
    Z^{-1} = 1+\gamma.
    \label{eq:Zstar2}
\end{equation}

%%%%%%%%%%%%%%%%%%%%%%%%%
%%%%%%%%%%%%%%%%%%%%%%%%%
%%%%%%%%%%%%%%%%%%%%%%%%%
%%%%%%%%%%%%%%%%%%%%%%%%%

\section{Dirac cones with an $n$-fold rotational axis, $n>2$, have isotropic velocity}
\label{sec:isotropic}

In this appendix, we prove that a surface Dirac cone with an $n$-fold rotation axis perpendicular to the surface, with $n>2$, has an isotropic velocity.

The most general linear Hamiltonian describing a Dirac cone on the surface of a 3D TI is $H_\text{Dirac}(\mathbf{k}) = c^\dagger_{\mathbf{k},s} h_{s,s'}(\mathbf{k})c_{\mathbf{k},s'}$, where
\begin{equation}
    h(\mathbf{k}) =  \sum_{i,j=x,y} k_i v_{ij} \sigma_j + \sum_{i=x,y} k_i w_i \sigma_z,
    \label{eq:Diracgeneral}
\end{equation}
the Pauli matrices act on spin ($s=\uparrow, \downarrow$) and $v_{ij}, w_i$ are real numbers that determine the dispersion of the Dirac cone to linear order.
Eq.~(\ref{eq:Diracgeneral}) is the most general linear Hamiltonian that satisfies time-reversal symmetry: $h(\mathbf{k}) = \sigma_y h^*(-\mathbf{k}) \sigma_y$.

A rotation by $\theta$ about the $\hat{n}$-axis is implemented by the operator $e^{-i\theta\hat{n}\cdot \sigma/2}$, where $\sigma/2$ is the angular momentum operator for a spin-$\frac{1}{2}$ object.
Therefore, a $n$-fold rotation about the $\hat{z}$ axis perpendicular to the plane imposes the additional constraint:
\begin{equation}
h(\mathbf{k}) = e^{\frac{2\pi i}{n} \frac{\sigma_z}{2}} h(R_n\mathbf{k})e^{-\frac{2\pi i}{n} \frac{\sigma_z}{2}} ,
\label{eq:definvariance}
\end{equation}
where
\begin{equation}
    R_n = \begin{pmatrix} \cos \frac{2\pi}{n} & -\sin \frac{2\pi}{n}  \\ \sin \frac{2\pi}{n}  & \cos \frac{2\pi}{n}  \end{pmatrix}
\end{equation}
Substituting Eq.~(\ref{eq:Diracgeneral}) into Eq.~(\ref{eq:definvariance}) yields the following two constraints:
\begin{align}
    k_i w_i \sigma_z &= \left( R_n \right)_{ij}k_j w_i\sigma_z \label{eq:zeq} \\
    k_i v_{ij} \sigma_j 
    &= k_{i} \left( R_n^T v R_n \right)_{ij}\sigma_{j}\label{eq:xyeq}
\end{align}
Eq.~(\ref{eq:zeq}) requires that $w_i = 0$ (for $n>1$), which still permits $H_\text{Dirac}$ to be anisotropic.
However, for $n>2$, Eq.~(\ref{eq:xyeq}) requires:
\begin{equation}
    v_{xx} = v_{yy}\equiv v\cos\alpha, \quad v_{yx} = -v_{xy}\equiv v\sin\alpha,
\end{equation}
Defining the rotated Pauli matrices $\tilde{\sigma} = R_n\sigma$,
the Hamiltonian takes the form:
\begin{equation}
    h(\mathbf{k}) = kv\cdot\tilde{\sigma}, \text{ for $n>2$},
\end{equation}
which is isotropic and has an emergent $O(2)$ symmetry.
(Higher order terms in the Hamiltonian will generically reduce this emergent symmetry to the appropriate crystal symmetry group.)

This completes the proof that an $(n>2)$-fold rotational symmetry enforces an isotropic velocity.
It is intuitive that a two-fold symmetry cannot enforce an isotropic velocity since it does not mix the $x$ and $y$ directions.
One might have thought that a $3$- or $4$-fold symmetry would allow for anisotropy between directions that are not related by symmetry, but the proof shows that such terms can only appear at quadratic or higher order in $k$, consistent with earlier observations of hexagonal warping of the surface Fermi surface Bi$_2$Te$_3$ due to cubic terms in the surface Hamiltonian \cite{fu2009hexagonal,Chen2009experimental}.

%%%%%%%%%%%%%%%%%%%%%%%%%%%%%%%%%
%%%%%%%%%%%%%%%%%%%%%%%%%%%%%%%%
\section{Velocity of satellite Dirac cones}

In Sec.~\ref{sec:satellite}, we derived the energy and dispersion for the SDCs nearest to the original Dirac cone in energy. When the 3D TI surface is invariant under a $2\pi/N$ rotation, the SDCs next-nearest in energy to the original Dirac cone occur from the superlattice potential coupling $N$ degenerate momenta.
The resulting Hamiltonian obtained by combining Eqs.~(\ref{eq:HN-degenpert}) and (\ref{eq:N-overlap}) in the main text:
\begin{equation}
    H_N =   \cos(\pi/N) \sum_{n=0}^{N-1} V_{\mathbf{Q}_n}  e^{-i\pi/N}  \ket{n+1}\bra{n} + \mathrm{H.c.},
    \label{eq:HN-degenpert-2}
\end{equation}
where 
\begin{equation}
    \ket{n} \equiv \ket{k_0(\cos\frac{2\pi n}{N}, \sin \frac{2\pi n}{N}),+}
    \label{eq:defn}
\end{equation} 
in the notation of Eq.~(\ref{eq:kspinor}).

In the main text, we considered the case when $N$ is even, which has the special property that there exists a basis where $V_{\mathbf{Q}_n}$ is real, due to the fact that $\mathbf{Q}_n$ and $\mathbf{Q}_{n+N/2}$ are time-reversed partners.
In Appendix~\ref{sec:Nodd}, we study the $N$ odd case.
In Appendix~\ref{sec:SDCvelocity} we derive the SDC dispersion stated in the main text for the $N$ even case.

\subsection{$N$ odd}
\label{sec:Nodd}

We can always choose a gauge in Eq.~(\ref{eq:HN-degenpert-2}) such that the phases of $V_{\mathbf{Q}_n}$ are evenly distributed, i.e., $V_{\mathbf{Q}_n} = We^{i\alpha}$, where $W = |V_{\mathbf{Q}_n}|$ (which is independent of $n$ due to the $2\pi/N$ rotational symmetry) and $\alpha$ satisfies $(We^{i\alpha})^N = \prod_n V_{\mathbf{Q}_n}$. In the $N$ even case, $\prod_n V_{\mathbf{Q}_n} = W^N$, allowing $\alpha = 0$. 
Other crystal symmetries, in addition to the $2\pi/N$ rotation, may also force $\alpha = 0$.
Here, we consider the generic case when $\alpha \neq 0$.

In this gauge, the Hamiltonian (\ref{eq:HN-degenpert-2}) is written as:
\begin{equation}
    H'_N =  W \cos \frac{\pi}{N} \sum_{n=0}^{N-1} e^{i\alpha-i\pi/N}  \ket{n+1}\bra{n} + \mathrm{H.c.},
    \label{eq:HN-degenpert-3}
\end{equation}

The eigenstates of (\ref{eq:HN-degenpert-3}) are the rotational eigenstates 
\begin{equation}
    \ket{j} = \frac1{\sqrt{N}}\sum_{n=0}^{N-1} e^{2\pi i j n/ N} \ket{n},
    \label{eq:defj-2}
\end{equation}
which have energy
\begin{equation}
    E^{\mathrm{sat}}_j  = v k_0 + 2W \cos \frac{\pi}{N} \cos \left( \frac{2\pi}{N} (j+\frac{1}{2}) - \alpha\right),
    \label{eq:EsatN-3}
\end{equation}
which would be identical to Eq.~(\ref{eq:EsatN}) in the main text if $\alpha = 0$.
(Although Eq.~(\ref{eq:EsatN}) describes a set of doubly-degenerate states, while Eq.~(\ref{eq:EsatN-3}) describes the generically non-degenerate eigenstates of Eq.~(\ref{eq:HN-degenpert-3}) for $N$ odd.)

The time-reversed partners of the states $\ket{j}$ are the eigenstates of the the time-reversed copy of the Hamiltonian in (\ref{eq:HN-degenpert-3}):
\begin{multline}
    H_N'' =\!  W\cos \frac{\pi}{N} \sum_{n=0}^{N-1}\! e^{-i\alpha-i\pi/N} \ket{n+1+N/2}\bra{n+N/2} \\ + \mathrm{H.c.},
    \label{eq:HN-degenpert-4}
\end{multline}
which is found by using the action of time-reversal in (\ref{eq:TRspinor}), which maps $\ket{n} \mapsto e^{-2\pi in /N} \ket{n+N/2}$, where the states $\ket{n+1/2}$ are defined using the same formula as for $\ket{n}$ in Eq.~(\ref{eq:defn}). 

The eigenstates of Eq.~(\ref{eq:HN-degenpert-4}) are
\begin{equation}
    \ket{j'} = \frac1{\sqrt{N}}\sum_{n=0}^{N-1} e^{-2\pi i (j'+1) n/ N} \ket{n+N/2},
    \label{eq:defj-3}
\end{equation}
so that $\ket{j}$ and $\ket{j'}$ are time-reversed partners and, consequently, share the energy eigenvalue $E^{\mathrm{sat}}_j$ in (\ref{eq:EsatN-3}).
Thus, $\ket{j}$ and $\ket{j'}$ remain degenerate to all orders in perturbation theory and form the band touching point of a SDC.

In this case, the two two tight binding models represented by $H_N'$ and $H_N''$ are decoupled due to the potential not having a term that connects them, so even in a small vicinity of $\mathbf{k}$ around the points, they remain decoupled and we expect no dispersion from these states.
The case of $N$ even does allow for dispersion though and we carry out this procedure in Appendix~\ref{sec:SDCvelocity} in the $N$ even case.

\subsection{$N>2$ even}
\label{sec:SDCvelocity}

When $N$ is even, $V_{\mathbf{Q}_n}$ can be chosen to be real in Eq.~(\ref{eq:HN-degenpert-2}) due to time-reversal symmetry mapping $\ket{n} \mapsto e^{-2\pi in /N} \ket{n+N/2}$.
The resulting Hamiltonian is Eq.~(\ref{eq:HN}) in the main text, which we repeat here for convenience:
\begin{equation}
    \tilde H_N = W\cos(\pi/N) \sum_{n=0}^{N-1} e^{-i\pi/N} \ket{n+1}\bra{n} + \mathrm{H.c.}.
    \label{eq:HN-2}
\end{equation}
The eigenvalues of Eq.~(\ref{eq:HN-2}), given by Eq.~(\ref{eq:EsatN}) in the main text, are 
\begin{equation}
    E^{\mathrm{sat}}_j  = v k_0 + 2W \cos(\pi/N) \cos(\tfrac{2\pi(j + 1/2)}{N}),
    \label{eq:EsatN-2}
\end{equation} 
which divide the $N$ points into $N/2$ degenerate pairs. Each pair forms a gapless SDC that is protected by time reversal symmetry, as discussed in the main text.

To find the dispersion of these SDCs, we must expand around each degenerate pair, which yields a correction, given to linear order in $|\mathbf{k}|$:
\begin{widetext}
\begin{equation}
    \Delta \tilde H_N = vk \sum_{n=0}^{N-1}\cos\left( \varphi_\mathbf{k} - 2\pi n/N \right)\ket{n}\bra{n}  + \left[ i W \sin\frac{\pi}{N} e^{-2\pi i/N}\frac{k}{k_0} \sum_{n=0}^{N-1}\cos\left( \varphi_\mathbf{k} - 2\pi (n+1/2)/N \right) \ket{n+1}\bra{n} + \mathrm{H.c.} \right]
\end{equation}
Using $\ket{j} = \frac1{\sqrt{N}}\sum_{n=0}^{N-1} e^{2\pi i j n/ N} \ket{n}$ (Eq.~(\ref{eq:defj})),
\begin{equation}
    \Delta \tilde H_N = (k_x - i k_y) \sum_{j=0}^{N-1}\left( \frac{v}{2} 
    + \frac{W}{k_0} \sin \frac{\pi}{N} \sin \left( \frac{2\pi \left(j + \frac{3}{2} \right)}{N}  \right)
    \right)\ket{j+1}\bra{j} +  \mathrm{H.c.}
\end{equation}
To determine the dispersion nearby the degenerate states $\ket{j}$ and $\ket{-1-j}$, we need to evaluate the matrix element of $\tilde H_N$ between these states:
\begin{equation}
    \braket{j | \Delta \tilde H_N | -1 - j} = \begin{cases}
      \left[\tfrac v2 + \tfrac{W}{k_0} \sin^2(\pi/N)\right](k_x + i k_y), & j=-1 ,\\
      \left[\tfrac v2 + \tfrac{W}{k_0} \sin^2(\pi/N)\right](k_x - i k_y), & j=0 ,\\
      \left[\tfrac v2 - \tfrac{W}{k_0} \sin^2(\pi/N)\right](k_x + i k_y), & j=N/2-1,\\
      \left[\tfrac v2 - \tfrac{W}{k_0} \sin^2(\pi/N)\right](k_x - i k_y), & j=N/2,\\
      0,& \text{otherwise.}
    \end{cases}
    \label{eq:vsatN}
\end{equation}
\end{widetext}
Thus, linear order perturbation theory yields two Dirac cones, one formed by the states at $j=-1,0$ and another by $j=N/2-1,N/2$.
These two Dirac cones are located at energies,
\begin{equation}
    E_{\pm}^\mathrm{sat} = v k_0 \pm 2W \cos^2(\pi/N),
    \label{eq:EsatGamma-2}
\end{equation}
which comes from evaluating Eq.~(\ref{eq:EsatN-2}) at $j=0$ and $j=N/2$,
and they have isotropic velocities
\begin{equation}
    v_{\pm}^\mathrm{sat} = \tfrac v2 \pm \tfrac{W}{k_0} \sin^2(\pi/N),
        \label{eq:vsatGamma-2}
\end{equation}
which comes from the coefficients of $k_x$ and $k_y$ in Eq.~(\ref{eq:vsatN}).

We have derived the dispersion of two Dirac cones. However, we started with $N$ degenerate states that split into $N/2$ degenerate pairs.
Therefore, if $N>4$, there are $(N-4)/2$ degenerate pairs that have no dispersion to linear order in $|\mathbf{k}|$.
We expect these states to have a non-zero dispersion at higher order in $\mathbf{k}$.  

%%%%%%%%%%%%%%%%%%%%%%%%%%%%%%%%%
%%%%%%%%%%%%%%%%%%%%%%%%%%%%%%%%
\section{Perturbation theory}
\label{sec:higherpert}

\subsection{General setup}\label{sec:generalpert}

To derive the perturbation theory used in Secs.~\ref{sec:satellite} and \ref{sec:ptlattice} to compute the energy and velocity of SDCs, we consider a Hamiltonian $H = H_0 + V$ and its corresponding Green's function, $\hat G_\omega = (\omega - H)^{-1}$ which can be expanded in $V$ as
\begin{equation}
 \hat G_\omega = \hat G_\omega^{(0)} + \hat G_\omega^{(0)} V \hat G_\omega^{(0)} + \hat G_\omega^{(0)} V \hat G_\omega^{(0)} V \hat G_\omega^{(0)} + \cdots,
\end{equation}
where $\hat G_\omega^{(0)} = (\omega - H_0)^{-1}$. 

To describe the SDCs, we are interested in the effect of $V$ on a particular set of states in a narrow energy range (specifically, the degenerate states that comprise the SDC and nearby states in $\mathbf{k}$).
Let $P_{\mathbf k}$ be the projector onto these eigenstates, which satisfies $[H_0,P_{\mathbf k}] = 0$.
Defining $G(\mathbf k,\omega) \equiv P_{\mathbf k} \hat G_\omega P_{\mathbf k}$, we can resum the series to obtain
\begin{equation}
  G(\mathbf k, \omega) = (\omega - H_0(\mathbf k) - \Sigma(\omega,\mathbf k))^{-1},
\end{equation}
where 
\begin{equation}
    \Sigma(\omega,\mathbf k) = P_{\mathbf k}V P_\mathbf{k} + P_{\mathbf k}V (1-P_\mathbf{k}) \hat G_\omega^{(0)} (1-P_{\mathbf k})V P_\mathbf{k} + \cdots.
\end{equation}

To calculate high orders in perturbation theory, first consider the case where $V$ couples two states we call $\ket{1}$ at $\mathbf k_1$ and $\ket{2}$ at $\mathbf k_2$ (we consider $\ket{1}$ and $\ket{2}$ to be two-component spinors; the full eigenstates, including momentum, would be $\ket{\mathbf k_j, j} = \ket{\mathbf k_j} \otimes \ket j$).
Our method, which we illustrate pictorially with examples below, starts by mapping out all paths in momentum space (up to the maximum order we are interested in computing perturbatively) that connect these states both to themselves and to each other.
These paths are used to evaluate matrix elements: we decorate each vertex with the appropriate Green's function 
\begin{equation}
G^{(0)}(\mathbf q, \omega) = \begin{pmatrix} \braket{\mathbf q, \uparrow | \hat G^{(0)}_\omega| \mathbf q, \uparrow} & \braket{\mathbf q, \uparrow | \hat G^{(0)}_\omega| \mathbf q, \downarrow} \\  \braket{\mathbf q, \downarrow | \hat G^{(0)}_\omega| \mathbf q, \uparrow} & \braket{\mathbf q, \downarrow | \hat G^{(0)}_\omega| \mathbf q, \downarrow}  \end{pmatrix},
\end{equation}
where if the vertex is at $\mathbf k_1$ ($\mathbf k_2$), then we must use the projected Green's function $G_{\perp}^{(0)}(\mathbf k_1, \omega) \equiv (\mathbb{1} - \ket{1}\bra{1})G^{(0)}(\mathbf k_1, \omega) $ (similarly for $G_{\perp}^{(0)}(\mathbf k_2, \omega)$).
When we decompose our potential as we have in the main text
\begin{equation}
    V = \sum_{\mathbf Q} V_{\mathbf Q} e^{i \mathbf Q \cdot \mathbf x},
\end{equation}
the operators $V_{\mathbf Q}$ are then associated with the legs of the path.
In particular, if we take a square lattice where $\mathbf Q = (Q,0),$ $(0,Q)$, $(-Q,0)$, or $(0,-Q)$, then we have the rules
\begin{equation}
    \begin{aligned}
    \tikz[baseline=-0.5ex, label distance = 5mm]{ 
    \draw[->,shorten >= 0.25ex] (0,0) -- (3ex,0);
    }& = V_{(Q,0)}
     & 
    \tikz[baseline=.5ex, label distance = 5mm]{ 
    \draw[->,shorten >= 0.25ex] (0,0) -- (0,3ex);
    }\;
     & = V_{(0,Q)} 
    \\
    \tikz[baseline=-0.5ex, label distance = 5mm]{ 
    \draw[->,shorten >= 0.25ex] (0,0) -- (-3ex,0);
    }
    & = V_{(-Q,0)} 
    &
    \tikz[baseline=-2.25ex, label distance = 5mm]{ 
    \draw[->,shorten >= 0.25ex] (0,0) -- (0,-3ex);
    }\; & = 
    V_{(0,-Q)} 
    \\
    \tikz[baseline=-1ex, label distance = 5mm]{ 
    \fill (0,0) circle (1pt) coordinate (A) node[below right = -0.5ex] {\tiny $\mathbf k$};
    }\! & = G^{(0)}( \mathbf k, \omega) 
    &
    \tikz[baseline=-1ex, label distance = 5mm]{ 
    \draw[fill=white] (0,0) circle (1pt) coordinate (A) node[below right = -0.5ex] {\tiny $\mathbf k_i$};
    }\! & = \begin{cases}
      \ket{i} &  \text{if }\tikz[baseline=-0.5ex, label distance = 5mm]{ 
       \draw[->,shorten >= 0.25ex] (0,0) -- (3ex,0);
    \draw[fill=white] (0,0) circle (1pt) coordinate (A) node[below right = -0.5ex] {\tiny $\mathbf k_i$};
     } \\
      G_\perp^{(0)}( \mathbf k_i, \omega)  & \text{if }\tikz[baseline=-0.5ex, label distance = 5mm]{ 
       \draw[->,shorten >= 0.25ex] (-3ex,0) -- (0,0);
       \draw[->,shorten >= 0.25ex] (0,0) -- (3ex,0);
    \draw[fill=white] (0,0) circle (1pt) coordinate (A) node[below right = -0.5ex] {\tiny $\mathbf k_i$};
     } \\
      \bra{i} & \text{if }\tikz[baseline=-0.5ex, label distance = 5mm]{ 
       \draw[->,shorten >= 0.25ex] (0,0) -- (3ex,0);
        \draw[fill=white] (3ex,0) circle (1pt) coordinate (A) node[below right = -0.5ex] {\tiny $\mathbf k_i$};
     } 
      \end{cases}
    \end{aligned}
\end{equation}
We give here a couple of examples with $\mathbf k_2 = \mathbf k_1 + (2Q,0)$. 
As a first example, consider a path from $\mathbf{k}_1$ to itself that does not pass through $\mathbf{k}_2$:
\begin{multline}
 \bra{1} V_{\mathbf Q_8} G^{(0)}( \mathbf q_7, \omega)V_{\mathbf Q_7} G^{(0)}( \mathbf q_6, \omega)V_{\mathbf Q_6}\\ \times G^{(0)}( \mathbf q_5, \omega)V_{\mathbf Q_5} G_\perp^{(0)}( \mathbf k_1, \omega)V_{\mathbf Q_4} G^{(0)}( \mathbf q_3, \omega)V_{\mathbf Q_3} \\ \times G^{(0)}( \mathbf q_2, \omega)V_{\mathbf Q_2} G^{(0)}( \mathbf q_1, \omega)V_{\mathbf Q_1} \ket{1} \\ =
    \tikz[baseline=.1ex, label distance = 5mm]{ 
    \draw[fill=white] (0,0) circle (1pt) coordinate (A) node[above right = -0.5ex] {\tiny $\mathbf k_1$}; 
    \draw[fill=white] (6ex,0) circle (1pt) coordinate (A2) node[above right = -0.5ex] {\tiny $\mathbf k_2$}; 
    \fill (0,3ex) circle (1pt) coordinate (B);
    \draw[->,shorten >= 0.25ex] (A) -- node[left,xshift=0.5ex] {\tiny 1} (B);
    \fill (3ex,3ex) circle (1pt) coordinate (C);
    \fill (3ex,0) circle (1pt) coordinate (D);
    \draw[->,shorten >= 0.25ex] (B) -- node[above,yshift=-0.5ex] {\tiny 2} (C);
    \draw[->,shorten >= 0.25ex] (C) -- node[right,xshift=-0.5ex] {\tiny 3} (D);
    \draw[->,shorten >= 0.25ex] (D) -- node[below=-0.5ex] {\tiny 4} (A);
    \fill (-3ex,0) circle (1pt) coordinate (E);
    \fill (-3ex,-3ex) circle (1pt) coordinate (F);
    \fill (0,-3ex) circle (1pt) coordinate (G);
    \draw[->,shorten >= 0.25ex] (A) -- node[above=-0.5ex] {\tiny 5} (E);
    \draw[->,shorten >= 0.25ex] (E) -- node[left=-0.5ex] {\tiny 6} (F);
    \draw[->,shorten >= 0.25ex] (F) -- node[below=-0.5ex] {\tiny 7} (G);
    \draw[->,shorten >= 0.25ex] (G) -- node[right=-0.5ex] {\tiny 8} (A);
    \draw[fill=white] (A) circle (1pt);
    },
\end{multline}
where here and in the subsequent examples, $\mathbf q_i = \mathbf k_1 + \sum_{j=1}^i \mathbf Q_j $.
As a second example, consider a path from $\mathbf{k}_1$ to itself that does pass through $\mathbf{k}_2$:
\begin{multline}
\bra{1} V_{\mathbf Q_6} G^{(0)}( \mathbf q_5, \omega)V_{\mathbf Q_5} G_\perp^{(0)}( \mathbf k_2, \omega)V_{\mathbf Q_4} \\ \times G^{(0)}( \mathbf q_3, \omega)V_{\mathbf Q_3} G^{(0)}( \mathbf q_2, \omega)V_{\mathbf Q_2} G^{(0)}( \mathbf q_1, \omega)V_{\mathbf Q_1} \ket{1}  \\
  =\tikz[baseline=.1ex]{
    \draw[fill=white] (0,0) circle (1pt) coordinate (A1) node[below left = -0.5ex] {\tiny $\mathbf k_1$}; 
    \draw[fill=white] (6ex,0) circle (1pt) coordinate (A2) node[below right = -0.5ex] {\tiny $\mathbf k_2$}; 
    \fill (0,3ex) circle (1pt) coordinate (B);
    \fill (3ex,3ex) circle (1pt) coordinate (C);
    \fill (6ex,3ex) circle (1pt) coordinate (D);
    \fill (3ex,0) circle (1pt) coordinate (E);
    \draw[->,shorten >= 0.25ex] (A1) -- node[left=-0.5ex] {\tiny 1} (B);
    \draw[->,shorten >= 0.25ex] (B) -- node[above=-0.5ex] {\tiny 2} (C);
    \draw[->,shorten >= 0.25ex] (C) -- node[above=-0.5ex] {\tiny 3} (D);
    \draw[->,shorten >= 0.25ex] (D) -- node[right=-0.5ex] {\tiny 4} (A2);
    \draw[->,shorten >= 0.25ex] (A2) -- node[below=-0.5ex] {\tiny 5} (E);
    \draw[->,shorten >= 0.25ex] (E) -- node[below=-0.5ex] {\tiny 6} (A1);
    \draw[fill=white] (A1) circle (1pt);
    \draw[fill=white] (A2) circle (1pt);
  }.
\end{multline}
As a third example, consider a path that starts at $\mathbf{k}_1$ and ends at $\mathbf{k}_2$:
\begin{multline}
  \bra{2}V_{\mathbf Q_4} G^{(0)}( \mathbf q_3, \omega)V_{\mathbf Q_3}G^{(0)}( \mathbf q_2, \omega)\\ \times V_{\mathbf Q_2} G^{(0)}( \mathbf q_1, \omega)V_{\mathbf Q_1} \ket{1}  =
  \tikz[baseline=.1ex]{
    \draw[fill=white] (0,0) circle (1pt) coordinate (A1) node[below left = -0.5ex] {\tiny $\mathbf k_1$}; 
    \draw[fill=white] (6ex,0) circle (1pt) coordinate (A2) node[below right = -0.5ex] {\tiny $\mathbf k_2$}; 
    \fill (0,3ex) circle (1pt) coordinate (B);
    \fill (3ex,3ex) circle (1pt) coordinate (C);
    \fill (6ex,3ex) circle (1pt) coordinate (D);
    \draw[->,shorten >= 0.25ex] (A1) -- node[left=-0.5ex] {\tiny 1} (B);
    \draw[->,shorten >= 0.25ex] (B) -- node[above=-0.5ex] {\tiny 2} (C);
    \draw[->,shorten >= 0.25ex] (C) -- node[above=-0.5ex] {\tiny 3} (D);
    \draw[->,shorten >= 0.25ex] (D) -- node[right=-0.5ex] {\tiny 4} (A2);
    \draw[fill=white] (A1) circle (1pt);
    \draw[fill=white] (A2) circle (1pt);
  }.
\end{multline}
A few details: 
1.\ Paths are allowed to retrace. 
2.\ If $\mathbf k_1 = \mathbf k_2$ but $\braket{2|1}=0$, then $G_{\perp}^{(0)}(\mathbf k_1, \omega) =  (\mathbb{1} - \ket{1}\bra{1}-\ket{2}\bra{2})G^{(0)}(\mathbf k_1, \omega)$ (which is identically zero if the on-site Hilbert space is dimension 2, as it is for us).
3.\ The method generalizes to considering $N$ degenerate points $\mathbf k_i$ straightforwardly.
4.\ The length of any particular path corresponds to the order of perturbation theory that it contributes to; by summing over all paths of a given length, we evaluate the self-energy to that order in perturbation theory.

With the self-energy evaluated we can expand it assuming that both $\omega$ is small and $\mathbf k_{1,2} = \mathbf K_{1,2} + \mathbf k$ for small $|\mathbf k|$:
\begin{equation}
    \Sigma(\omega,\mathbf k) \approx \Sigma_0 + \Sigma_\omega \omega + \bm\Sigma_\mathbf{k} \cdot \mathbf k.
\end{equation}
Meanwhile, we assume that when $\mathbf k=0$ the states are degenerate, so the bare Hamiltonian in the basis $\{\ket 1, \ket 2 \}$ takes the form
\begin{equation}
H_0(\mathbf k) = \begin{pmatrix} E_1(\mathbf k_1) & 0 \\ 0 & E_2(\mathbf k_2) \end{pmatrix} \approx \begin{pmatrix} E_0 + \mathbf v_1 \cdot \mathbf k & 0 \\ 0 & E_0 + \mathbf v_2 \cdot \mathbf k \end{pmatrix} ,
\end{equation}
where $E_j(\mathbf k_j) =\braket{\mathbf k_j, j | H_0 | \mathbf k_j , j}$, the energy of the eigenstate represented by $\ket{j}$, and $\mathbf v_j$ represents the expansion coeffcients for small $\mathbf k$ of the respective energies.
To compactly represent this, we say $H_0(\mathbf k) = E_0 + \hat{\mathbf v}_0 \cdot \mathbf k$ with $\hat{\mathbf v}_0$ is a vector of diagonal matrices with $\mathbf v_1$ and $\mathbf v_2$ on the diagonal, so that the Green's function takes the form
\begin{equation}
    G(\mathbf k,\omega) = [ \omega(1+\Sigma_\omega) - \Sigma_0 - (\hat{\mathbf v}_0+\bm\Sigma_{\mathbf k})\cdot \mathbf k]^{-1},
\end{equation}
from which we can read off, in the usual manner, the quasiparticle residue
\begin{equation}
    Z = (1+ \Sigma_\omega)^{-1},
\end{equation}
and the effective Hamiltonian
\begin{equation}
    H_{\mathrm{eff}}(\mathbf k) = E_0 + Z [ \Sigma_0 + (\hat{\mathbf v}_0 + \bm \Sigma_{\mathbf k})\cdot \mathbf k],
\end{equation}
such that $G(\mathbf k,\omega) = Z/(\omega - H_{\mathrm{eff}}(\mathbf k) - E_0)$.
This formulation allows for the pertubation theory to be easily implemented by a computer algebra system such as Mathematica, producing the results in Sec.~\ref{sec:ptlatticeA}.

\subsection{Tight-binding models in $k$-space at arbitrary order}

In Sec.~\ref{sec:satellite} and Appendix~\ref{sec:SDCvelocity} we showed how to obtain tight binding models near a SDC at lowest order. 
Here we extend these models to higher orders in perturbation theory, as is necessary, for example, to obtain the fits in Fig.~\ref{fig:SatelliteEnergies}.

Utilizing the spinor notation $\ket{e^{i\varphi}}$, defined in Eq.~\eqref{eq:spinor},
and assuming that the potential is invariant under the discrete rotational symmetry of the interface
($V_{\mathbf Q} = W$ for all $\mathbf Q$ of the same magnitude), %however, to obtain 
the hopping amplitude from the tight-binding state $\ket{e^{i\varphi_0}}$ to another state $\ket{e^{i\phi_j}}$, is determined by a sum over all paths:
\begin{equation}
    t_{j0} = \sum_{\ell}  W^{|\ell|}\braket{e^{i\varphi_j}| \prod_{i} G^{(0)}( \mathbf q_i, \omega) | e^{i\varphi_0}},
    \label{eq:hopPT}
\end{equation}
where $\ell$ represents a path in momentum space and $\{ \mathbf q_i \}$, the momenta along that path; this expression assumes that if any of the $\mathbf q_i$'s correspond to the states represented by $\varphi_j$, $G^{(0)}$ ought to be projected onto the orthogonal state ($G^{(0)}_\perp$ as previously defined).

Decomposing the Green's function into a sum of Pauli marices,
$G^{(0)}(\mathbf q_i,\omega) = g_j^0 + g_j^x \sigma_x + g_j^y \sigma_y$, for real $g_j^{0,x,y}$,
it follows that
\begin{equation}
G^{(0)}(\mathbf q_2,\omega) G^{(0)}(\mathbf q_1, \omega) = g_{21}^0 + g_{21}^x\sigma_x + g_{21}^y\sigma_y - i g_{21}^z\sigma_z  
\end{equation}
for real $g_{21}^\mu$.
Proceeding inductively, 
\begin{equation}
    \prod_{i} G^{(0)}(\mathbf q_i,\omega)  = g_{\{ \mathbf q_i \}}^0 + g_{\{ \mathbf q_i \}}^x\sigma_x + g_{\{ \mathbf q_i \}}^y\sigma_y - i g_{\{ \mathbf q_i \}}^z\sigma_z  
\end{equation}
with $g_{\{ \mathbf q_i \}}^\mu$ all real.
Using this result to evaluate the matrix element in Eq.~(\ref{eq:hopPT}) yields
\begin{multline}
\braket{e^{i\varphi_j}| \prod_{i} G^{(0)}(\mathbf q_i,\omega) | e^{i\varphi_0}} 
= [(g_{\{ \mathbf q_i \}}^0 + g_{\{ \mathbf q_i \}}^x)\cos(\varphi_j/2) \\ + (g_{\{ \mathbf q_i \}}^y + g_{\{ \mathbf q_i \}}^z)\sin(\varphi_j/2)] e^{-i\varphi_j/2} ,
\end{multline}
which shows that the phase of the matrix element is independent of its path (although its amplitude may be path-dependent).
Using $\varphi_j = 2\pi j/n$, it follows that the hopping amplitude $t_{j0}$ defined in Eq.~(\ref{eq:hopPT}) satisfies
\begin{equation}
    t_{j0} = | t_{j0} | e^{-i\pi j/n}.
\end{equation}
Extending the same logic to the entire tight-binding basis,
$t_{jj'}=| t_{jj'} | e^{-i\pi (j-j')/n}$, where $C_n$ symmetry requires $|t_{jj'}| = |t_{j-j',0}|$.
Therefore, the $\pi$-flux Hamiltonian derived in Sec.~\ref{sec:satellite} remains robust to higher orders in perturbation theory.
It follows that the eigenstates of this Hamilonian are unchanged to arbitrary order in perturbation theory; in particular, its energies are doubly-degenerate for even $n$.

\subsection{High order perturbative results on the lattice model}
\label{sec:ptlatticeA}

In Sec.~\ref{sec:ptlattice} we used perturbation theory to fit the energy and velocity of SDCs obtained from a lattice model.
We now derive these results by applying the perturbation theory derived in Appendix~\ref{sec:generalpert}. 
We obtain the renormalized velocity of the original Dirac cone at the $\Gamma$ point [denoted as $\Gamma_0$ in Fig.~\ref{fig:satelliteQ} (b)]
to fifth order in terms of the parameter $\alpha \equiv W^2/(\Delta v_0 Q)$ 
\begin{equation}
    v_{\Gamma_0}/v_0 = \left(1 + \frac{25401}{80}\alpha^4\right)Z_{\Gamma_0},
\end{equation}
in the vicinity of the Dirac node energy
\begin{equation}
    E_{\Gamma_0}  = \mu_V + \left(80\alpha^3 + \frac{699904}{160}\alpha^5\right)Z_{\Gamma_0},
\end{equation}
with a quasiparticle residue
\begin{equation}
    Z_{\Gamma_0}^{-1}=1+25\alpha^2 + \frac{153117}{80}\alpha^4
\end{equation}
for our parameter choice $v_0 = t$ and $\mu_V=5W^2$.
As we demonstrate numerically below, the  satellite peak we observe that possesses a true pseudogap and semimetalic behaviour (i.e. has no other bands passing through the SDC energy) is the fourth closest SDC to the original Dirac cone at the $\Gamma$ point, which is labelled $\Gamma_1$ in Fig~\ref{fig:satelliteQ} (b).
Focusing on $\Gamma_1$ we find: its location in energy 
\begin{multline}
    E_{\Gamma_{1}}^{\mathrm{sat}}=-v_0\sin(Q) + \mu_V 
    \\
    + Qv_0Z_{\Gamma_1}\Big(2\alpha + 31\alpha^2 + \frac{497}{2}\alpha^3 + \frac{2997103}{1008}\alpha^4 + \frac{186925}{6}\alpha^5\Big),
\end{multline}
its velocity 
\begin{multline}
    v_{\Gamma_{1}}^{\mathrm{sat}}/v_0=Z_{\Gamma_1}\Big(\cos(Q)/2 - 3\alpha/2 + 195\alpha^2/4 + 2175\alpha^3/4 
    \\
    + \frac{8726507491}{1016064}\alpha^4 + \frac{8888576021}{80640}\alpha^5\Big),
\end{multline}
and the quasiparticle residue
\begin{multline}
    Z_{\Gamma_{1}}^{-1} = 1 + 183\alpha^2/2 + 1977\alpha^3/2 \\
    +\frac{8306436637}{508032}\alpha^4 + \frac{423661547}{2016}\alpha^5.
\end{multline}

We also list the perturbative expressions we have obtained for $\Gamma_{-1}$ and $\Gamma_2$ that are plotted in Figs~\ref{fig:SatelliteEnergies}.
Focusing on the  SDC at $\Gamma_{-1}$ we obtain: its location in energy 
\begin{multline}
    E_{\Gamma_{-1}}^{\mathrm{sat}}=v_0\sin(Q) + \mu_V 
    \\
    - Qv_0\Big( 2\alpha + 11\alpha^2 - \frac{87}{2}\alpha^3 + \frac{424411}{1008}\alpha^4 
    -\frac{1405877}{504}\alpha^5\Big)Z_{\Gamma_{-1}},
\end{multline}
its velocity 
\begin{multline}
    v_{\Gamma_{-1}}^{\mathrm{sat}}/v_0=\Big((\cos(Q) - \alpha)/2 - 13(\alpha^2 + \alpha^3)/4
    \\
    - \frac{260278001}{1016064}\alpha^4 + \frac{4381849513}{5080320}\alpha^5\Big)Z_{\Gamma_{-1}},
\end{multline}
and its a quasiparticle residue
\begin{equation}
    Z_{\Gamma_{-1}}^{-1} = 1 + \frac{39}{2}\alpha^2 - \frac{215}{2}\alpha^3 
    + \frac{843139273}{508032}\alpha^4.
\end{equation}
Last, we turn to our results at $\Gamma_2$, where we obtain: its location in energy 
\begin{multline}
    E_{\Gamma_{2}}^{\mathrm{sat}}=v_0\sin(Q) + \mu_V 
    \\
    - Qv_0\Big( 2\alpha - 11\alpha^2 - \frac{87}{2}\alpha^3 - \frac{424411}{1008}\alpha^4 
    -\frac{1405877}{504}\alpha^5\Big)Z_{\Gamma_{2}},
\end{multline}
its velocity 
\begin{multline}
    v_{\Gamma_{2}}^{\mathrm{sat}}/v_0=\Big(\cos(Q)/2 + \alpha/2 - 13\alpha^2/4 + 13\alpha^3/4
    \\
    - \frac{260278001}{1016064}\alpha^4 - \frac{4381849513}{5080320}\alpha^5\Big)Z_{\Gamma_{2}},
\end{multline}
and its quasiparticle residue
\begin{multline}
    Z_{\Gamma_{2}}^{-1} = 1 + 39\alpha^2/2 + 215\alpha^3/2 
    \\
    + \frac{843139273}{508032}\alpha^4 + \frac{1546113283}{127008}\alpha^5.
\end{multline}

%%%%%%%%%%%%%%%%%%%%%%%%%
%%%%%%%%%%%%%%%%%%%%%%%%%
%%%%%%%%%%%%%%%%%%%%%%%%%
%%%%%%%%%%%%%%%%%%%%%%%%%

\bibliography{moire}

\begin{thebibliography}{116}%
\makeatletter
\providecommand \@ifxundefined [1]{%
 \@ifx{#1\undefined}
}%
\providecommand \@ifnum [1]{%
 \ifnum #1\expandafter \@firstoftwo
 \else \expandafter \@secondoftwo
 \fi
}%
\providecommand \@ifx [1]{%
 \ifx #1\expandafter \@firstoftwo
 \else \expandafter \@secondoftwo
 \fi
}%
\providecommand \natexlab [1]{#1}%
\providecommand \enquote  [1]{``#1''}%
\providecommand \bibnamefont  [1]{#1}%
\providecommand \bibfnamefont [1]{#1}%
\providecommand \citenamefont [1]{#1}%
\providecommand \href@noop [0]{\@secondoftwo}%
\providecommand \href [0]{\begingroup \@sanitize@url \@href}%
\providecommand \@href[1]{\@@startlink{#1}\@@href}%
\providecommand \@@href[1]{\endgroup#1\@@endlink}%
\providecommand \@sanitize@url [0]{\catcode `\\12\catcode `\$12\catcode
  `\&12\catcode `\#12\catcode `\^12\catcode `\_12\catcode `\%12\relax}%
\providecommand \@@startlink[1]{}%
\providecommand \@@endlink[0]{}%
\providecommand \url  [0]{\begingroup\@sanitize@url \@url }%
\providecommand \@url [1]{\endgroup\@href {#1}{\urlprefix }}%
\providecommand \urlprefix  [0]{URL }%
\providecommand \Eprint [0]{\href }%
\providecommand \doibase [0]{https://doi.org/}%
\providecommand \selectlanguage [0]{\@gobble}%
\providecommand \bibinfo  [0]{\@secondoftwo}%
\providecommand \bibfield  [0]{\@secondoftwo}%
\providecommand \translation [1]{[#1]}%
\providecommand \BibitemOpen [0]{}%
\providecommand \bibitemStop [0]{}%
\providecommand \bibitemNoStop [0]{.\EOS\space}%
\providecommand \EOS [0]{\spacefactor3000\relax}%
\providecommand \BibitemShut  [1]{\csname bibitem#1\endcsname}%
\let\auto@bib@innerbib\@empty
%</preamble>
\bibitem [{\citenamefont {Geim}\ and\ \citenamefont
  {Novoselov}(2007)}]{geim2007}%
  \BibitemOpen
  \bibfield  {author} {\bibinfo {author} {\bibfnamefont {A.~K.}\ \bibnamefont
  {Geim}}\ and\ \bibinfo {author} {\bibfnamefont {K.~S.}\ \bibnamefont
  {Novoselov}},\ }\bibfield  {title} {\bibinfo {title} {The rise of graphene},\
  }\href {https://doi.org/10.1038/nmat1849} {\bibfield  {journal} {\bibinfo
  {journal} {Nat. Mater.}\ }\textbf {\bibinfo {volume} {6}},\ \bibinfo {pages}
  {183} (\bibinfo {year} {2007})}\BibitemShut {NoStop}%
\bibitem [{\citenamefont {Wan}\ \emph {et~al.}(2011)\citenamefont {Wan},
  \citenamefont {Turner}, \citenamefont {Vishwanath},\ and\ \citenamefont
  {Savrasov}}]{Wan11}%
  \BibitemOpen
  \bibfield  {author} {\bibinfo {author} {\bibfnamefont {X.}~\bibnamefont
  {Wan}}, \bibinfo {author} {\bibfnamefont {A.~M.}\ \bibnamefont {Turner}},
  \bibinfo {author} {\bibfnamefont {A.}~\bibnamefont {Vishwanath}},\ and\
  \bibinfo {author} {\bibfnamefont {S.~Y.}\ \bibnamefont {Savrasov}},\
  }\bibfield  {title} {\bibinfo {title} {Topological semimetal and {Fermi}-arc
  surface states in the electronic structure of pyrochlore iridates},\ }\href
  {https://doi.org/10.1103/PhysRevB.83.205101} {\bibfield  {journal} {\bibinfo
  {journal} {Phys. Rev. B}\ }\textbf {\bibinfo {volume} {83}},\ \bibinfo
  {pages} {205101} (\bibinfo {year} {2011})}\BibitemShut {NoStop}%
\bibitem [{\citenamefont {Weng}\ \emph {et~al.}(2015)\citenamefont {Weng},
  \citenamefont {Fang}, \citenamefont {Fang}, \citenamefont {Bernevig},\ and\
  \citenamefont {Dai}}]{Weng15}%
  \BibitemOpen
  \bibfield  {author} {\bibinfo {author} {\bibfnamefont {H.}~\bibnamefont
  {Weng}}, \bibinfo {author} {\bibfnamefont {C.}~\bibnamefont {Fang}}, \bibinfo
  {author} {\bibfnamefont {Z.}~\bibnamefont {Fang}}, \bibinfo {author}
  {\bibfnamefont {B.~A.}\ \bibnamefont {Bernevig}},\ and\ \bibinfo {author}
  {\bibfnamefont {X.}~\bibnamefont {Dai}},\ }\bibfield  {title} {\bibinfo
  {title} {{Weyl} semimetal phase in noncentrosymmetric transition-metal
  monophosphides},\ }\href@noop {} {\bibfield  {journal} {\bibinfo  {journal}
  {Phys. Rev. X}\ }\textbf {\bibinfo {volume} {5}},\ \bibinfo {pages} {011029}
  (\bibinfo {year} {2015})}\BibitemShut {NoStop}%
\bibitem [{\citenamefont {Huang}\ \emph {et~al.}(2015)\citenamefont {Huang},
  \citenamefont {Xu}, \citenamefont {Belopolski}, \citenamefont {Lee},
  \citenamefont {Chang}, \citenamefont {Wang}, \citenamefont {Alidoust},
  \citenamefont {Bian}, \citenamefont {Neupane}, \citenamefont {Zhang} \emph
  {et~al.}}]{Huang15}%
  \BibitemOpen
  \bibfield  {author} {\bibinfo {author} {\bibfnamefont {S.-M.}\ \bibnamefont
  {Huang}}, \bibinfo {author} {\bibfnamefont {S.-Y.}\ \bibnamefont {Xu}},
  \bibinfo {author} {\bibfnamefont {I.}~\bibnamefont {Belopolski}}, \bibinfo
  {author} {\bibfnamefont {C.-C.}\ \bibnamefont {Lee}}, \bibinfo {author}
  {\bibfnamefont {G.}~\bibnamefont {Chang}}, \bibinfo {author} {\bibfnamefont
  {B.}~\bibnamefont {Wang}}, \bibinfo {author} {\bibfnamefont {N.}~\bibnamefont
  {Alidoust}}, \bibinfo {author} {\bibfnamefont {G.}~\bibnamefont {Bian}},
  \bibinfo {author} {\bibfnamefont {M.}~\bibnamefont {Neupane}}, \bibinfo
  {author} {\bibfnamefont {C.}~\bibnamefont {Zhang}}, \emph {et~al.},\
  }\bibfield  {title} {\bibinfo {title} {A {Weyl} fermion semimetal with
  surface {Fermi} arcs in the transition metal monopnictide {TaAs} class},\
  }\href {https://doi.org/10.1038/ncomms8373} {\bibfield  {journal} {\bibinfo
  {journal} {Nat. Commun.}\ }\textbf {\bibinfo {volume} {6}},\ \bibinfo {pages}
  {7373} (\bibinfo {year} {2015})}\BibitemShut {NoStop}%
\bibitem [{\citenamefont {Xu}\ \emph {et~al.}(2015{\natexlab{a}})\citenamefont
  {Xu}, \citenamefont {Alidoust}, \citenamefont {Belopolski}, \citenamefont
  {Yuan}, \citenamefont {Bian}, \citenamefont {Chang}, \citenamefont {Zheng},
  \citenamefont {Strocov}, \citenamefont {Sanchez}, \citenamefont {Chang} \emph
  {et~al.}}]{Xu15}%
  \BibitemOpen
  \bibfield  {author} {\bibinfo {author} {\bibfnamefont {S.-Y.}\ \bibnamefont
  {Xu}}, \bibinfo {author} {\bibfnamefont {N.}~\bibnamefont {Alidoust}},
  \bibinfo {author} {\bibfnamefont {I.}~\bibnamefont {Belopolski}}, \bibinfo
  {author} {\bibfnamefont {Z.}~\bibnamefont {Yuan}}, \bibinfo {author}
  {\bibfnamefont {G.}~\bibnamefont {Bian}}, \bibinfo {author} {\bibfnamefont
  {T.-R.}\ \bibnamefont {Chang}}, \bibinfo {author} {\bibfnamefont
  {H.}~\bibnamefont {Zheng}}, \bibinfo {author} {\bibfnamefont {V.~N.}\
  \bibnamefont {Strocov}}, \bibinfo {author} {\bibfnamefont {D.~S.}\
  \bibnamefont {Sanchez}}, \bibinfo {author} {\bibfnamefont {G.}~\bibnamefont
  {Chang}}, \emph {et~al.},\ }\bibfield  {title} {\bibinfo {title} {Discovery
  of a {Weyl} fermion state with {Fermi} arcs in niobium arsenide},\
  }\href@noop {} {\bibfield  {journal} {\bibinfo  {journal} {Nat. Phys.}\
  }\textbf {\bibinfo {volume} {11}},\ \bibinfo {pages} {748} (\bibinfo {year}
  {2015}{\natexlab{a}})}\BibitemShut {NoStop}%
\bibitem [{\citenamefont {Lv}\ \emph {et~al.}(2015{\natexlab{a}})\citenamefont
  {Lv}, \citenamefont {Xu}, \citenamefont {Weng}, \citenamefont {Ma},
  \citenamefont {Richard}, \citenamefont {Huang}, \citenamefont {Zhao},
  \citenamefont {Chen}, \citenamefont {Matt}, \citenamefont {Bisti} \emph
  {et~al.}}]{Lv15}%
  \BibitemOpen
  \bibfield  {author} {\bibinfo {author} {\bibfnamefont {B.}~\bibnamefont
  {Lv}}, \bibinfo {author} {\bibfnamefont {N.}~\bibnamefont {Xu}}, \bibinfo
  {author} {\bibfnamefont {H.}~\bibnamefont {Weng}}, \bibinfo {author}
  {\bibfnamefont {J.}~\bibnamefont {Ma}}, \bibinfo {author} {\bibfnamefont
  {P.}~\bibnamefont {Richard}}, \bibinfo {author} {\bibfnamefont
  {X.}~\bibnamefont {Huang}}, \bibinfo {author} {\bibfnamefont
  {L.}~\bibnamefont {Zhao}}, \bibinfo {author} {\bibfnamefont {G.}~\bibnamefont
  {Chen}}, \bibinfo {author} {\bibfnamefont {C.}~\bibnamefont {Matt}}, \bibinfo
  {author} {\bibfnamefont {F.}~\bibnamefont {Bisti}}, \emph {et~al.},\
  }\bibfield  {title} {\bibinfo {title} {Observation of {Weyl} nodes in
  {TaAs}},\ }\href {https://doi.org/10.1038/nphys3426} {\bibfield  {journal}
  {\bibinfo  {journal} {Nat. Phys.}\ }\textbf {\bibinfo {volume} {11}},\
  \bibinfo {pages} {724} (\bibinfo {year} {2015}{\natexlab{a}})}\BibitemShut
  {NoStop}%
\bibitem [{\citenamefont {Xu}\ \emph {et~al.}(2015{\natexlab{b}})\citenamefont
  {Xu}, \citenamefont {Belopolski}, \citenamefont {Alidoust}, \citenamefont
  {Neupane}, \citenamefont {Bian}, \citenamefont {Zhang}, \citenamefont
  {Sankar}, \citenamefont {Chang}, \citenamefont {Yuan}, \citenamefont {Lee}
  \emph {et~al.}}]{Xu15a}%
  \BibitemOpen
  \bibfield  {author} {\bibinfo {author} {\bibfnamefont {S.-Y.}\ \bibnamefont
  {Xu}}, \bibinfo {author} {\bibfnamefont {I.}~\bibnamefont {Belopolski}},
  \bibinfo {author} {\bibfnamefont {N.}~\bibnamefont {Alidoust}}, \bibinfo
  {author} {\bibfnamefont {M.}~\bibnamefont {Neupane}}, \bibinfo {author}
  {\bibfnamefont {G.}~\bibnamefont {Bian}}, \bibinfo {author} {\bibfnamefont
  {C.}~\bibnamefont {Zhang}}, \bibinfo {author} {\bibfnamefont
  {R.}~\bibnamefont {Sankar}}, \bibinfo {author} {\bibfnamefont
  {G.}~\bibnamefont {Chang}}, \bibinfo {author} {\bibfnamefont
  {Z.}~\bibnamefont {Yuan}}, \bibinfo {author} {\bibfnamefont {C.-C.}\
  \bibnamefont {Lee}}, \emph {et~al.},\ }\bibfield  {title} {\bibinfo {title}
  {Discovery of a {Weyl} fermion semimetal and topological {Fermi} arcs},\
  }\href@noop {} {\bibfield  {journal} {\bibinfo  {journal} {Science}\ }\textbf
  {\bibinfo {volume} {349}},\ \bibinfo {pages} {613} (\bibinfo {year}
  {2015}{\natexlab{b}})}\BibitemShut {NoStop}%
\bibitem [{\citenamefont {Lv}\ \emph {et~al.}(2015{\natexlab{b}})\citenamefont
  {Lv}, \citenamefont {Weng}, \citenamefont {Fu}, \citenamefont {Wang},
  \citenamefont {Miao}, \citenamefont {Ma}, \citenamefont {Richard},
  \citenamefont {Huang}, \citenamefont {Zhao}, \citenamefont {Chen} \emph
  {et~al.}}]{Lv15a}%
  \BibitemOpen
  \bibfield  {author} {\bibinfo {author} {\bibfnamefont {B.}~\bibnamefont
  {Lv}}, \bibinfo {author} {\bibfnamefont {H.}~\bibnamefont {Weng}}, \bibinfo
  {author} {\bibfnamefont {B.}~\bibnamefont {Fu}}, \bibinfo {author}
  {\bibfnamefont {X.}~\bibnamefont {Wang}}, \bibinfo {author} {\bibfnamefont
  {H.}~\bibnamefont {Miao}}, \bibinfo {author} {\bibfnamefont {J.}~\bibnamefont
  {Ma}}, \bibinfo {author} {\bibfnamefont {P.}~\bibnamefont {Richard}},
  \bibinfo {author} {\bibfnamefont {X.}~\bibnamefont {Huang}}, \bibinfo
  {author} {\bibfnamefont {L.}~\bibnamefont {Zhao}}, \bibinfo {author}
  {\bibfnamefont {G.}~\bibnamefont {Chen}}, \emph {et~al.},\ }\bibfield
  {title} {\bibinfo {title} {Experimental discovery of {Weyl} semimetal
  {TaAs}},\ }\href@noop {} {\bibfield  {journal} {\bibinfo  {journal} {Phys.
  Rev. X}\ }\textbf {\bibinfo {volume} {5}},\ \bibinfo {pages} {031013}
  (\bibinfo {year} {2015}{\natexlab{b}})}\BibitemShut {NoStop}%
\bibitem [{\citenamefont {Xiong}\ \emph {et~al.}(2015)\citenamefont {Xiong},
  \citenamefont {Kushwaha}, \citenamefont {Liang}, \citenamefont {Krizan},
  \citenamefont {Hirschberger}, \citenamefont {Wang}, \citenamefont {Cava},\
  and\ \citenamefont {Ong}}]{Xiong2015}%
  \BibitemOpen
  \bibfield  {author} {\bibinfo {author} {\bibfnamefont {J.}~\bibnamefont
  {Xiong}}, \bibinfo {author} {\bibfnamefont {S.~K.}\ \bibnamefont {Kushwaha}},
  \bibinfo {author} {\bibfnamefont {T.}~\bibnamefont {Liang}}, \bibinfo
  {author} {\bibfnamefont {J.~W.}\ \bibnamefont {Krizan}}, \bibinfo {author}
  {\bibfnamefont {M.}~\bibnamefont {Hirschberger}}, \bibinfo {author}
  {\bibfnamefont {W.}~\bibnamefont {Wang}}, \bibinfo {author} {\bibfnamefont
  {R.~J.}\ \bibnamefont {Cava}},\ and\ \bibinfo {author} {\bibfnamefont
  {N.~P.}\ \bibnamefont {Ong}},\ }\bibfield  {title} {\bibinfo {title}
  {Evidence for the chiral anomaly in the {Dirac} semimetal {Na$_3$Bi}},\
  }\href@noop {} {\bibfield  {journal} {\bibinfo  {journal} {Science}\ }\textbf
  {\bibinfo {volume} {350}},\ \bibinfo {pages} {413} (\bibinfo {year}
  {2015})}\BibitemShut {NoStop}%
\bibitem [{\citenamefont {Young}\ \emph {et~al.}(2012)\citenamefont {Young},
  \citenamefont {Zaheer}, \citenamefont {Teo}, \citenamefont {Kane},
  \citenamefont {Mele},\ and\ \citenamefont {Rappe}}]{Young12}%
  \BibitemOpen
  \bibfield  {author} {\bibinfo {author} {\bibfnamefont {S.~M.}\ \bibnamefont
  {Young}}, \bibinfo {author} {\bibfnamefont {S.}~\bibnamefont {Zaheer}},
  \bibinfo {author} {\bibfnamefont {J.~C.~Y.}\ \bibnamefont {Teo}}, \bibinfo
  {author} {\bibfnamefont {C.~L.}\ \bibnamefont {Kane}}, \bibinfo {author}
  {\bibfnamefont {E.~J.}\ \bibnamefont {Mele}},\ and\ \bibinfo {author}
  {\bibfnamefont {A.~M.}\ \bibnamefont {Rappe}},\ }\bibfield  {title} {\bibinfo
  {title} {{Dirac} semimetal in three dimensions},\ }\href
  {https://doi.org/10.1103/PhysRevLett.108.140405} {\bibfield  {journal}
  {\bibinfo  {journal} {Phys. Rev. Lett.}\ }\textbf {\bibinfo {volume} {108}},\
  \bibinfo {pages} {140405} (\bibinfo {year} {2012})}\BibitemShut {NoStop}%
\bibitem [{\citenamefont {Wang}\ \emph {et~al.}(2012)\citenamefont {Wang},
  \citenamefont {Sun}, \citenamefont {Chen}, \citenamefont {Franchini},
  \citenamefont {Xu}, \citenamefont {Weng}, \citenamefont {Dai},\ and\
  \citenamefont {Fang}}]{Wang12}%
  \BibitemOpen
  \bibfield  {author} {\bibinfo {author} {\bibfnamefont {Z.}~\bibnamefont
  {Wang}}, \bibinfo {author} {\bibfnamefont {Y.}~\bibnamefont {Sun}}, \bibinfo
  {author} {\bibfnamefont {X.-Q.}\ \bibnamefont {Chen}}, \bibinfo {author}
  {\bibfnamefont {C.}~\bibnamefont {Franchini}}, \bibinfo {author}
  {\bibfnamefont {G.}~\bibnamefont {Xu}}, \bibinfo {author} {\bibfnamefont
  {H.}~\bibnamefont {Weng}}, \bibinfo {author} {\bibfnamefont {X.}~\bibnamefont
  {Dai}},\ and\ \bibinfo {author} {\bibfnamefont {Z.}~\bibnamefont {Fang}},\
  }\bibfield  {title} {\bibinfo {title} {{Dirac} semimetal and topological
  phase transitions in {${A}_{3}$Bi ($A=\text{Na}$, K, Rb)}},\ }\href
  {https://doi.org/10.1103/PhysRevB.85.195320} {\bibfield  {journal} {\bibinfo
  {journal} {Phys. Rev. B}\ }\textbf {\bibinfo {volume} {85}},\ \bibinfo
  {pages} {195320} (\bibinfo {year} {2012})}\BibitemShut {NoStop}%
\bibitem [{\citenamefont {Liu}\ \emph {et~al.}(2014{\natexlab{a}})\citenamefont
  {Liu}, \citenamefont {Zhou}, \citenamefont {Zhang}, \citenamefont {Wang},
  \citenamefont {Weng}, \citenamefont {Prabhakaran}, \citenamefont {Mo},
  \citenamefont {Shen}, \citenamefont {Fang}, \citenamefont {Dai} \emph
  {et~al.}}]{Liu14}%
  \BibitemOpen
  \bibfield  {author} {\bibinfo {author} {\bibfnamefont {Z.}~\bibnamefont
  {Liu}}, \bibinfo {author} {\bibfnamefont {B.}~\bibnamefont {Zhou}}, \bibinfo
  {author} {\bibfnamefont {Y.}~\bibnamefont {Zhang}}, \bibinfo {author}
  {\bibfnamefont {Z.}~\bibnamefont {Wang}}, \bibinfo {author} {\bibfnamefont
  {H.}~\bibnamefont {Weng}}, \bibinfo {author} {\bibfnamefont {D.}~\bibnamefont
  {Prabhakaran}}, \bibinfo {author} {\bibfnamefont {S.-K.}\ \bibnamefont {Mo}},
  \bibinfo {author} {\bibfnamefont {Z.}~\bibnamefont {Shen}}, \bibinfo {author}
  {\bibfnamefont {Z.}~\bibnamefont {Fang}}, \bibinfo {author} {\bibfnamefont
  {X.}~\bibnamefont {Dai}}, \emph {et~al.},\ }\bibfield  {title} {\bibinfo
  {title} {Discovery of a three-dimensional topological {Dirac} semimetal,
  {Na$_3$Bi}},\ }\href@noop {} {\bibfield  {journal} {\bibinfo  {journal}
  {Science}\ }\textbf {\bibinfo {volume} {343}},\ \bibinfo {pages} {864}
  (\bibinfo {year} {2014}{\natexlab{a}})}\BibitemShut {NoStop}%
\bibitem [{\citenamefont {Liu}\ \emph {et~al.}(2014{\natexlab{b}})\citenamefont
  {Liu}, \citenamefont {Jiang}, \citenamefont {Zhou}, \citenamefont {Wang},
  \citenamefont {Zhang}, \citenamefont {Weng}, \citenamefont {Prabhakaran},
  \citenamefont {Mo}, \citenamefont {Peng}, \citenamefont {Dudin} \emph
  {et~al.}}]{Liu14a}%
  \BibitemOpen
  \bibfield  {author} {\bibinfo {author} {\bibfnamefont {Z.}~\bibnamefont
  {Liu}}, \bibinfo {author} {\bibfnamefont {J.}~\bibnamefont {Jiang}}, \bibinfo
  {author} {\bibfnamefont {B.}~\bibnamefont {Zhou}}, \bibinfo {author}
  {\bibfnamefont {Z.}~\bibnamefont {Wang}}, \bibinfo {author} {\bibfnamefont
  {Y.}~\bibnamefont {Zhang}}, \bibinfo {author} {\bibfnamefont
  {H.}~\bibnamefont {Weng}}, \bibinfo {author} {\bibfnamefont {D.}~\bibnamefont
  {Prabhakaran}}, \bibinfo {author} {\bibfnamefont {S.}~\bibnamefont {Mo}},
  \bibinfo {author} {\bibfnamefont {H.}~\bibnamefont {Peng}}, \bibinfo {author}
  {\bibfnamefont {P.}~\bibnamefont {Dudin}}, \emph {et~al.},\ }\bibfield
  {title} {\bibinfo {title} {A stable three-dimensional topological {Dirac}
  semimetal {Cd$_3$As$_2$}},\ }\href@noop {} {\bibfield  {journal} {\bibinfo
  {journal} {Nat. Mater.}\ }\textbf {\bibinfo {volume} {13}},\ \bibinfo {pages}
  {677} (\bibinfo {year} {2014}{\natexlab{b}})}\BibitemShut {NoStop}%
\bibitem [{\citenamefont {Steinberg}\ \emph {et~al.}(2014)\citenamefont
  {Steinberg}, \citenamefont {Young}, \citenamefont {Zaheer}, \citenamefont
  {Kane}, \citenamefont {Mele},\ and\ \citenamefont {Rappe}}]{Steinberg14}%
  \BibitemOpen
  \bibfield  {author} {\bibinfo {author} {\bibfnamefont {J.~A.}\ \bibnamefont
  {Steinberg}}, \bibinfo {author} {\bibfnamefont {S.~M.}\ \bibnamefont
  {Young}}, \bibinfo {author} {\bibfnamefont {S.}~\bibnamefont {Zaheer}},
  \bibinfo {author} {\bibfnamefont {C.~L.}\ \bibnamefont {Kane}}, \bibinfo
  {author} {\bibfnamefont {E.~J.}\ \bibnamefont {Mele}},\ and\ \bibinfo
  {author} {\bibfnamefont {A.~M.}\ \bibnamefont {Rappe}},\ }\bibfield  {title}
  {\bibinfo {title} {Bulk {Dirac} points in distorted spinels},\ }\href
  {https://doi.org/10.1103/PhysRevLett.112.036403} {\bibfield  {journal}
  {\bibinfo  {journal} {Phys. Rev. Lett.}\ }\textbf {\bibinfo {volume} {112}},\
  \bibinfo {pages} {036403} (\bibinfo {year} {2014})}\BibitemShut {NoStop}%
\bibitem [{\citenamefont {Bradlyn}\ \emph {et~al.}(2016)\citenamefont
  {Bradlyn}, \citenamefont {Cano}, \citenamefont {Wang}, \citenamefont
  {Vergniory}, \citenamefont {Felser}, \citenamefont {Cava},\ and\
  \citenamefont {Bernevig}}]{Bradlyn2016}%
  \BibitemOpen
  \bibfield  {author} {\bibinfo {author} {\bibfnamefont {B.}~\bibnamefont
  {Bradlyn}}, \bibinfo {author} {\bibfnamefont {J.}~\bibnamefont {Cano}},
  \bibinfo {author} {\bibfnamefont {Z.}~\bibnamefont {Wang}}, \bibinfo {author}
  {\bibfnamefont {M.~G.}\ \bibnamefont {Vergniory}}, \bibinfo {author}
  {\bibfnamefont {C.}~\bibnamefont {Felser}}, \bibinfo {author} {\bibfnamefont
  {R.~J.}\ \bibnamefont {Cava}},\ and\ \bibinfo {author} {\bibfnamefont
  {B.~A.}\ \bibnamefont {Bernevig}},\ }\bibfield  {title} {\bibinfo {title}
  {Beyond {Dirac} and {Weyl} fermions: Unconventional quasiparticles in
  conventional crystals},\ }\href {https://doi.org/10.1126/science.aaf5037}
  {\bibfield  {journal} {\bibinfo  {journal} {Science}\ }\textbf {\bibinfo
  {volume} {353}},\ \bibinfo {pages} {aaf5037} (\bibinfo {year}
  {2016})}\BibitemShut {NoStop}%
\bibitem [{\citenamefont {Chang}\ \emph {et~al.}(2017)\citenamefont {Chang},
  \citenamefont {Xu}, \citenamefont {Wieder}, \citenamefont {Sanchez},
  \citenamefont {Huang}, \citenamefont {Belopolski}, \citenamefont {Chang},
  \citenamefont {Zhang}, \citenamefont {Bansil}, \citenamefont {Lin},\ and\
  \citenamefont {Hasan}}]{Chang2017}%
  \BibitemOpen
  \bibfield  {author} {\bibinfo {author} {\bibfnamefont {G.}~\bibnamefont
  {Chang}}, \bibinfo {author} {\bibfnamefont {S.-Y.}\ \bibnamefont {Xu}},
  \bibinfo {author} {\bibfnamefont {B.~J.}\ \bibnamefont {Wieder}}, \bibinfo
  {author} {\bibfnamefont {D.~S.}\ \bibnamefont {Sanchez}}, \bibinfo {author}
  {\bibfnamefont {S.-M.}\ \bibnamefont {Huang}}, \bibinfo {author}
  {\bibfnamefont {I.}~\bibnamefont {Belopolski}}, \bibinfo {author}
  {\bibfnamefont {T.-R.}\ \bibnamefont {Chang}}, \bibinfo {author}
  {\bibfnamefont {S.}~\bibnamefont {Zhang}}, \bibinfo {author} {\bibfnamefont
  {A.}~\bibnamefont {Bansil}}, \bibinfo {author} {\bibfnamefont
  {H.}~\bibnamefont {Lin}},\ and\ \bibinfo {author} {\bibfnamefont {M.~Z.}\
  \bibnamefont {Hasan}},\ }\bibfield  {title} {\bibinfo {title} {Unconventional
  chiral fermions and large topological {Fermi} arcs in {RhSi}},\ }\href
  {https://doi.org/10.1103/PhysRevLett.119.206401} {\bibfield  {journal}
  {\bibinfo  {journal} {Phys. Rev. Lett.}\ }\textbf {\bibinfo {volume} {119}},\
  \bibinfo {pages} {206401} (\bibinfo {year} {2017})}\BibitemShut {NoStop}%
\bibitem [{\citenamefont {{Schr{\"o}ter}}\ \emph {et~al.}(2019)\citenamefont
  {{Schr{\"o}ter}}, \citenamefont {{Pei}}, \citenamefont {{Vergniory}},
  \citenamefont {{Sun}}, \citenamefont {{Manna}}, \citenamefont {{de Juan}},
  \citenamefont {{Krieger}}, \citenamefont {{S{\"u}ss}}, \citenamefont
  {{Schmidt}}, \citenamefont {{Dudin}}, \citenamefont {{Bradlyn}},
  \citenamefont {{Kim}}, \citenamefont {{Schmitt}}, \citenamefont {{Cacho}},
  \citenamefont {{Felser}}, \citenamefont {{Strocov}},\ and\ \citenamefont
  {{Chen}}}]{Schroter2018}%
  \BibitemOpen
  \bibfield  {author} {\bibinfo {author} {\bibfnamefont {N.~B.~M.}\
  \bibnamefont {{Schr{\"o}ter}}}, \bibinfo {author} {\bibfnamefont
  {D.}~\bibnamefont {{Pei}}}, \bibinfo {author} {\bibfnamefont {M.~G.}\
  \bibnamefont {{Vergniory}}}, \bibinfo {author} {\bibfnamefont
  {Y.}~\bibnamefont {{Sun}}}, \bibinfo {author} {\bibfnamefont
  {K.}~\bibnamefont {{Manna}}}, \bibinfo {author} {\bibfnamefont
  {F.}~\bibnamefont {{de Juan}}}, \bibinfo {author} {\bibfnamefont {J.~A.}\
  \bibnamefont {{Krieger}}}, \bibinfo {author} {\bibfnamefont {V.}~\bibnamefont
  {{S{\"u}ss}}}, \bibinfo {author} {\bibfnamefont {M.}~\bibnamefont
  {{Schmidt}}}, \bibinfo {author} {\bibfnamefont {P.}~\bibnamefont {{Dudin}}},
  \bibinfo {author} {\bibfnamefont {B.}~\bibnamefont {{Bradlyn}}}, \bibinfo
  {author} {\bibfnamefont {T.~K.}\ \bibnamefont {{Kim}}}, \bibinfo {author}
  {\bibfnamefont {T.}~\bibnamefont {{Schmitt}}}, \bibinfo {author}
  {\bibfnamefont {C.}~\bibnamefont {{Cacho}}}, \bibinfo {author} {\bibfnamefont
  {C.}~\bibnamefont {{Felser}}}, \bibinfo {author} {\bibfnamefont {V.~N.}\
  \bibnamefont {{Strocov}}},\ and\ \bibinfo {author} {\bibfnamefont
  {Y.}~\bibnamefont {{Chen}}},\ }\bibfield  {title} {\bibinfo {title} {Chiral
  topological semimetal with multifold band crossings and long {Fermi} arcs},\
  }\href {https://doi.org/10.1038/s41567-019-0511-y} {\bibfield  {journal}
  {\bibinfo  {journal} {Nat. Phys.}\ }\textbf {\bibinfo {volume} {15}},\
  \bibinfo {pages} {759} (\bibinfo {year} {2019})}\BibitemShut {NoStop}%
\bibitem [{\citenamefont {Rao}\ \emph {et~al.}(2019)\citenamefont {Rao},
  \citenamefont {Li}, \citenamefont {Zhang}, \citenamefont {Tian},
  \citenamefont {Li}, \citenamefont {Fu}, \citenamefont {Tang}, \citenamefont
  {Wang}, \citenamefont {Li}, \citenamefont {Fan}, \citenamefont {Li},
  \citenamefont {Huang}, \citenamefont {Liu}, \citenamefont {Long},
  \citenamefont {Fang}, \citenamefont {Weng}, \citenamefont {Shi},
  \citenamefont {Lei}, \citenamefont {Sun}, \citenamefont {Qian},\ and\
  \citenamefont {Ding}}]{rao2019observation}%
  \BibitemOpen
  \bibfield  {author} {\bibinfo {author} {\bibfnamefont {Z.}~\bibnamefont
  {Rao}}, \bibinfo {author} {\bibfnamefont {H.}~\bibnamefont {Li}}, \bibinfo
  {author} {\bibfnamefont {T.}~\bibnamefont {Zhang}}, \bibinfo {author}
  {\bibfnamefont {S.}~\bibnamefont {Tian}}, \bibinfo {author} {\bibfnamefont
  {C.}~\bibnamefont {Li}}, \bibinfo {author} {\bibfnamefont {B.}~\bibnamefont
  {Fu}}, \bibinfo {author} {\bibfnamefont {C.}~\bibnamefont {Tang}}, \bibinfo
  {author} {\bibfnamefont {L.}~\bibnamefont {Wang}}, \bibinfo {author}
  {\bibfnamefont {Z.}~\bibnamefont {Li}}, \bibinfo {author} {\bibfnamefont
  {W.}~\bibnamefont {Fan}}, \bibinfo {author} {\bibfnamefont {J.}~\bibnamefont
  {Li}}, \bibinfo {author} {\bibfnamefont {Y.}~\bibnamefont {Huang}}, \bibinfo
  {author} {\bibfnamefont {Z.}~\bibnamefont {Liu}}, \bibinfo {author}
  {\bibfnamefont {Y.}~\bibnamefont {Long}}, \bibinfo {author} {\bibfnamefont
  {C.}~\bibnamefont {Fang}}, \bibinfo {author} {\bibfnamefont {H.}~\bibnamefont
  {Weng}}, \bibinfo {author} {\bibfnamefont {Y.}~\bibnamefont {Shi}}, \bibinfo
  {author} {\bibfnamefont {H.}~\bibnamefont {Lei}}, \bibinfo {author}
  {\bibfnamefont {Y.}~\bibnamefont {Sun}}, \bibinfo {author} {\bibfnamefont
  {T.}~\bibnamefont {Qian}},\ and\ \bibinfo {author} {\bibfnamefont
  {H.}~\bibnamefont {Ding}},\ }\bibfield  {title} {\bibinfo {title}
  {Observation of unconventional chiral fermions with long {Fermi} arcs in
  {CoSi}},\ }\href {https://doi.org/10.1038/s41586-019-1031-8} {\bibfield
  {journal} {\bibinfo  {journal} {Nature}\ }\textbf {\bibinfo {volume} {567}},\
  \bibinfo {pages} {496} (\bibinfo {year} {2019})}\BibitemShut {NoStop}%
\bibitem [{\citenamefont {Sanchez}\ \emph {et~al.}(2019)\citenamefont
  {Sanchez}, \citenamefont {Belopolski}, \citenamefont {Cochran}, \citenamefont
  {Xu}, \citenamefont {Yin}, \citenamefont {Chang}, \citenamefont {Xie},
  \citenamefont {Manna}, \citenamefont {S{\"u}{\ss}}, \citenamefont {Huang},
  \citenamefont {Alidoust}, \citenamefont {Multer}, \citenamefont {Zhang},
  \citenamefont {Shumiya}, \citenamefont {Wang}, \citenamefont {Wang},
  \citenamefont {Chang}, \citenamefont {Felser}, \citenamefont {Xu},
  \citenamefont {Jia}, \citenamefont {Lin},\ and\ \citenamefont
  {Hasan}}]{sanchez2019topological}%
  \BibitemOpen
  \bibfield  {author} {\bibinfo {author} {\bibfnamefont {D.~S.}\ \bibnamefont
  {Sanchez}}, \bibinfo {author} {\bibfnamefont {I.}~\bibnamefont {Belopolski}},
  \bibinfo {author} {\bibfnamefont {T.~A.}\ \bibnamefont {Cochran}}, \bibinfo
  {author} {\bibfnamefont {X.}~\bibnamefont {Xu}}, \bibinfo {author}
  {\bibfnamefont {J.-X.}\ \bibnamefont {Yin}}, \bibinfo {author} {\bibfnamefont
  {G.}~\bibnamefont {Chang}}, \bibinfo {author} {\bibfnamefont
  {W.}~\bibnamefont {Xie}}, \bibinfo {author} {\bibfnamefont {K.}~\bibnamefont
  {Manna}}, \bibinfo {author} {\bibfnamefont {V.}~\bibnamefont {S{\"u}{\ss}}},
  \bibinfo {author} {\bibfnamefont {C.-Y.}\ \bibnamefont {Huang}}, \bibinfo
  {author} {\bibfnamefont {N.}~\bibnamefont {Alidoust}}, \bibinfo {author}
  {\bibfnamefont {D.}~\bibnamefont {Multer}}, \bibinfo {author} {\bibfnamefont
  {S.~S.}\ \bibnamefont {Zhang}}, \bibinfo {author} {\bibfnamefont
  {N.}~\bibnamefont {Shumiya}}, \bibinfo {author} {\bibfnamefont
  {X.}~\bibnamefont {Wang}}, \bibinfo {author} {\bibfnamefont {G.-Q.}\
  \bibnamefont {Wang}}, \bibinfo {author} {\bibfnamefont {T.-R.}\ \bibnamefont
  {Chang}}, \bibinfo {author} {\bibfnamefont {C.}~\bibnamefont {Felser}},
  \bibinfo {author} {\bibfnamefont {S.-Y.}\ \bibnamefont {Xu}}, \bibinfo
  {author} {\bibfnamefont {S.}~\bibnamefont {Jia}}, \bibinfo {author}
  {\bibfnamefont {H.}~\bibnamefont {Lin}},\ and\ \bibinfo {author}
  {\bibfnamefont {M.~Z.}\ \bibnamefont {Hasan}},\ }\bibfield  {title} {\bibinfo
  {title} {Topological chiral crystals with helicoid-arc quantum states},\
  }\href {https://doi.org/10.1038/s41586-019-1037-2} {\bibfield  {journal}
  {\bibinfo  {journal} {Nature}\ }\textbf {\bibinfo {volume} {567}},\ \bibinfo
  {pages} {500} (\bibinfo {year} {2019})}\BibitemShut {NoStop}%
\bibitem [{\citenamefont {Cano}\ \emph {et~al.}(2019)\citenamefont {Cano},
  \citenamefont {Bradlyn},\ and\ \citenamefont
  {Vergniory}}]{cano2019multifold}%
  \BibitemOpen
  \bibfield  {author} {\bibinfo {author} {\bibfnamefont {J.}~\bibnamefont
  {Cano}}, \bibinfo {author} {\bibfnamefont {B.}~\bibnamefont {Bradlyn}},\ and\
  \bibinfo {author} {\bibfnamefont {M.}~\bibnamefont {Vergniory}},\ }\bibfield
  {title} {\bibinfo {title} {Multifold nodal points in magnetic materials},\
  }\href@noop {} {\bibfield  {journal} {\bibinfo  {journal} {APL Mater.}\
  }\textbf {\bibinfo {volume} {7}},\ \bibinfo {pages} {101125} (\bibinfo {year}
  {2019})}\BibitemShut {NoStop}%
\bibitem [{\citenamefont {Klemenz}\ \emph {et~al.}(2020)\citenamefont
  {Klemenz}, \citenamefont {Schoop},\ and\ \citenamefont
  {Cano}}]{klemenz2020systematic}%
  \BibitemOpen
  \bibfield  {author} {\bibinfo {author} {\bibfnamefont {S.}~\bibnamefont
  {Klemenz}}, \bibinfo {author} {\bibfnamefont {L.}~\bibnamefont {Schoop}},\
  and\ \bibinfo {author} {\bibfnamefont {J.}~\bibnamefont {Cano}},\ }\bibfield
  {title} {\bibinfo {title} {Systematic study of stacked square nets: From
  {Dirac} fermions to material realizations},\ }\href@noop {} {\bibfield
  {journal} {\bibinfo  {journal} {Phys. Rev. B}\ }\textbf {\bibinfo {volume}
  {101}},\ \bibinfo {pages} {165121} (\bibinfo {year} {2020})}\BibitemShut
  {NoStop}%
\bibitem [{\citenamefont {Fu}\ \emph {et~al.}(2007)\citenamefont {Fu},
  \citenamefont {Kane},\ and\ \citenamefont {Mele}}]{fu2007topological}%
  \BibitemOpen
  \bibfield  {author} {\bibinfo {author} {\bibfnamefont {L.}~\bibnamefont
  {Fu}}, \bibinfo {author} {\bibfnamefont {C.~L.}\ \bibnamefont {Kane}},\ and\
  \bibinfo {author} {\bibfnamefont {E.~J.}\ \bibnamefont {Mele}},\ }\bibfield
  {title} {\bibinfo {title} {Topological insulators in three dimensions},\
  }\href {https://doi.org/10.1103/PhysRevLett.98.106803} {\bibfield  {journal}
  {\bibinfo  {journal} {Phys. Rev. Lett.}\ }\textbf {\bibinfo {volume} {98}},\
  \bibinfo {pages} {106803} (\bibinfo {year} {2007})}\BibitemShut {NoStop}%
\bibitem [{\citenamefont {Moore}\ and\ \citenamefont
  {Balents}(2007)}]{moore2007topological}%
  \BibitemOpen
  \bibfield  {author} {\bibinfo {author} {\bibfnamefont {J.~E.}\ \bibnamefont
  {Moore}}\ and\ \bibinfo {author} {\bibfnamefont {L.}~\bibnamefont
  {Balents}},\ }\bibfield  {title} {\bibinfo {title} {Topological invariants of
  time-reversal-invariant band structures},\ }\href
  {https://doi.org/10.1103/PhysRevB.75.121306} {\bibfield  {journal} {\bibinfo
  {journal} {Phys. Rev. B}\ }\textbf {\bibinfo {volume} {75}},\ \bibinfo
  {pages} {121306} (\bibinfo {year} {2007})}\BibitemShut {NoStop}%
\bibitem [{\citenamefont {Roy}(2009)}]{roy2009topological}%
  \BibitemOpen
  \bibfield  {author} {\bibinfo {author} {\bibfnamefont {R.}~\bibnamefont
  {Roy}},\ }\bibfield  {title} {\bibinfo {title} {Topological phases and the
  quantum spin {Hall} effect in three dimensions},\ }\href
  {https://doi.org/10.1103/PhysRevB.79.195322} {\bibfield  {journal} {\bibinfo
  {journal} {Phys. Rev. B}\ }\textbf {\bibinfo {volume} {79}},\ \bibinfo
  {pages} {195322} (\bibinfo {year} {2009})}\BibitemShut {NoStop}%
\bibitem [{\citenamefont {Qi}\ \emph {et~al.}(2008)\citenamefont {Qi},
  \citenamefont {Hughes},\ and\ \citenamefont {Zhang}}]{qi2008topological}%
  \BibitemOpen
  \bibfield  {author} {\bibinfo {author} {\bibfnamefont {X.-L.}\ \bibnamefont
  {Qi}}, \bibinfo {author} {\bibfnamefont {T.~L.}\ \bibnamefont {Hughes}},\
  and\ \bibinfo {author} {\bibfnamefont {S.-C.}\ \bibnamefont {Zhang}},\
  }\bibfield  {title} {\bibinfo {title} {Topological field theory of
  time-reversal invariant insulators},\ }\href
  {https://doi.org/10.1103/PhysRevB.78.195424} {\bibfield  {journal} {\bibinfo
  {journal} {Phys. Rev. B}\ }\textbf {\bibinfo {volume} {78}},\ \bibinfo
  {pages} {195424} (\bibinfo {year} {2008})}\BibitemShut {NoStop}%
\bibitem [{\citenamefont {Schnyder}\ \emph {et~al.}(2008)\citenamefont
  {Schnyder}, \citenamefont {Ryu}, \citenamefont {Furusaki},\ and\
  \citenamefont {Ludwig}}]{schnyder2008classification}%
  \BibitemOpen
  \bibfield  {author} {\bibinfo {author} {\bibfnamefont {A.~P.}\ \bibnamefont
  {Schnyder}}, \bibinfo {author} {\bibfnamefont {S.}~\bibnamefont {Ryu}},
  \bibinfo {author} {\bibfnamefont {A.}~\bibnamefont {Furusaki}},\ and\
  \bibinfo {author} {\bibfnamefont {A.~W.}\ \bibnamefont {Ludwig}},\ }\bibfield
   {title} {\bibinfo {title} {Classification of topological insulators and
  superconductors in three spatial dimensions},\ }\href@noop {} {\bibfield
  {journal} {\bibinfo  {journal} {Phys. Rev. B}\ }\textbf {\bibinfo {volume}
  {78}},\ \bibinfo {pages} {195125} (\bibinfo {year} {2008})}\BibitemShut
  {NoStop}%
\bibitem [{\citenamefont {Xia}\ \emph {et~al.}(2009)\citenamefont {Xia},
  \citenamefont {Qian}, \citenamefont {Hsieh}, \citenamefont {Wray},
  \citenamefont {Pal}, \citenamefont {Lin}, \citenamefont {Bansil},
  \citenamefont {Grauer}, \citenamefont {Hor}, \citenamefont {Cava} \emph
  {et~al.}}]{xia2009observation}%
  \BibitemOpen
  \bibfield  {author} {\bibinfo {author} {\bibfnamefont {Y.}~\bibnamefont
  {Xia}}, \bibinfo {author} {\bibfnamefont {D.}~\bibnamefont {Qian}}, \bibinfo
  {author} {\bibfnamefont {D.}~\bibnamefont {Hsieh}}, \bibinfo {author}
  {\bibfnamefont {L.}~\bibnamefont {Wray}}, \bibinfo {author} {\bibfnamefont
  {A.}~\bibnamefont {Pal}}, \bibinfo {author} {\bibfnamefont {H.}~\bibnamefont
  {Lin}}, \bibinfo {author} {\bibfnamefont {A.}~\bibnamefont {Bansil}},
  \bibinfo {author} {\bibfnamefont {D.}~\bibnamefont {Grauer}}, \bibinfo
  {author} {\bibfnamefont {Y.~S.}\ \bibnamefont {Hor}}, \bibinfo {author}
  {\bibfnamefont {R.~J.}\ \bibnamefont {Cava}}, \emph {et~al.},\ }\bibfield
  {title} {\bibinfo {title} {Observation of a large-gap topological-insulator
  class with a single {Dirac} cone on the surface},\ }\href@noop {} {\bibfield
  {journal} {\bibinfo  {journal} {Nat. Phys.}\ }\textbf {\bibinfo {volume}
  {5}},\ \bibinfo {pages} {398} (\bibinfo {year} {2009})}\BibitemShut {NoStop}%
\bibitem [{\citenamefont {Zhang}\ \emph {et~al.}(2009)\citenamefont {Zhang},
  \citenamefont {Liu}, \citenamefont {Qi}, \citenamefont {Dai}, \citenamefont
  {Fang},\ and\ \citenamefont {Zhang}}]{zhang2009topological}%
  \BibitemOpen
  \bibfield  {author} {\bibinfo {author} {\bibfnamefont {H.}~\bibnamefont
  {Zhang}}, \bibinfo {author} {\bibfnamefont {C.-X.}\ \bibnamefont {Liu}},
  \bibinfo {author} {\bibfnamefont {X.-L.}\ \bibnamefont {Qi}}, \bibinfo
  {author} {\bibfnamefont {X.}~\bibnamefont {Dai}}, \bibinfo {author}
  {\bibfnamefont {Z.}~\bibnamefont {Fang}},\ and\ \bibinfo {author}
  {\bibfnamefont {S.-C.}\ \bibnamefont {Zhang}},\ }\bibfield  {title} {\bibinfo
  {title} {Topological insulators in {Bi$_2$Se$_3$, Bi$_2$Te$_3$ and
  Sb$_2$Te$_3$} with a single {Dirac} cone on the surface},\ }\href
  {https://doi.org/10.1038/nphys1270} {\bibfield  {journal} {\bibinfo
  {journal} {Nat. Phys.}\ }\textbf {\bibinfo {volume} {5}},\ \bibinfo {pages}
  {438} (\bibinfo {year} {2009})}\BibitemShut {NoStop}%
\bibitem [{\citenamefont {Chen}\ \emph {et~al.}(2009)\citenamefont {Chen},
  \citenamefont {Analytis}, \citenamefont {Chu}, \citenamefont {Liu},
  \citenamefont {Mo}, \citenamefont {Qi}, \citenamefont {Zhang}, \citenamefont
  {Lu}, \citenamefont {Dai}, \citenamefont {Fang}, \citenamefont {Zhang},
  \citenamefont {Fisher}, \citenamefont {Hussain},\ and\ \citenamefont
  {Shen}}]{Chen2009experimental}%
  \BibitemOpen
  \bibfield  {author} {\bibinfo {author} {\bibfnamefont {Y.~L.}\ \bibnamefont
  {Chen}}, \bibinfo {author} {\bibfnamefont {J.~G.}\ \bibnamefont {Analytis}},
  \bibinfo {author} {\bibfnamefont {J.-H.}\ \bibnamefont {Chu}}, \bibinfo
  {author} {\bibfnamefont {Z.~K.}\ \bibnamefont {Liu}}, \bibinfo {author}
  {\bibfnamefont {S.-K.}\ \bibnamefont {Mo}}, \bibinfo {author} {\bibfnamefont
  {X.~L.}\ \bibnamefont {Qi}}, \bibinfo {author} {\bibfnamefont {H.~J.}\
  \bibnamefont {Zhang}}, \bibinfo {author} {\bibfnamefont {D.~H.}\ \bibnamefont
  {Lu}}, \bibinfo {author} {\bibfnamefont {X.}~\bibnamefont {Dai}}, \bibinfo
  {author} {\bibfnamefont {Z.}~\bibnamefont {Fang}}, \bibinfo {author}
  {\bibfnamefont {S.~C.}\ \bibnamefont {Zhang}}, \bibinfo {author}
  {\bibfnamefont {I.~R.}\ \bibnamefont {Fisher}}, \bibinfo {author}
  {\bibfnamefont {Z.}~\bibnamefont {Hussain}},\ and\ \bibinfo {author}
  {\bibfnamefont {Z.-X.}\ \bibnamefont {Shen}},\ }\bibfield  {title} {\bibinfo
  {title} {Experimental realization of a three-dimensional topological
  insulator, {Bi$_2$Te$_3$}},\ }\href {https://doi.org/10.1126/science.1173034}
  {\bibfield  {journal} {\bibinfo  {journal} {Science}\ }\textbf {\bibinfo
  {volume} {325}},\ \bibinfo {pages} {178} (\bibinfo {year}
  {2009})}\BibitemShut {NoStop}%
\bibitem [{\citenamefont {Hsieh}\ \emph {et~al.}(2009)\citenamefont {Hsieh},
  \citenamefont {Xia}, \citenamefont {Qian}, \citenamefont {Wray},
  \citenamefont {Meier}, \citenamefont {Dil}, \citenamefont {Osterwalder},
  \citenamefont {Patthey}, \citenamefont {Fedorov}, \citenamefont {Lin},
  \citenamefont {Bansil}, \citenamefont {Grauer}, \citenamefont {Hor},
  \citenamefont {Cava},\ and\ \citenamefont {Hasan}}]{hsieh2009observation}%
  \BibitemOpen
  \bibfield  {author} {\bibinfo {author} {\bibfnamefont {D.}~\bibnamefont
  {Hsieh}}, \bibinfo {author} {\bibfnamefont {Y.}~\bibnamefont {Xia}}, \bibinfo
  {author} {\bibfnamefont {D.}~\bibnamefont {Qian}}, \bibinfo {author}
  {\bibfnamefont {L.}~\bibnamefont {Wray}}, \bibinfo {author} {\bibfnamefont
  {F.}~\bibnamefont {Meier}}, \bibinfo {author} {\bibfnamefont {J.~H.}\
  \bibnamefont {Dil}}, \bibinfo {author} {\bibfnamefont {J.}~\bibnamefont
  {Osterwalder}}, \bibinfo {author} {\bibfnamefont {L.}~\bibnamefont
  {Patthey}}, \bibinfo {author} {\bibfnamefont {A.~V.}\ \bibnamefont
  {Fedorov}}, \bibinfo {author} {\bibfnamefont {H.}~\bibnamefont {Lin}},
  \bibinfo {author} {\bibfnamefont {A.}~\bibnamefont {Bansil}}, \bibinfo
  {author} {\bibfnamefont {D.}~\bibnamefont {Grauer}}, \bibinfo {author}
  {\bibfnamefont {Y.~S.}\ \bibnamefont {Hor}}, \bibinfo {author} {\bibfnamefont
  {R.~J.}\ \bibnamefont {Cava}},\ and\ \bibinfo {author} {\bibfnamefont
  {M.~Z.}\ \bibnamefont {Hasan}},\ }\bibfield  {title} {\bibinfo {title}
  {Observation of time-reversal-protected single-{Dirac}-cone
  topological-insulator states in {Bi$_2$Te$_3$} and {Sb$_2$Te$_3$}},\ }\href
  {https://doi.org/10.1103/PhysRevLett.103.146401} {\bibfield  {journal}
  {\bibinfo  {journal} {Phys. Rev. Lett.}\ }\textbf {\bibinfo {volume} {103}},\
  \bibinfo {pages} {146401} (\bibinfo {year} {2009})}\BibitemShut {NoStop}%
\bibitem [{\citenamefont {Hasan}\ and\ \citenamefont
  {Kane}(2010)}]{hasan2010colloquium}%
  \BibitemOpen
  \bibfield  {author} {\bibinfo {author} {\bibfnamefont {M.~Z.}\ \bibnamefont
  {Hasan}}\ and\ \bibinfo {author} {\bibfnamefont {C.~L.}\ \bibnamefont
  {Kane}},\ }\bibfield  {title} {\bibinfo {title} {Colloquium: Topological
  insulators},\ }\href {https://doi.org/10.1103/RevModPhys.82.3045} {\bibfield
  {journal} {\bibinfo  {journal} {Rev. Mod. Phys.}\ }\textbf {\bibinfo {volume}
  {82}},\ \bibinfo {pages} {3045} (\bibinfo {year} {2010})}\BibitemShut
  {NoStop}%
\bibitem [{\citenamefont {Andrei}\ and\ \citenamefont
  {MacDonald}(2020)}]{andrei2020graphene}%
  \BibitemOpen
  \bibfield  {author} {\bibinfo {author} {\bibfnamefont {E.~Y.}\ \bibnamefont
  {Andrei}}\ and\ \bibinfo {author} {\bibfnamefont {A.~H.}\ \bibnamefont
  {MacDonald}},\ }\href@noop {} {\bibinfo {title} {Graphene bilayers with a
  twist}} (\bibinfo {year} {2020}),\ \Eprint {https://arxiv.org/abs/2008.08129}
  {arXiv:2008.08129 [cond-mat.mes-hall]} \BibitemShut {NoStop}%
\bibitem [{\citenamefont {Bistritzer}\ and\ \citenamefont
  {MacDonald}(2011)}]{bistritzer2011moire}%
  \BibitemOpen
  \bibfield  {author} {\bibinfo {author} {\bibfnamefont {R.}~\bibnamefont
  {Bistritzer}}\ and\ \bibinfo {author} {\bibfnamefont {A.~H.}\ \bibnamefont
  {MacDonald}},\ }\bibfield  {title} {\bibinfo {title} {Moir{\'e} bands in
  twisted double-layer graphene},\ }\href@noop {} {\bibfield  {journal}
  {\bibinfo  {journal} {P. Natl. Acad. Sci. USA}\ }\textbf {\bibinfo {volume}
  {108}},\ \bibinfo {pages} {12233} (\bibinfo {year} {2011})}\BibitemShut
  {NoStop}%
\bibitem [{\citenamefont {Lopes~dos Santos}\ \emph
  {et~al.}(2012{\natexlab{a}})\citenamefont {Lopes~dos Santos}, \citenamefont
  {Peres},\ and\ \citenamefont {Castro~Neto}}]{Santos-2012}%
  \BibitemOpen
  \bibfield  {author} {\bibinfo {author} {\bibfnamefont {J.~M.~B.}\
  \bibnamefont {Lopes~dos Santos}}, \bibinfo {author} {\bibfnamefont
  {N.~M.~R.}\ \bibnamefont {Peres}},\ and\ \bibinfo {author} {\bibfnamefont
  {A.~H.}\ \bibnamefont {Castro~Neto}},\ }\bibfield  {title} {\bibinfo {title}
  {Continuum model of the twisted graphene bilayer},\ }\href
  {https://doi.org/10.1103/PhysRevB.86.155449} {\bibfield  {journal} {\bibinfo
  {journal} {Phys. Rev. B}\ }\textbf {\bibinfo {volume} {86}},\ \bibinfo
  {pages} {155449} (\bibinfo {year} {2012}{\natexlab{a}})}\BibitemShut
  {NoStop}%
\bibitem [{\citenamefont {Cao}\ \emph {et~al.}(2018{\natexlab{a}})\citenamefont
  {Cao}, \citenamefont {Fatemi}, \citenamefont {Fang}, \citenamefont
  {Watanabe}, \citenamefont {Taniguchi}, \citenamefont {Kaxiras},\ and\
  \citenamefont {Jarillo-Herrero}}]{cao2018unconventional}%
  \BibitemOpen
  \bibfield  {author} {\bibinfo {author} {\bibfnamefont {Y.}~\bibnamefont
  {Cao}}, \bibinfo {author} {\bibfnamefont {V.}~\bibnamefont {Fatemi}},
  \bibinfo {author} {\bibfnamefont {S.}~\bibnamefont {Fang}}, \bibinfo {author}
  {\bibfnamefont {K.}~\bibnamefont {Watanabe}}, \bibinfo {author}
  {\bibfnamefont {T.}~\bibnamefont {Taniguchi}}, \bibinfo {author}
  {\bibfnamefont {E.}~\bibnamefont {Kaxiras}},\ and\ \bibinfo {author}
  {\bibfnamefont {P.}~\bibnamefont {Jarillo-Herrero}},\ }\bibfield  {title}
  {\bibinfo {title} {Unconventional superconductivity in magic-angle graphene
  superlattices},\ }\href@noop {} {\bibfield  {journal} {\bibinfo  {journal}
  {Nature}\ }\textbf {\bibinfo {volume} {556}},\ \bibinfo {pages} {43}
  (\bibinfo {year} {2018}{\natexlab{a}})}\BibitemShut {NoStop}%
\bibitem [{\citenamefont {Cao}\ \emph {et~al.}(2018{\natexlab{b}})\citenamefont
  {Cao}, \citenamefont {Fatemi}, \citenamefont {Demir}, \citenamefont {Fang},
  \citenamefont {Tomarken}, \citenamefont {Luo}, \citenamefont
  {Sanchez-Yamagishi}, \citenamefont {Watanabe}, \citenamefont {Taniguchi},
  \citenamefont {Kaxiras}, \citenamefont {Ashoori},\ and\ \citenamefont
  {Jarillo-Herrero}}]{cao2018correlated}%
  \BibitemOpen
  \bibfield  {author} {\bibinfo {author} {\bibfnamefont {Y.}~\bibnamefont
  {Cao}}, \bibinfo {author} {\bibfnamefont {V.}~\bibnamefont {Fatemi}},
  \bibinfo {author} {\bibfnamefont {A.}~\bibnamefont {Demir}}, \bibinfo
  {author} {\bibfnamefont {S.}~\bibnamefont {Fang}}, \bibinfo {author}
  {\bibfnamefont {S.~L.}\ \bibnamefont {Tomarken}}, \bibinfo {author}
  {\bibfnamefont {J.~Y.}\ \bibnamefont {Luo}}, \bibinfo {author} {\bibfnamefont
  {J.~D.}\ \bibnamefont {Sanchez-Yamagishi}}, \bibinfo {author} {\bibfnamefont
  {K.}~\bibnamefont {Watanabe}}, \bibinfo {author} {\bibfnamefont
  {T.}~\bibnamefont {Taniguchi}}, \bibinfo {author} {\bibfnamefont
  {E.}~\bibnamefont {Kaxiras}}, \bibinfo {author} {\bibfnamefont {R.~C.}\
  \bibnamefont {Ashoori}},\ and\ \bibinfo {author} {\bibfnamefont
  {P.}~\bibnamefont {Jarillo-Herrero}},\ }\bibfield  {title} {\bibinfo {title}
  {Correlated insulator behaviour at half-filling in magic-angle graphene
  superlattices},\ }\href@noop {} {\bibfield  {journal} {\bibinfo  {journal}
  {Nature}\ }\textbf {\bibinfo {volume} {556}},\ \bibinfo {pages} {80}
  (\bibinfo {year} {2018}{\natexlab{b}})}\BibitemShut {NoStop}%
\bibitem [{\citenamefont {Yankowitz}\ \emph {et~al.}(2019)\citenamefont
  {Yankowitz}, \citenamefont {Chen}, \citenamefont {Polshyn}, \citenamefont
  {Zhang}, \citenamefont {Watanabe}, \citenamefont {Taniguchi}, \citenamefont
  {Graf}, \citenamefont {Young},\ and\ \citenamefont
  {Dean}}]{yankowitz2019tuning}%
  \BibitemOpen
  \bibfield  {author} {\bibinfo {author} {\bibfnamefont {M.}~\bibnamefont
  {Yankowitz}}, \bibinfo {author} {\bibfnamefont {S.}~\bibnamefont {Chen}},
  \bibinfo {author} {\bibfnamefont {H.}~\bibnamefont {Polshyn}}, \bibinfo
  {author} {\bibfnamefont {Y.}~\bibnamefont {Zhang}}, \bibinfo {author}
  {\bibfnamefont {K.}~\bibnamefont {Watanabe}}, \bibinfo {author}
  {\bibfnamefont {T.}~\bibnamefont {Taniguchi}}, \bibinfo {author}
  {\bibfnamefont {D.}~\bibnamefont {Graf}}, \bibinfo {author} {\bibfnamefont
  {A.~F.}\ \bibnamefont {Young}},\ and\ \bibinfo {author} {\bibfnamefont
  {C.~R.}\ \bibnamefont {Dean}},\ }\bibfield  {title} {\bibinfo {title} {Tuning
  superconductivity in twisted bilayer graphene},\ }\href@noop {} {\bibfield
  {journal} {\bibinfo  {journal} {Science}\ }\textbf {\bibinfo {volume}
  {363}},\ \bibinfo {pages} {1059} (\bibinfo {year} {2019})}\BibitemShut
  {NoStop}%
\bibitem [{\citenamefont {Lu}\ \emph {et~al.}(2019)\citenamefont {Lu},
  \citenamefont {Stepanov}, \citenamefont {Yang}, \citenamefont {Xie},
  \citenamefont {Aamir}, \citenamefont {Das}, \citenamefont {Urgell},
  \citenamefont {Watanabe}, \citenamefont {Taniguchi}, \citenamefont {Zhang},
  \citenamefont {Bachtold}, \citenamefont {MacDonald},\ and\ \citenamefont
  {Efetov}}]{lu2019superconductors}%
  \BibitemOpen
  \bibfield  {author} {\bibinfo {author} {\bibfnamefont {X.}~\bibnamefont
  {Lu}}, \bibinfo {author} {\bibfnamefont {P.}~\bibnamefont {Stepanov}},
  \bibinfo {author} {\bibfnamefont {W.}~\bibnamefont {Yang}}, \bibinfo {author}
  {\bibfnamefont {M.}~\bibnamefont {Xie}}, \bibinfo {author} {\bibfnamefont
  {M.~A.}\ \bibnamefont {Aamir}}, \bibinfo {author} {\bibfnamefont
  {I.}~\bibnamefont {Das}}, \bibinfo {author} {\bibfnamefont {C.}~\bibnamefont
  {Urgell}}, \bibinfo {author} {\bibfnamefont {K.}~\bibnamefont {Watanabe}},
  \bibinfo {author} {\bibfnamefont {T.}~\bibnamefont {Taniguchi}}, \bibinfo
  {author} {\bibfnamefont {G.}~\bibnamefont {Zhang}}, \bibinfo {author}
  {\bibfnamefont {A.}~\bibnamefont {Bachtold}}, \bibinfo {author}
  {\bibfnamefont {A.~H.}\ \bibnamefont {MacDonald}},\ and\ \bibinfo {author}
  {\bibfnamefont {D.~K.}\ \bibnamefont {Efetov}},\ }\bibfield  {title}
  {\bibinfo {title} {Superconductors, orbital magnets and correlated states in
  magic-angle bilayer graphene},\ }\href@noop {} {\bibfield  {journal}
  {\bibinfo  {journal} {Nature}\ }\textbf {\bibinfo {volume} {574}},\ \bibinfo
  {pages} {653} (\bibinfo {year} {2019})}\BibitemShut {NoStop}%
\bibitem [{\citenamefont {Sharpe}\ \emph {et~al.}(2019)\citenamefont {Sharpe},
  \citenamefont {Fox}, \citenamefont {Barnard}, \citenamefont {Finney},
  \citenamefont {Watanabe}, \citenamefont {Taniguchi}, \citenamefont
  {Kastner},\ and\ \citenamefont {Goldhaber-Gordon}}]{sharpe2019emergent}%
  \BibitemOpen
  \bibfield  {author} {\bibinfo {author} {\bibfnamefont {A.~L.}\ \bibnamefont
  {Sharpe}}, \bibinfo {author} {\bibfnamefont {E.~J.}\ \bibnamefont {Fox}},
  \bibinfo {author} {\bibfnamefont {A.~W.}\ \bibnamefont {Barnard}}, \bibinfo
  {author} {\bibfnamefont {J.}~\bibnamefont {Finney}}, \bibinfo {author}
  {\bibfnamefont {K.}~\bibnamefont {Watanabe}}, \bibinfo {author}
  {\bibfnamefont {T.}~\bibnamefont {Taniguchi}}, \bibinfo {author}
  {\bibfnamefont {M.}~\bibnamefont {Kastner}},\ and\ \bibinfo {author}
  {\bibfnamefont {D.}~\bibnamefont {Goldhaber-Gordon}},\ }\bibfield  {title}
  {\bibinfo {title} {Emergent ferromagnetism near three-quarters filling in
  twisted bilayer graphene},\ }\href@noop {} {\bibfield  {journal} {\bibinfo
  {journal} {Science}\ }\textbf {\bibinfo {volume} {365}},\ \bibinfo {pages}
  {605} (\bibinfo {year} {2019})}\BibitemShut {NoStop}%
\bibitem [{\citenamefont {Serlin}\ \emph {et~al.}(2020)\citenamefont {Serlin},
  \citenamefont {Tschirhart}, \citenamefont {Polshyn}, \citenamefont {Zhang},
  \citenamefont {Zhu}, \citenamefont {Watanabe}, \citenamefont {Taniguchi},
  \citenamefont {Balents},\ and\ \citenamefont {Young}}]{serlin2019intrinsic}%
  \BibitemOpen
  \bibfield  {author} {\bibinfo {author} {\bibfnamefont {M.}~\bibnamefont
  {Serlin}}, \bibinfo {author} {\bibfnamefont {C.}~\bibnamefont {Tschirhart}},
  \bibinfo {author} {\bibfnamefont {H.}~\bibnamefont {Polshyn}}, \bibinfo
  {author} {\bibfnamefont {Y.}~\bibnamefont {Zhang}}, \bibinfo {author}
  {\bibfnamefont {J.}~\bibnamefont {Zhu}}, \bibinfo {author} {\bibfnamefont
  {K.}~\bibnamefont {Watanabe}}, \bibinfo {author} {\bibfnamefont
  {T.}~\bibnamefont {Taniguchi}}, \bibinfo {author} {\bibfnamefont
  {L.}~\bibnamefont {Balents}},\ and\ \bibinfo {author} {\bibfnamefont
  {A.}~\bibnamefont {Young}},\ }\bibfield  {title} {\bibinfo {title} {Intrinsic
  quantized anomalous {Hall} effect in a moir{\'e} heterostructure},\ }\href
  {https://doi.org/10.1126/science.aay5533} {\bibfield  {journal} {\bibinfo
  {journal} {Science}\ }\textbf {\bibinfo {volume} {367}},\ \bibinfo {pages}
  {900} (\bibinfo {year} {2020})}\BibitemShut {NoStop}%
\bibitem [{\citenamefont {Kerelsky}\ \emph {et~al.}(2019)\citenamefont
  {Kerelsky}, \citenamefont {McGilly}, \citenamefont {Kennes}, \citenamefont
  {Xian}, \citenamefont {Yankowitz}, \citenamefont {Chen}, \citenamefont
  {Watanabe}, \citenamefont {Taniguchi}, \citenamefont {Hone}, \citenamefont
  {Dean}, \citenamefont {Rubio},\ and\ \citenamefont
  {Pasupathy}}]{kerelsky2019maximized}%
  \BibitemOpen
  \bibfield  {author} {\bibinfo {author} {\bibfnamefont {A.}~\bibnamefont
  {Kerelsky}}, \bibinfo {author} {\bibfnamefont {L.~J.}\ \bibnamefont
  {McGilly}}, \bibinfo {author} {\bibfnamefont {D.~M.}\ \bibnamefont {Kennes}},
  \bibinfo {author} {\bibfnamefont {L.}~\bibnamefont {Xian}}, \bibinfo {author}
  {\bibfnamefont {M.}~\bibnamefont {Yankowitz}}, \bibinfo {author}
  {\bibfnamefont {S.}~\bibnamefont {Chen}}, \bibinfo {author} {\bibfnamefont
  {K.}~\bibnamefont {Watanabe}}, \bibinfo {author} {\bibfnamefont
  {T.}~\bibnamefont {Taniguchi}}, \bibinfo {author} {\bibfnamefont
  {J.}~\bibnamefont {Hone}}, \bibinfo {author} {\bibfnamefont {C.}~\bibnamefont
  {Dean}}, \bibinfo {author} {\bibfnamefont {A.}~\bibnamefont {Rubio}},\ and\
  \bibinfo {author} {\bibfnamefont {A.~N.}\ \bibnamefont {Pasupathy}},\
  }\bibfield  {title} {\bibinfo {title} {Maximized electron interactions at the
  magic angle in twisted bilayer graphene},\ }\href@noop {} {\bibfield
  {journal} {\bibinfo  {journal} {Nature}\ }\textbf {\bibinfo {volume} {572}},\
  \bibinfo {pages} {95} (\bibinfo {year} {2019})}\BibitemShut {NoStop}%
\bibitem [{\citenamefont {Xie}\ \emph {et~al.}(2019)\citenamefont {Xie},
  \citenamefont {Lian}, \citenamefont {J{\"a}ck}, \citenamefont {Liu},
  \citenamefont {Chiu}, \citenamefont {Watanabe}, \citenamefont {Taniguchi},
  \citenamefont {Bernevig},\ and\ \citenamefont
  {Yazdani}}]{xie2019spectroscopic}%
  \BibitemOpen
  \bibfield  {author} {\bibinfo {author} {\bibfnamefont {Y.}~\bibnamefont
  {Xie}}, \bibinfo {author} {\bibfnamefont {B.}~\bibnamefont {Lian}}, \bibinfo
  {author} {\bibfnamefont {B.}~\bibnamefont {J{\"a}ck}}, \bibinfo {author}
  {\bibfnamefont {X.}~\bibnamefont {Liu}}, \bibinfo {author} {\bibfnamefont
  {C.-L.}\ \bibnamefont {Chiu}}, \bibinfo {author} {\bibfnamefont
  {K.}~\bibnamefont {Watanabe}}, \bibinfo {author} {\bibfnamefont
  {T.}~\bibnamefont {Taniguchi}}, \bibinfo {author} {\bibfnamefont {B.~A.}\
  \bibnamefont {Bernevig}},\ and\ \bibinfo {author} {\bibfnamefont
  {A.}~\bibnamefont {Yazdani}},\ }\bibfield  {title} {\bibinfo {title}
  {Spectroscopic signatures of many-body correlations in magic-angle twisted
  bilayer graphene},\ }\href@noop {} {\bibfield  {journal} {\bibinfo  {journal}
  {Nature}\ }\textbf {\bibinfo {volume} {572}},\ \bibinfo {pages} {101}
  (\bibinfo {year} {2019})}\BibitemShut {NoStop}%
\bibitem [{\citenamefont {Jiang}\ \emph {et~al.}(2019)\citenamefont {Jiang},
  \citenamefont {Lai}, \citenamefont {Watanabe}, \citenamefont {Taniguchi},
  \citenamefont {Haule}, \citenamefont {Mao},\ and\ \citenamefont
  {Andrei}}]{jiang2019charge}%
  \BibitemOpen
  \bibfield  {author} {\bibinfo {author} {\bibfnamefont {Y.}~\bibnamefont
  {Jiang}}, \bibinfo {author} {\bibfnamefont {X.}~\bibnamefont {Lai}}, \bibinfo
  {author} {\bibfnamefont {K.}~\bibnamefont {Watanabe}}, \bibinfo {author}
  {\bibfnamefont {T.}~\bibnamefont {Taniguchi}}, \bibinfo {author}
  {\bibfnamefont {K.}~\bibnamefont {Haule}}, \bibinfo {author} {\bibfnamefont
  {J.}~\bibnamefont {Mao}},\ and\ \bibinfo {author} {\bibfnamefont {E.~Y.}\
  \bibnamefont {Andrei}},\ }\bibfield  {title} {\bibinfo {title} {Charge order
  and broken rotational symmetry in magic-angle twisted bilayer graphene},\
  }\href@noop {} {\bibfield  {journal} {\bibinfo  {journal} {Nature}\ }\textbf
  {\bibinfo {volume} {573}},\ \bibinfo {pages} {91} (\bibinfo {year}
  {2019})}\BibitemShut {NoStop}%
\bibitem [{\citenamefont {Tang}\ \emph {et~al.}(2020)\citenamefont {Tang},
  \citenamefont {Li}, \citenamefont {Li}, \citenamefont {Xu}, \citenamefont
  {Liu}, \citenamefont {Barmak}, \citenamefont {Watanabe}, \citenamefont
  {Taniguchi}, \citenamefont {MacDonald}, \citenamefont {Shan},\ and\
  \citenamefont {Mak}}]{Tang2020simulation}%
  \BibitemOpen
  \bibfield  {author} {\bibinfo {author} {\bibfnamefont {Y.}~\bibnamefont
  {Tang}}, \bibinfo {author} {\bibfnamefont {L.}~\bibnamefont {Li}}, \bibinfo
  {author} {\bibfnamefont {T.}~\bibnamefont {Li}}, \bibinfo {author}
  {\bibfnamefont {Y.}~\bibnamefont {Xu}}, \bibinfo {author} {\bibfnamefont
  {S.}~\bibnamefont {Liu}}, \bibinfo {author} {\bibfnamefont {K.}~\bibnamefont
  {Barmak}}, \bibinfo {author} {\bibfnamefont {K.}~\bibnamefont {Watanabe}},
  \bibinfo {author} {\bibfnamefont {T.}~\bibnamefont {Taniguchi}}, \bibinfo
  {author} {\bibfnamefont {A.~H.}\ \bibnamefont {MacDonald}}, \bibinfo {author}
  {\bibfnamefont {J.}~\bibnamefont {Shan}},\ and\ \bibinfo {author}
  {\bibfnamefont {K.~F.}\ \bibnamefont {Mak}},\ }\bibfield  {title} {\bibinfo
  {title} {Simulation of {Hubbard} model physics in {WSe$_2$/WS$_2$} moir{\'e}
  superlattices},\ }\href {https://doi.org/10.1038/s41586-020-2085-3}
  {\bibfield  {journal} {\bibinfo  {journal} {Nature}\ }\textbf {\bibinfo
  {volume} {579}},\ \bibinfo {pages} {353} (\bibinfo {year}
  {2020})}\BibitemShut {NoStop}%
\bibitem [{\citenamefont {Regan}\ \emph {et~al.}(2020)\citenamefont {Regan},
  \citenamefont {Wang}, \citenamefont {Jin}, \citenamefont {Bakti~Utama},
  \citenamefont {Gao}, \citenamefont {Wei}, \citenamefont {Zhao}, \citenamefont
  {Zhao}, \citenamefont {Zhang}, \citenamefont {Yumigeta}, \citenamefont
  {Blei}, \citenamefont {Carlstr{\"o}m}, \citenamefont {Watanabe},
  \citenamefont {Taniguchi}, \citenamefont {Tongay}, \citenamefont {Crommie},
  \citenamefont {Zettl},\ and\ \citenamefont {Wang}}]{regan20mott}%
  \BibitemOpen
  \bibfield  {author} {\bibinfo {author} {\bibfnamefont {E.~C.}\ \bibnamefont
  {Regan}}, \bibinfo {author} {\bibfnamefont {D.}~\bibnamefont {Wang}},
  \bibinfo {author} {\bibfnamefont {C.}~\bibnamefont {Jin}}, \bibinfo {author}
  {\bibfnamefont {M.~I.}\ \bibnamefont {Bakti~Utama}}, \bibinfo {author}
  {\bibfnamefont {B.}~\bibnamefont {Gao}}, \bibinfo {author} {\bibfnamefont
  {X.}~\bibnamefont {Wei}}, \bibinfo {author} {\bibfnamefont {S.}~\bibnamefont
  {Zhao}}, \bibinfo {author} {\bibfnamefont {W.}~\bibnamefont {Zhao}}, \bibinfo
  {author} {\bibfnamefont {Z.}~\bibnamefont {Zhang}}, \bibinfo {author}
  {\bibfnamefont {K.}~\bibnamefont {Yumigeta}}, \bibinfo {author}
  {\bibfnamefont {M.}~\bibnamefont {Blei}}, \bibinfo {author} {\bibfnamefont
  {J.~D.}\ \bibnamefont {Carlstr{\"o}m}}, \bibinfo {author} {\bibfnamefont
  {K.}~\bibnamefont {Watanabe}}, \bibinfo {author} {\bibfnamefont
  {T.}~\bibnamefont {Taniguchi}}, \bibinfo {author} {\bibfnamefont
  {S.}~\bibnamefont {Tongay}}, \bibinfo {author} {\bibfnamefont
  {M.}~\bibnamefont {Crommie}}, \bibinfo {author} {\bibfnamefont
  {A.}~\bibnamefont {Zettl}},\ and\ \bibinfo {author} {\bibfnamefont
  {F.}~\bibnamefont {Wang}},\ }\bibfield  {title} {\bibinfo {title} {{Mott} and
  generalized {Wigner} crystal states in {WSe$_2$/WS$_2$} moir{\'e}
  superlattices},\ }\href {https://doi.org/10.1038/s41586-020-2092-4}
  {\bibfield  {journal} {\bibinfo  {journal} {Nature}\ }\textbf {\bibinfo
  {volume} {579}},\ \bibinfo {pages} {359} (\bibinfo {year}
  {2020})}\BibitemShut {NoStop}%
\bibitem [{\citenamefont {Shimazaki}\ \emph {et~al.}(2020)\citenamefont
  {Shimazaki}, \citenamefont {Schwartz}, \citenamefont {Watanabe},
  \citenamefont {Taniguchi}, \citenamefont {Kroner},\ and\ \citenamefont
  {Imamo{\u g}lu}}]{shimazaki2020strongly}%
  \BibitemOpen
  \bibfield  {author} {\bibinfo {author} {\bibfnamefont {Y.}~\bibnamefont
  {Shimazaki}}, \bibinfo {author} {\bibfnamefont {I.}~\bibnamefont {Schwartz}},
  \bibinfo {author} {\bibfnamefont {K.}~\bibnamefont {Watanabe}}, \bibinfo
  {author} {\bibfnamefont {T.}~\bibnamefont {Taniguchi}}, \bibinfo {author}
  {\bibfnamefont {M.}~\bibnamefont {Kroner}},\ and\ \bibinfo {author}
  {\bibfnamefont {A.}~\bibnamefont {Imamo{\u g}lu}},\ }\bibfield  {title}
  {\bibinfo {title} {Strongly correlated electrons and hybrid excitons in a
  moir{\'e} heterostructure},\ }\href
  {https://doi.org/10.1038/s41586-020-2191-2} {\bibfield  {journal} {\bibinfo
  {journal} {Nature}\ }\textbf {\bibinfo {volume} {580}},\ \bibinfo {pages}
  {472} (\bibinfo {year} {2020})}\BibitemShut {NoStop}%
\bibitem [{\citenamefont {Wang}\ \emph
  {et~al.}(2020{\natexlab{a}})\citenamefont {Wang}, \citenamefont {Shih},
  \citenamefont {Ghiotto}, \citenamefont {Xian}, \citenamefont {Rhodes},
  \citenamefont {Tan}, \citenamefont {Claassen}, \citenamefont {Kennes},
  \citenamefont {Bai}, \citenamefont {Kim}, \citenamefont {Watanabe},
  \citenamefont {Taniguchi}, \citenamefont {Zhu}, \citenamefont {Hone},
  \citenamefont {Rubio}, \citenamefont {Pasupathy},\ and\ \citenamefont
  {Dean}}]{wang2019magic}%
  \BibitemOpen
  \bibfield  {author} {\bibinfo {author} {\bibfnamefont {L.}~\bibnamefont
  {Wang}}, \bibinfo {author} {\bibfnamefont {E.-M.}\ \bibnamefont {Shih}},
  \bibinfo {author} {\bibfnamefont {A.}~\bibnamefont {Ghiotto}}, \bibinfo
  {author} {\bibfnamefont {L.}~\bibnamefont {Xian}}, \bibinfo {author}
  {\bibfnamefont {D.~A.}\ \bibnamefont {Rhodes}}, \bibinfo {author}
  {\bibfnamefont {C.}~\bibnamefont {Tan}}, \bibinfo {author} {\bibfnamefont
  {M.}~\bibnamefont {Claassen}}, \bibinfo {author} {\bibfnamefont {D.~M.}\
  \bibnamefont {Kennes}}, \bibinfo {author} {\bibfnamefont {Y.}~\bibnamefont
  {Bai}}, \bibinfo {author} {\bibfnamefont {B.}~\bibnamefont {Kim}}, \bibinfo
  {author} {\bibfnamefont {K.}~\bibnamefont {Watanabe}}, \bibinfo {author}
  {\bibfnamefont {T.}~\bibnamefont {Taniguchi}}, \bibinfo {author}
  {\bibfnamefont {X.}~\bibnamefont {Zhu}}, \bibinfo {author} {\bibfnamefont
  {J.}~\bibnamefont {Hone}}, \bibinfo {author} {\bibfnamefont {A.}~\bibnamefont
  {Rubio}}, \bibinfo {author} {\bibfnamefont {A.}~\bibnamefont {Pasupathy}},\
  and\ \bibinfo {author} {\bibfnamefont {C.~R.}\ \bibnamefont {Dean}},\
  }\bibfield  {title} {\bibinfo {title} {Correlated electronic phases in
  twisted bilayer transition metal dichalcogenides},\ }\href
  {https://doi.org/10.1038/s41563-020-0708-6} {\bibfield  {journal} {\bibinfo
  {journal} {Nat. Mater.}\ }\textbf {\bibinfo {volume} {19}},\ \bibinfo {pages}
  {861} (\bibinfo {year} {2020}{\natexlab{a}})}\BibitemShut {NoStop}%
\bibitem [{\citenamefont {Tsai}\ \emph {et~al.}(2019)\citenamefont {Tsai},
  \citenamefont {Zhang}, \citenamefont {Zhu}, \citenamefont {Luo},
  \citenamefont {Carr}, \citenamefont {Luskin}, \citenamefont {Kaxiras},\ and\
  \citenamefont {Wang}}]{tsai2019correlated}%
  \BibitemOpen
  \bibfield  {author} {\bibinfo {author} {\bibfnamefont {K.-T.}\ \bibnamefont
  {Tsai}}, \bibinfo {author} {\bibfnamefont {X.}~\bibnamefont {Zhang}},
  \bibinfo {author} {\bibfnamefont {Z.}~\bibnamefont {Zhu}}, \bibinfo {author}
  {\bibfnamefont {Y.}~\bibnamefont {Luo}}, \bibinfo {author} {\bibfnamefont
  {S.}~\bibnamefont {Carr}}, \bibinfo {author} {\bibfnamefont {M.}~\bibnamefont
  {Luskin}}, \bibinfo {author} {\bibfnamefont {E.}~\bibnamefont {Kaxiras}},\
  and\ \bibinfo {author} {\bibfnamefont {K.}~\bibnamefont {Wang}},\ }\bibfield
  {title} {\bibinfo {title} {Correlated superconducting and insulating states
  in twisted trilayer graphene moire of moire superlattices},\ }\href@noop {}
  {\bibfield  {journal} {\bibinfo  {journal} {arXiv preprint arXiv:1912.03375}\
  } (\bibinfo {year} {2019})}\BibitemShut {NoStop}%
\bibitem [{\citenamefont {Cao}\ \emph {et~al.}(2020)\citenamefont {Cao},
  \citenamefont {Rodan-Legrain}, \citenamefont {Rubies-Bigorda}, \citenamefont
  {Park}, \citenamefont {Watanabe}, \citenamefont {Taniguchi},\ and\
  \citenamefont {Jarillo-Herrero}}]{cao2019electric}%
  \BibitemOpen
  \bibfield  {author} {\bibinfo {author} {\bibfnamefont {Y.}~\bibnamefont
  {Cao}}, \bibinfo {author} {\bibfnamefont {D.}~\bibnamefont {Rodan-Legrain}},
  \bibinfo {author} {\bibfnamefont {O.}~\bibnamefont {Rubies-Bigorda}},
  \bibinfo {author} {\bibfnamefont {J.~M.}\ \bibnamefont {Park}}, \bibinfo
  {author} {\bibfnamefont {K.}~\bibnamefont {Watanabe}}, \bibinfo {author}
  {\bibfnamefont {T.}~\bibnamefont {Taniguchi}},\ and\ \bibinfo {author}
  {\bibfnamefont {P.}~\bibnamefont {Jarillo-Herrero}},\ }\bibfield  {title}
  {\bibinfo {title} {Tunable correlated states and spin-polarized phases in
  twisted bilayer--bilayer graphene},\ }\href
  {https://doi.org/10.1038/s41586-020-2260-6} {\bibfield  {journal} {\bibinfo
  {journal} {Nature}\ }\textbf {\bibinfo {volume} {583}},\ \bibinfo {pages}
  {215} (\bibinfo {year} {2020})}\BibitemShut {NoStop}%
\bibitem [{\citenamefont {Burg}\ \emph {et~al.}(2019)\citenamefont {Burg},
  \citenamefont {Zhu}, \citenamefont {Taniguchi}, \citenamefont {Watanabe},
  \citenamefont {MacDonald},\ and\ \citenamefont {Tutuc}}]{burg2019correlated}%
  \BibitemOpen
  \bibfield  {author} {\bibinfo {author} {\bibfnamefont {G.~W.}\ \bibnamefont
  {Burg}}, \bibinfo {author} {\bibfnamefont {J.}~\bibnamefont {Zhu}}, \bibinfo
  {author} {\bibfnamefont {T.}~\bibnamefont {Taniguchi}}, \bibinfo {author}
  {\bibfnamefont {K.}~\bibnamefont {Watanabe}}, \bibinfo {author}
  {\bibfnamefont {A.~H.}\ \bibnamefont {MacDonald}},\ and\ \bibinfo {author}
  {\bibfnamefont {E.}~\bibnamefont {Tutuc}},\ }\bibfield  {title} {\bibinfo
  {title} {Correlated insulating states in twisted double bilayer graphene},\
  }\href {https://doi.org/10.1103/PhysRevLett.123.197702} {\bibfield  {journal}
  {\bibinfo  {journal} {Phys. Rev. Lett.}\ }\textbf {\bibinfo {volume} {123}},\
  \bibinfo {pages} {197702} (\bibinfo {year} {2019})}\BibitemShut {NoStop}%
\bibitem [{\citenamefont {Shen}\ \emph {et~al.}(2020)\citenamefont {Shen},
  \citenamefont {Chu}, \citenamefont {Wu}, \citenamefont {Li}, \citenamefont
  {Wang}, \citenamefont {Zhao}, \citenamefont {Tang}, \citenamefont {Liu},
  \citenamefont {Tian}, \citenamefont {Watanabe}, \citenamefont {Taniguchi},
  \citenamefont {Yang}, \citenamefont {Meng}, \citenamefont {Shi},
  \citenamefont {Yazyev},\ and\ \citenamefont {Zhang}}]{shen2020correlated}%
  \BibitemOpen
  \bibfield  {author} {\bibinfo {author} {\bibfnamefont {C.}~\bibnamefont
  {Shen}}, \bibinfo {author} {\bibfnamefont {Y.}~\bibnamefont {Chu}}, \bibinfo
  {author} {\bibfnamefont {Q.}~\bibnamefont {Wu}}, \bibinfo {author}
  {\bibfnamefont {N.}~\bibnamefont {Li}}, \bibinfo {author} {\bibfnamefont
  {S.}~\bibnamefont {Wang}}, \bibinfo {author} {\bibfnamefont {Y.}~\bibnamefont
  {Zhao}}, \bibinfo {author} {\bibfnamefont {J.}~\bibnamefont {Tang}}, \bibinfo
  {author} {\bibfnamefont {J.}~\bibnamefont {Liu}}, \bibinfo {author}
  {\bibfnamefont {J.}~\bibnamefont {Tian}}, \bibinfo {author} {\bibfnamefont
  {K.}~\bibnamefont {Watanabe}}, \bibinfo {author} {\bibfnamefont
  {T.}~\bibnamefont {Taniguchi}}, \bibinfo {author} {\bibfnamefont
  {R.}~\bibnamefont {Yang}}, \bibinfo {author} {\bibfnamefont {Z.~Y.}\
  \bibnamefont {Meng}}, \bibinfo {author} {\bibfnamefont {D.}~\bibnamefont
  {Shi}}, \bibinfo {author} {\bibfnamefont {O.~V.}\ \bibnamefont {Yazyev}},\
  and\ \bibinfo {author} {\bibfnamefont {G.}~\bibnamefont {Zhang}},\ }\bibfield
   {title} {\bibinfo {title} {Correlated states in twisted double bilayer
  graphene},\ }\href@noop {} {\bibfield  {journal} {\bibinfo  {journal} {Nat.
  Phys.}\ }\textbf {\bibinfo {volume} {16}},\ \bibinfo {pages} {520} (\bibinfo
  {year} {2020})}\BibitemShut {NoStop}%
\bibitem [{\citenamefont {Liu}\ \emph {et~al.}(2020)\citenamefont {Liu},
  \citenamefont {Hao}, \citenamefont {Khalaf}, \citenamefont {Lee},
  \citenamefont {Ronen}, \citenamefont {Yoo}, \citenamefont {Haei~Najafabadi},
  \citenamefont {Watanabe}, \citenamefont {Taniguchi}, \citenamefont
  {Vishwanath},\ and\ \citenamefont {Kim}}]{liu2020tunable}%
  \BibitemOpen
  \bibfield  {author} {\bibinfo {author} {\bibfnamefont {X.}~\bibnamefont
  {Liu}}, \bibinfo {author} {\bibfnamefont {Z.}~\bibnamefont {Hao}}, \bibinfo
  {author} {\bibfnamefont {E.}~\bibnamefont {Khalaf}}, \bibinfo {author}
  {\bibfnamefont {J.~Y.}\ \bibnamefont {Lee}}, \bibinfo {author} {\bibfnamefont
  {Y.}~\bibnamefont {Ronen}}, \bibinfo {author} {\bibfnamefont
  {H.}~\bibnamefont {Yoo}}, \bibinfo {author} {\bibfnamefont {D.}~\bibnamefont
  {Haei~Najafabadi}}, \bibinfo {author} {\bibfnamefont {K.}~\bibnamefont
  {Watanabe}}, \bibinfo {author} {\bibfnamefont {T.}~\bibnamefont {Taniguchi}},
  \bibinfo {author} {\bibfnamefont {A.}~\bibnamefont {Vishwanath}},\ and\
  \bibinfo {author} {\bibfnamefont {P.}~\bibnamefont {Kim}},\ }\bibfield
  {title} {\bibinfo {title} {Tunable spin-polarized correlated states in
  twisted double bilayer graphene},\ }\href
  {https://doi.org/10.1038/s41586-020-2458-7} {\bibfield  {journal} {\bibinfo
  {journal} {Nature}\ }\textbf {\bibinfo {volume} {583}},\ \bibinfo {pages}
  {221} (\bibinfo {year} {2020})}\BibitemShut {NoStop}%
\bibitem [{\citenamefont {Chen}\ \emph
  {et~al.}(2019{\natexlab{a}})\citenamefont {Chen}, \citenamefont {Jiang},
  \citenamefont {Wu}, \citenamefont {Lyu}, \citenamefont {Li}, \citenamefont
  {Chittari}, \citenamefont {Watanabe}, \citenamefont {Taniguchi},
  \citenamefont {Shi}, \citenamefont {Jung}, \citenamefont {Zhang},\ and\
  \citenamefont {Wang}}]{chen2019evidence}%
  \BibitemOpen
  \bibfield  {author} {\bibinfo {author} {\bibfnamefont {G.}~\bibnamefont
  {Chen}}, \bibinfo {author} {\bibfnamefont {L.}~\bibnamefont {Jiang}},
  \bibinfo {author} {\bibfnamefont {S.}~\bibnamefont {Wu}}, \bibinfo {author}
  {\bibfnamefont {B.}~\bibnamefont {Lyu}}, \bibinfo {author} {\bibfnamefont
  {H.}~\bibnamefont {Li}}, \bibinfo {author} {\bibfnamefont {B.~L.}\
  \bibnamefont {Chittari}}, \bibinfo {author} {\bibfnamefont {K.}~\bibnamefont
  {Watanabe}}, \bibinfo {author} {\bibfnamefont {T.}~\bibnamefont {Taniguchi}},
  \bibinfo {author} {\bibfnamefont {Z.}~\bibnamefont {Shi}}, \bibinfo {author}
  {\bibfnamefont {J.}~\bibnamefont {Jung}}, \bibinfo {author} {\bibfnamefont
  {Y.}~\bibnamefont {Zhang}},\ and\ \bibinfo {author} {\bibfnamefont
  {F.}~\bibnamefont {Wang}},\ }\bibfield  {title} {\bibinfo {title} {Evidence
  of a gate-tunable {Mott} insulator in a trilayer graphene moir{\'e}
  superlattice},\ }\href@noop {} {\bibfield  {journal} {\bibinfo  {journal}
  {Nat. Phys.}\ }\textbf {\bibinfo {volume} {15}},\ \bibinfo {pages} {237}
  (\bibinfo {year} {2019}{\natexlab{a}})}\BibitemShut {NoStop}%
\bibitem [{\citenamefont {Chen}\ \emph
  {et~al.}(2019{\natexlab{b}})\citenamefont {Chen}, \citenamefont {Sharpe},
  \citenamefont {Gallagher}, \citenamefont {Rosen}, \citenamefont {Fox},
  \citenamefont {Jiang}, \citenamefont {Lyu}, \citenamefont {Li}, \citenamefont
  {Watanabe}, \citenamefont {Taniguchi}, \citenamefont {Jung}, \citenamefont
  {Shi}, \citenamefont {Goldhaber-Gordon}, \citenamefont {Zhang},\ and\
  \citenamefont {Wang}}]{chen2019signatures}%
  \BibitemOpen
  \bibfield  {author} {\bibinfo {author} {\bibfnamefont {G.}~\bibnamefont
  {Chen}}, \bibinfo {author} {\bibfnamefont {A.~L.}\ \bibnamefont {Sharpe}},
  \bibinfo {author} {\bibfnamefont {P.}~\bibnamefont {Gallagher}}, \bibinfo
  {author} {\bibfnamefont {I.~T.}\ \bibnamefont {Rosen}}, \bibinfo {author}
  {\bibfnamefont {E.~J.}\ \bibnamefont {Fox}}, \bibinfo {author} {\bibfnamefont
  {L.}~\bibnamefont {Jiang}}, \bibinfo {author} {\bibfnamefont
  {B.}~\bibnamefont {Lyu}}, \bibinfo {author} {\bibfnamefont {H.}~\bibnamefont
  {Li}}, \bibinfo {author} {\bibfnamefont {K.}~\bibnamefont {Watanabe}},
  \bibinfo {author} {\bibfnamefont {T.}~\bibnamefont {Taniguchi}}, \bibinfo
  {author} {\bibfnamefont {J.}~\bibnamefont {Jung}}, \bibinfo {author}
  {\bibfnamefont {Z.}~\bibnamefont {Shi}}, \bibinfo {author} {\bibfnamefont
  {D.}~\bibnamefont {Goldhaber-Gordon}}, \bibinfo {author} {\bibfnamefont
  {Y.}~\bibnamefont {Zhang}},\ and\ \bibinfo {author} {\bibfnamefont
  {F.}~\bibnamefont {Wang}},\ }\bibfield  {title} {\bibinfo {title} {Signatures
  of tunable superconductivity in a trilayer graphene moir{\'e} superlattice},\
  }\href {https://doi.org/10.1038/s41586-019-1393-y} {\bibfield  {journal}
  {\bibinfo  {journal} {Nature}\ }\textbf {\bibinfo {volume} {572}},\ \bibinfo
  {pages} {215} (\bibinfo {year} {2019}{\natexlab{b}})}\BibitemShut {NoStop}%
\bibitem [{\citenamefont {Chen}\ \emph
  {et~al.}(2020{\natexlab{a}})\citenamefont {Chen}, \citenamefont {Sharpe},
  \citenamefont {Fox}, \citenamefont {Zhang}, \citenamefont {Wang},
  \citenamefont {Jiang}, \citenamefont {Lyu}, \citenamefont {Li}, \citenamefont
  {Watanabe}, \citenamefont {Taniguchi}, \citenamefont {Shi}, \citenamefont
  {Senthil}, \citenamefont {Goldhaber-Gordon}, \citenamefont {Zhang},\ and\
  \citenamefont {Wang}}]{chen2020tunable}%
  \BibitemOpen
  \bibfield  {author} {\bibinfo {author} {\bibfnamefont {G.}~\bibnamefont
  {Chen}}, \bibinfo {author} {\bibfnamefont {A.~L.}\ \bibnamefont {Sharpe}},
  \bibinfo {author} {\bibfnamefont {E.~J.}\ \bibnamefont {Fox}}, \bibinfo
  {author} {\bibfnamefont {Y.-H.}\ \bibnamefont {Zhang}}, \bibinfo {author}
  {\bibfnamefont {S.}~\bibnamefont {Wang}}, \bibinfo {author} {\bibfnamefont
  {L.}~\bibnamefont {Jiang}}, \bibinfo {author} {\bibfnamefont
  {B.}~\bibnamefont {Lyu}}, \bibinfo {author} {\bibfnamefont {H.}~\bibnamefont
  {Li}}, \bibinfo {author} {\bibfnamefont {K.}~\bibnamefont {Watanabe}},
  \bibinfo {author} {\bibfnamefont {T.}~\bibnamefont {Taniguchi}}, \bibinfo
  {author} {\bibfnamefont {Z.}~\bibnamefont {Shi}}, \bibinfo {author}
  {\bibfnamefont {T.}~\bibnamefont {Senthil}}, \bibinfo {author} {\bibfnamefont
  {D.}~\bibnamefont {Goldhaber-Gordon}}, \bibinfo {author} {\bibfnamefont
  {Y.}~\bibnamefont {Zhang}},\ and\ \bibinfo {author} {\bibfnamefont
  {F.}~\bibnamefont {Wang}},\ }\bibfield  {title} {\bibinfo {title} {Tunable
  correlated {Chern} insulator and ferromagnetism in a moir{\'e}
  superlattice},\ }\href {https://doi.org/10.1038/s41586-020-2049-7} {\bibfield
   {journal} {\bibinfo  {journal} {Nature}\ }\textbf {\bibinfo {volume}
  {579}},\ \bibinfo {pages} {56} (\bibinfo {year}
  {2020}{\natexlab{a}})}\BibitemShut {NoStop}%
\bibitem [{\citenamefont {Chen}\ \emph
  {et~al.}(2020{\natexlab{b}})\citenamefont {Chen}, \citenamefont {He},
  \citenamefont {Zhang}, \citenamefont {Hsieh}, \citenamefont {Fei},
  \citenamefont {Watanabe}, \citenamefont {Taniguchi}, \citenamefont {Cobden},
  \citenamefont {Xu}, \citenamefont {Dean} \emph {et~al.}}]{Chen2020}%
  \BibitemOpen
  \bibfield  {author} {\bibinfo {author} {\bibfnamefont {S.}~\bibnamefont
  {Chen}}, \bibinfo {author} {\bibfnamefont {M.}~\bibnamefont {He}}, \bibinfo
  {author} {\bibfnamefont {Y.-H.}\ \bibnamefont {Zhang}}, \bibinfo {author}
  {\bibfnamefont {V.}~\bibnamefont {Hsieh}}, \bibinfo {author} {\bibfnamefont
  {Z.}~\bibnamefont {Fei}}, \bibinfo {author} {\bibfnamefont {K.}~\bibnamefont
  {Watanabe}}, \bibinfo {author} {\bibfnamefont {T.}~\bibnamefont {Taniguchi}},
  \bibinfo {author} {\bibfnamefont {D.~H.}\ \bibnamefont {Cobden}}, \bibinfo
  {author} {\bibfnamefont {X.}~\bibnamefont {Xu}}, \bibinfo {author}
  {\bibfnamefont {C.~R.}\ \bibnamefont {Dean}}, \emph {et~al.},\ }\bibfield
  {title} {\bibinfo {title} {Electrically tunable correlated and topological
  states in twisted monolayer-bilayer graphene},\ }\href@noop {} {\bibfield
  {journal} {\bibinfo  {journal} {arXiv preprint arXiv:2004.11340}\ } (\bibinfo
  {year} {2020}{\natexlab{b}})}\BibitemShut {NoStop}%
\bibitem [{\citenamefont {Fu}\ \emph {et~al.}(2020{\natexlab{a}})\citenamefont
  {Fu}, \citenamefont {K{\"o}nig}, \citenamefont {Wilson}, \citenamefont
  {Chou},\ and\ \citenamefont {Pixley}}]{fuMagicangleSemimetals2020}%
  \BibitemOpen
  \bibfield  {author} {\bibinfo {author} {\bibfnamefont {Y.}~\bibnamefont
  {Fu}}, \bibinfo {author} {\bibfnamefont {E.~J.}\ \bibnamefont {K{\"o}nig}},
  \bibinfo {author} {\bibfnamefont {J.~H.}\ \bibnamefont {Wilson}}, \bibinfo
  {author} {\bibfnamefont {Y.-Z.}\ \bibnamefont {Chou}},\ and\ \bibinfo
  {author} {\bibfnamefont {J.~H.}\ \bibnamefont {Pixley}},\ }\bibfield  {title}
  {\bibinfo {title} {Magic-angle semimetals},\ }\href
  {https://doi.org/10.1038/s41535-020-00271-9} {\bibfield  {journal} {\bibinfo
  {journal} {npj Quantum Mater.}\ }\textbf {\bibinfo {volume} {5}},\ \bibinfo
  {pages} {1} (\bibinfo {year} {2020}{\natexlab{a}})}\BibitemShut {NoStop}%
\bibitem [{\citenamefont {Gonz\'alez-Tudela}\ and\ \citenamefont
  {Cirac}(2019)}]{PhysRevA.100.053604}%
  \BibitemOpen
  \bibfield  {author} {\bibinfo {author} {\bibfnamefont {A.}~\bibnamefont
  {Gonz\'alez-Tudela}}\ and\ \bibinfo {author} {\bibfnamefont {J.~I.}\
  \bibnamefont {Cirac}},\ }\bibfield  {title} {\bibinfo {title} {Cold atoms in
  twisted-bilayer optical potentials},\ }\href
  {https://doi.org/10.1103/PhysRevA.100.053604} {\bibfield  {journal} {\bibinfo
   {journal} {Phys. Rev. A}\ }\textbf {\bibinfo {volume} {100}},\ \bibinfo
  {pages} {053604} (\bibinfo {year} {2019})}\BibitemShut {NoStop}%
\bibitem [{\citenamefont {Salamon}\ \emph {et~al.}(2020)\citenamefont
  {Salamon}, \citenamefont {Celi}, \citenamefont {Chhajlany}, \citenamefont
  {Fr\'erot}, \citenamefont {Lewenstein}, \citenamefont {Tarruell},\ and\
  \citenamefont {Rakshit}}]{PhysRevLett.125.030504}%
  \BibitemOpen
  \bibfield  {author} {\bibinfo {author} {\bibfnamefont {T.}~\bibnamefont
  {Salamon}}, \bibinfo {author} {\bibfnamefont {A.}~\bibnamefont {Celi}},
  \bibinfo {author} {\bibfnamefont {R.~W.}\ \bibnamefont {Chhajlany}}, \bibinfo
  {author} {\bibfnamefont {I.}~\bibnamefont {Fr\'erot}}, \bibinfo {author}
  {\bibfnamefont {M.}~\bibnamefont {Lewenstein}}, \bibinfo {author}
  {\bibfnamefont {L.}~\bibnamefont {Tarruell}},\ and\ \bibinfo {author}
  {\bibfnamefont {D.}~\bibnamefont {Rakshit}},\ }\bibfield  {title} {\bibinfo
  {title} {Simulating twistronics without a twist},\ }\href
  {https://doi.org/10.1103/PhysRevLett.125.030504} {\bibfield  {journal}
  {\bibinfo  {journal} {Phys. Rev. Lett.}\ }\textbf {\bibinfo {volume} {125}},\
  \bibinfo {pages} {030504} (\bibinfo {year} {2020})}\BibitemShut {NoStop}%
\bibitem [{\citenamefont {Luo}\ and\ \citenamefont
  {Zhang}(2020)}]{luo2020spin}%
  \BibitemOpen
  \bibfield  {author} {\bibinfo {author} {\bibfnamefont {X.-W.}\ \bibnamefont
  {Luo}}\ and\ \bibinfo {author} {\bibfnamefont {C.}~\bibnamefont {Zhang}},\
  }\bibfield  {title} {\bibinfo {title} {Spin-twisted optical lattices: Tunable
  flat bands and larkin-ovchinnikov superfluids},\ }\href@noop {} {\bibfield
  {journal} {\bibinfo  {journal} {arXiv preprint arXiv:2008.01351}\ } (\bibinfo
  {year} {2020})}\BibitemShut {NoStop}%
\bibitem [{\citenamefont {Pixley}\ \emph {et~al.}(2018)\citenamefont {Pixley},
  \citenamefont {Wilson}, \citenamefont {Huse},\ and\ \citenamefont
  {Gopalakrishnan}}]{Pixley-2018}%
  \BibitemOpen
  \bibfield  {author} {\bibinfo {author} {\bibfnamefont {J.~H.}\ \bibnamefont
  {Pixley}}, \bibinfo {author} {\bibfnamefont {J.~H.}\ \bibnamefont {Wilson}},
  \bibinfo {author} {\bibfnamefont {D.~A.}\ \bibnamefont {Huse}},\ and\
  \bibinfo {author} {\bibfnamefont {S.}~\bibnamefont {Gopalakrishnan}},\
  }\bibfield  {title} {\bibinfo {title} {{Weyl} semimetal to metal phase
  transitions driven by quasiperiodic potentials},\ }\href
  {https://doi.org/10.1103/PhysRevLett.120.207604} {\bibfield  {journal}
  {\bibinfo  {journal} {Phys. Rev. Lett.}\ }\textbf {\bibinfo {volume} {120}},\
  \bibinfo {pages} {207604} (\bibinfo {year} {2018})}\BibitemShut {NoStop}%
\bibitem [{\citenamefont {Chou}\ \emph {et~al.}(2020)\citenamefont {Chou},
  \citenamefont {Fu}, \citenamefont {Wilson}, \citenamefont {K{\"o}nig},\ and\
  \citenamefont {Pixley}}]{chouMagicangleSemimetalsChiral2020}%
  \BibitemOpen
  \bibfield  {author} {\bibinfo {author} {\bibfnamefont {Y.-Z.}\ \bibnamefont
  {Chou}}, \bibinfo {author} {\bibfnamefont {Y.}~\bibnamefont {Fu}}, \bibinfo
  {author} {\bibfnamefont {J.~H.}\ \bibnamefont {Wilson}}, \bibinfo {author}
  {\bibfnamefont {E.~J.}\ \bibnamefont {K{\"o}nig}},\ and\ \bibinfo {author}
  {\bibfnamefont {J.~H.}\ \bibnamefont {Pixley}},\ }\bibfield  {title}
  {\bibinfo {title} {Magic-angle semimetals with chiral symmetry},\ }\href
  {https://doi.org/10.1103/PhysRevB.101.235121} {\bibfield  {journal} {\bibinfo
   {journal} {Phys. Rev. B}\ }\textbf {\bibinfo {volume} {101}},\ \bibinfo
  {pages} {235121} (\bibinfo {year} {2020})}\BibitemShut {NoStop}%
\bibitem [{\citenamefont {Fu}\ \emph {et~al.}(2020{\natexlab{b}})\citenamefont
  {Fu}, \citenamefont {Wilson},\ and\ \citenamefont
  {Pixley}}]{fuFlatTopologicalBands2020}%
  \BibitemOpen
  \bibfield  {author} {\bibinfo {author} {\bibfnamefont {Y.}~\bibnamefont
  {Fu}}, \bibinfo {author} {\bibfnamefont {J.~H.}\ \bibnamefont {Wilson}},\
  and\ \bibinfo {author} {\bibfnamefont {J.~H.}\ \bibnamefont {Pixley}},\
  }\bibfield  {title} {\bibinfo {title} {Flat topological bands and eigenstate
  criticality in a quasiperiodic insulator},\ }\href@noop {} {\bibfield
  {journal} {\bibinfo  {journal} {arXiv:2003.00027 [cond-mat]}\ } (\bibinfo
  {year} {2020}{\natexlab{b}})},\ \Eprint {https://arxiv.org/abs/2003.00027}
  {arXiv:2003.00027 [cond-mat]} \BibitemShut {NoStop}%
\bibitem [{\citenamefont {Gon{\c{c}}alves}\ \emph {et~al.}(2020)\citenamefont
  {Gon{\c{c}}alves}, \citenamefont {Olyaei}, \citenamefont {Amorim},
  \citenamefont {Mondaini}, \citenamefont {Ribeiro},\ and\ \citenamefont
  {Castro}}]{gonccalves2020incommensurability}%
  \BibitemOpen
  \bibfield  {author} {\bibinfo {author} {\bibfnamefont {M.}~\bibnamefont
  {Gon{\c{c}}alves}}, \bibinfo {author} {\bibfnamefont {H.~Z.}\ \bibnamefont
  {Olyaei}}, \bibinfo {author} {\bibfnamefont {B.}~\bibnamefont {Amorim}},
  \bibinfo {author} {\bibfnamefont {R.}~\bibnamefont {Mondaini}}, \bibinfo
  {author} {\bibfnamefont {P.}~\bibnamefont {Ribeiro}},\ and\ \bibinfo {author}
  {\bibfnamefont {E.~V.}\ \bibnamefont {Castro}},\ }\bibfield  {title}
  {\bibinfo {title} {Incommensurability-induced sub-ballistic
  narrow-band-states in twisted bilayer graphene},\ }\href@noop {} {\bibfield
  {journal} {\bibinfo  {journal} {arXiv preprint arXiv:2008.07542}\ } (\bibinfo
  {year} {2020})}\BibitemShut {NoStop}%
\bibitem [{\citenamefont {Baum}\ and\ \citenamefont
  {Stern}(2012{\natexlab{a}})}]{baum2012magnetic}%
  \BibitemOpen
  \bibfield  {author} {\bibinfo {author} {\bibfnamefont {Y.}~\bibnamefont
  {Baum}}\ and\ \bibinfo {author} {\bibfnamefont {A.}~\bibnamefont {Stern}},\
  }\bibfield  {title} {\bibinfo {title} {Magnetic instability on the surface of
  topological insulators},\ }\href {https://doi.org/10.1103/PhysRevB.85.121105}
  {\bibfield  {journal} {\bibinfo  {journal} {Phys. Rev. B}\ }\textbf {\bibinfo
  {volume} {85}},\ \bibinfo {pages} {121105} (\bibinfo {year}
  {2012}{\natexlab{a}})}\BibitemShut {NoStop}%
\bibitem [{\citenamefont {Baum}\ and\ \citenamefont
  {Stern}(2012{\natexlab{b}})}]{baum2012density}%
  \BibitemOpen
  \bibfield  {author} {\bibinfo {author} {\bibfnamefont {Y.}~\bibnamefont
  {Baum}}\ and\ \bibinfo {author} {\bibfnamefont {A.}~\bibnamefont {Stern}},\
  }\bibfield  {title} {\bibinfo {title} {Density-waves instability and a
  skyrmion lattice on the surface of strong topological insulators},\ }\href
  {https://doi.org/10.1103/PhysRevB.86.195116} {\bibfield  {journal} {\bibinfo
  {journal} {Phys. Rev. B}\ }\textbf {\bibinfo {volume} {86}},\ \bibinfo
  {pages} {195116} (\bibinfo {year} {2012}{\natexlab{b}})}\BibitemShut
  {NoStop}%
\bibitem [{\citenamefont {Marchand}\ and\ \citenamefont
  {Franz}(2012)}]{marchand2012lattice}%
  \BibitemOpen
  \bibfield  {author} {\bibinfo {author} {\bibfnamefont {D.~J.~J.}\
  \bibnamefont {Marchand}}\ and\ \bibinfo {author} {\bibfnamefont
  {M.}~\bibnamefont {Franz}},\ }\bibfield  {title} {\bibinfo {title} {Lattice
  model for the surface states of a topological insulator with applications to
  magnetic and exciton instabilities},\ }\href
  {https://doi.org/10.1103/PhysRevB.86.155146} {\bibfield  {journal} {\bibinfo
  {journal} {Phys. Rev. B}\ }\textbf {\bibinfo {volume} {86}},\ \bibinfo
  {pages} {155146} (\bibinfo {year} {2012})}\BibitemShut {NoStop}%
\bibitem [{\citenamefont {Schmidt}(2012)}]{schmidt2012strong}%
  \BibitemOpen
  \bibfield  {author} {\bibinfo {author} {\bibfnamefont {M.~J.}\ \bibnamefont
  {Schmidt}},\ }\bibfield  {title} {\bibinfo {title} {Strong correlations at
  topological insulator surfaces and the breakdown of the bulk-boundary
  correspondence},\ }\href {https://doi.org/10.1103/PhysRevB.86.161110}
  {\bibfield  {journal} {\bibinfo  {journal} {Phys. Rev. B}\ }\textbf {\bibinfo
  {volume} {86}},\ \bibinfo {pages} {161110} (\bibinfo {year}
  {2012})}\BibitemShut {NoStop}%
\bibitem [{\citenamefont {Sitte}\ \emph {et~al.}(2013)\citenamefont {Sitte},
  \citenamefont {Rosch},\ and\ \citenamefont {Fritz}}]{sitte2013interaction}%
  \BibitemOpen
  \bibfield  {author} {\bibinfo {author} {\bibfnamefont {M.}~\bibnamefont
  {Sitte}}, \bibinfo {author} {\bibfnamefont {A.}~\bibnamefont {Rosch}},\ and\
  \bibinfo {author} {\bibfnamefont {L.}~\bibnamefont {Fritz}},\ }\bibfield
  {title} {\bibinfo {title} {Interaction effects on almost flat surface bands
  in topological insulators},\ }\href
  {https://doi.org/10.1103/PhysRevB.88.205107} {\bibfield  {journal} {\bibinfo
  {journal} {Phys. Rev. B}\ }\textbf {\bibinfo {volume} {88}},\ \bibinfo
  {pages} {205107} (\bibinfo {year} {2013})}\BibitemShut {NoStop}%
\bibitem [{\citenamefont {Mendler}\ \emph {et~al.}(2015)\citenamefont
  {Mendler}, \citenamefont {Kotetes},\ and\ \citenamefont
  {Sch\"on}}]{mendler2015magnetic}%
  \BibitemOpen
  \bibfield  {author} {\bibinfo {author} {\bibfnamefont {D.}~\bibnamefont
  {Mendler}}, \bibinfo {author} {\bibfnamefont {P.}~\bibnamefont {Kotetes}},\
  and\ \bibinfo {author} {\bibfnamefont {G.}~\bibnamefont {Sch\"on}},\
  }\bibfield  {title} {\bibinfo {title} {Magnetic order on a topological
  insulator surface with warping and proximity-induced superconductivity},\
  }\href {https://doi.org/10.1103/PhysRevB.91.155405} {\bibfield  {journal}
  {\bibinfo  {journal} {Phys. Rev. B}\ }\textbf {\bibinfo {volume} {91}},\
  \bibinfo {pages} {155405} (\bibinfo {year} {2015})}\BibitemShut {NoStop}%
\bibitem [{\citenamefont {Fu}\ and\ \citenamefont
  {Kane}(2008)}]{Fu2008superconducting}%
  \BibitemOpen
  \bibfield  {author} {\bibinfo {author} {\bibfnamefont {L.}~\bibnamefont
  {Fu}}\ and\ \bibinfo {author} {\bibfnamefont {C.~L.}\ \bibnamefont {Kane}},\
  }\bibfield  {title} {\bibinfo {title} {Superconducting proximity effect and
  majorana fermions at the surface of a topological insulator},\ }\href
  {https://doi.org/10.1103/PhysRevLett.100.096407} {\bibfield  {journal}
  {\bibinfo  {journal} {Phys. Rev. Lett.}\ }\textbf {\bibinfo {volume} {100}},\
  \bibinfo {pages} {096407} (\bibinfo {year} {2008})}\BibitemShut {NoStop}%
\bibitem [{\citenamefont {Santos}\ \emph {et~al.}(2010)\citenamefont {Santos},
  \citenamefont {Neupert}, \citenamefont {Chamon},\ and\ \citenamefont
  {Mudry}}]{Santos2010}%
  \BibitemOpen
  \bibfield  {author} {\bibinfo {author} {\bibfnamefont {L.}~\bibnamefont
  {Santos}}, \bibinfo {author} {\bibfnamefont {T.}~\bibnamefont {Neupert}},
  \bibinfo {author} {\bibfnamefont {C.}~\bibnamefont {Chamon}},\ and\ \bibinfo
  {author} {\bibfnamefont {C.}~\bibnamefont {Mudry}},\ }\bibfield  {title}
  {\bibinfo {title} {Superconductivity on the surface of topological insulators
  and in two-dimensional noncentrosymmetric materials},\ }\href
  {https://doi.org/10.1103/PhysRevB.81.184502} {\bibfield  {journal} {\bibinfo
  {journal} {Phys. Rev. B}\ }\textbf {\bibinfo {volume} {81}},\ \bibinfo
  {pages} {184502} (\bibinfo {year} {2010})}\BibitemShut {NoStop}%
\bibitem [{\citenamefont {Read}\ and\ \citenamefont
  {Green}(2000)}]{read2000paired}%
  \BibitemOpen
  \bibfield  {author} {\bibinfo {author} {\bibfnamefont {N.}~\bibnamefont
  {Read}}\ and\ \bibinfo {author} {\bibfnamefont {D.}~\bibnamefont {Green}},\
  }\bibfield  {title} {\bibinfo {title} {Paired states of fermions in two
  dimensions with breaking of parity and time-reversal symmetries and the
  fractional quantum {Hall} effect},\ }\href@noop {} {\bibfield  {journal}
  {\bibinfo  {journal} {Phys. Rev. B}\ }\textbf {\bibinfo {volume} {61}},\
  \bibinfo {pages} {10267} (\bibinfo {year} {2000})}\BibitemShut {NoStop}%
\bibitem [{\citenamefont {Kitaev}(2001)}]{kitaev2001unpaired}%
  \BibitemOpen
  \bibfield  {author} {\bibinfo {author} {\bibfnamefont {A.~Y.}\ \bibnamefont
  {Kitaev}},\ }\bibfield  {title} {\bibinfo {title} {Unpaired {Majorana}
  fermions in quantum wires},\ }\href@noop {} {\bibfield  {journal} {\bibinfo
  {journal} {Phys.-Usp.}\ }\textbf {\bibinfo {volume} {44}},\ \bibinfo {pages}
  {131} (\bibinfo {year} {2001})}\BibitemShut {NoStop}%
\bibitem [{\citenamefont {Volovik}(1999)}]{volovik1999fermion}%
  \BibitemOpen
  \bibfield  {author} {\bibinfo {author} {\bibfnamefont {G.}~\bibnamefont
  {Volovik}},\ }\bibfield  {title} {\bibinfo {title} {Fermion zero modes on
  vortices in chiral superconductors},\ }\href@noop {} {\bibfield  {journal}
  {\bibinfo  {journal} {J. Exp. Theor. Lett.}\ }\textbf {\bibinfo {volume}
  {70}},\ \bibinfo {pages} {609} (\bibinfo {year} {1999})}\BibitemShut
  {NoStop}%
\bibitem [{\citenamefont {Lopes Dos~Santos}\ \emph {et~al.}(2007)\citenamefont
  {Lopes Dos~Santos}, \citenamefont {Peres},\ and\ \citenamefont
  {Neto}}]{lopes2007graphene}%
  \BibitemOpen
  \bibfield  {author} {\bibinfo {author} {\bibfnamefont {J.}~\bibnamefont
  {Lopes Dos~Santos}}, \bibinfo {author} {\bibfnamefont {N.}~\bibnamefont
  {Peres}},\ and\ \bibinfo {author} {\bibfnamefont {A.~C.}\ \bibnamefont
  {Neto}},\ }\bibfield  {title} {\bibinfo {title} {Graphene bilayer with a
  twist: Electronic structure},\ }\href@noop {} {\bibfield  {journal} {\bibinfo
   {journal} {Phys. Rev. Lett.}\ }\textbf {\bibinfo {volume} {99}},\ \bibinfo
  {pages} {256802} (\bibinfo {year} {2007})}\BibitemShut {NoStop}%
\bibitem [{\citenamefont {Shallcross}\ \emph {et~al.}(2010)\citenamefont
  {Shallcross}, \citenamefont {Sharma}, \citenamefont {Kandelaki},\ and\
  \citenamefont {Pankratov}}]{Shallcross2010electronic}%
  \BibitemOpen
  \bibfield  {author} {\bibinfo {author} {\bibfnamefont {S.}~\bibnamefont
  {Shallcross}}, \bibinfo {author} {\bibfnamefont {S.}~\bibnamefont {Sharma}},
  \bibinfo {author} {\bibfnamefont {E.}~\bibnamefont {Kandelaki}},\ and\
  \bibinfo {author} {\bibfnamefont {O.~A.}\ \bibnamefont {Pankratov}},\
  }\bibfield  {title} {\bibinfo {title} {Electronic structure of turbostratic
  graphene},\ }\href {https://doi.org/10.1103/PhysRevB.81.165105} {\bibfield
  {journal} {\bibinfo  {journal} {Phys. Rev. B}\ }\textbf {\bibinfo {volume}
  {81}},\ \bibinfo {pages} {165105} (\bibinfo {year} {2010})}\BibitemShut
  {NoStop}%
\bibitem [{\citenamefont {Morell}\ \emph {et~al.}(2010)\citenamefont {Morell},
  \citenamefont {Correa}, \citenamefont {Vargas}, \citenamefont {Pacheco},\
  and\ \citenamefont {Barticevic}}]{morell2010flat}%
  \BibitemOpen
  \bibfield  {author} {\bibinfo {author} {\bibfnamefont {E.~S.}\ \bibnamefont
  {Morell}}, \bibinfo {author} {\bibfnamefont {J.}~\bibnamefont {Correa}},
  \bibinfo {author} {\bibfnamefont {P.}~\bibnamefont {Vargas}}, \bibinfo
  {author} {\bibfnamefont {M.}~\bibnamefont {Pacheco}},\ and\ \bibinfo {author}
  {\bibfnamefont {Z.}~\bibnamefont {Barticevic}},\ }\bibfield  {title}
  {\bibinfo {title} {Flat bands in slightly twisted bilayer graphene:
  Tight-binding calculations},\ }\href@noop {} {\bibfield  {journal} {\bibinfo
  {journal} {Phys. Rev. B}\ }\textbf {\bibinfo {volume} {82}},\ \bibinfo
  {pages} {121407} (\bibinfo {year} {2010})}\BibitemShut {NoStop}%
\bibitem [{\citenamefont {Lopes~dos Santos}\ \emph
  {et~al.}(2012{\natexlab{b}})\citenamefont {Lopes~dos Santos}, \citenamefont
  {Peres},\ and\ \citenamefont {Castro~Neto}}]{Lopes2012continuum}%
  \BibitemOpen
  \bibfield  {author} {\bibinfo {author} {\bibfnamefont {J.~M.~B.}\
  \bibnamefont {Lopes~dos Santos}}, \bibinfo {author} {\bibfnamefont
  {N.~M.~R.}\ \bibnamefont {Peres}},\ and\ \bibinfo {author} {\bibfnamefont
  {A.~H.}\ \bibnamefont {Castro~Neto}},\ }\bibfield  {title} {\bibinfo {title}
  {Continuum model of the twisted graphene bilayer},\ }\href
  {https://doi.org/10.1103/PhysRevB.86.155449} {\bibfield  {journal} {\bibinfo
  {journal} {Phys. Rev. B}\ }\textbf {\bibinfo {volume} {86}},\ \bibinfo
  {pages} {155449} (\bibinfo {year} {2012}{\natexlab{b}})}\BibitemShut
  {NoStop}%
\bibitem [{\citenamefont {Forsythe}\ \emph {et~al.}(2018)\citenamefont
  {Forsythe}, \citenamefont {Zhou}, \citenamefont {Watanabe}, \citenamefont
  {Taniguchi}, \citenamefont {Pasupathy}, \citenamefont {Moon}, \citenamefont
  {Koshino}, \citenamefont {Kim},\ and\ \citenamefont
  {Dean}}]{dielectric_patterning}%
  \BibitemOpen
  \bibfield  {author} {\bibinfo {author} {\bibfnamefont {C.}~\bibnamefont
  {Forsythe}}, \bibinfo {author} {\bibfnamefont {X.}~\bibnamefont {Zhou}},
  \bibinfo {author} {\bibfnamefont {K.}~\bibnamefont {Watanabe}}, \bibinfo
  {author} {\bibfnamefont {T.}~\bibnamefont {Taniguchi}}, \bibinfo {author}
  {\bibfnamefont {A.}~\bibnamefont {Pasupathy}}, \bibinfo {author}
  {\bibfnamefont {P.}~\bibnamefont {Moon}}, \bibinfo {author} {\bibfnamefont
  {M.}~\bibnamefont {Koshino}}, \bibinfo {author} {\bibfnamefont
  {P.}~\bibnamefont {Kim}},\ and\ \bibinfo {author} {\bibfnamefont {C.~R.}\
  \bibnamefont {Dean}},\ }\bibfield  {title} {\bibinfo {title} {Band structure
  engineering of {2D} materials using patterned dielectric superlattices},\
  }\href {https://doi.org/10.1038/s41565-018-0138-7} {\bibfield  {journal}
  {\bibinfo  {journal} {Nat. Nanotechnol.}\ }\textbf {\bibinfo {volume} {13}},\
  \bibinfo {pages} {566} (\bibinfo {year} {2018})}\BibitemShut {NoStop}%
\bibitem [{\citenamefont {Shi}\ \emph {et~al.}(2019)\citenamefont {Shi},
  \citenamefont {Ma},\ and\ \citenamefont {Song}}]{shi2019gate}%
  \BibitemOpen
  \bibfield  {author} {\bibinfo {author} {\bibfnamefont {L.-k.}\ \bibnamefont
  {Shi}}, \bibinfo {author} {\bibfnamefont {J.}~\bibnamefont {Ma}},\ and\
  \bibinfo {author} {\bibfnamefont {J.~C.}\ \bibnamefont {Song}},\ }\bibfield
  {title} {\bibinfo {title} {Gate-tunable flat bands in van der {Waals}
  patterned dielectric superlattices},\ }\href@noop {} {\bibfield  {journal}
  {\bibinfo  {journal} {2D Mater.}\ }\textbf {\bibinfo {volume} {7}},\ \bibinfo
  {pages} {015028} (\bibinfo {year} {2019})}\BibitemShut {NoStop}%
\bibitem [{\citenamefont {Li}\ \emph {et~al.}(2020)\citenamefont {Li},
  \citenamefont {Dietrich}, \citenamefont {Forsythe}, \citenamefont
  {Taniguchi}, \citenamefont {Watanabe},\ and\ \citenamefont
  {Moon}}]{li2020anisotropic}%
  \BibitemOpen
  \bibfield  {author} {\bibinfo {author} {\bibfnamefont {Y.}~\bibnamefont
  {Li}}, \bibinfo {author} {\bibfnamefont {S.}~\bibnamefont {Dietrich}},
  \bibinfo {author} {\bibfnamefont {C.}~\bibnamefont {Forsythe}}, \bibinfo
  {author} {\bibfnamefont {T.}~\bibnamefont {Taniguchi}}, \bibinfo {author}
  {\bibfnamefont {K.}~\bibnamefont {Watanabe}},\ and\ \bibinfo {author}
  {\bibfnamefont {P.}~\bibnamefont {Moon}},\ }\bibfield  {title} {\bibinfo
  {title} {Anisotropic band flattening in graphene with 1d superlattices},\
  }\href@noop {} {\bibfield  {journal} {\bibinfo  {journal} {arXiv preprint
  arXiv:2006.08868}\ } (\bibinfo {year} {2020})}\BibitemShut {NoStop}%
\bibitem [{\citenamefont {Jin}\ and\ \citenamefont
  {Jhi}(2013)}]{jin2013proximity}%
  \BibitemOpen
  \bibfield  {author} {\bibinfo {author} {\bibfnamefont {K.-H.}\ \bibnamefont
  {Jin}}\ and\ \bibinfo {author} {\bibfnamefont {S.-H.}\ \bibnamefont {Jhi}},\
  }\bibfield  {title} {\bibinfo {title} {Proximity-induced giant spin-orbit
  interaction in epitaxial graphene on a topological insulator},\ }\href@noop
  {} {\bibfield  {journal} {\bibinfo  {journal} {Phys. Rev. B}\ }\textbf
  {\bibinfo {volume} {87}},\ \bibinfo {pages} {075442} (\bibinfo {year}
  {2013})}\BibitemShut {NoStop}%
\bibitem [{\citenamefont {Zhang}\ \emph {et~al.}(2014)\citenamefont {Zhang},
  \citenamefont {Triola},\ and\ \citenamefont {Rossi}}]{zhang2014proximity}%
  \BibitemOpen
  \bibfield  {author} {\bibinfo {author} {\bibfnamefont {J.}~\bibnamefont
  {Zhang}}, \bibinfo {author} {\bibfnamefont {C.}~\bibnamefont {Triola}},\ and\
  \bibinfo {author} {\bibfnamefont {E.}~\bibnamefont {Rossi}},\ }\bibfield
  {title} {\bibinfo {title} {Proximity effect in
  graphene--topological-insulator heterostructures},\ }\href@noop {} {\bibfield
   {journal} {\bibinfo  {journal} {Phys. Rev. Lett.}\ }\textbf {\bibinfo
  {volume} {112}},\ \bibinfo {pages} {096802} (\bibinfo {year}
  {2014})}\BibitemShut {NoStop}%
\bibitem [{\citenamefont {Cao}\ \emph {et~al.}(2016)\citenamefont {Cao},
  \citenamefont {Zhang}, \citenamefont {Tang}, \citenamefont {Yang},
  \citenamefont {Sofo}, \citenamefont {Duan},\ and\ \citenamefont
  {Liu}}]{cao2016heavy}%
  \BibitemOpen
  \bibfield  {author} {\bibinfo {author} {\bibfnamefont {W.}~\bibnamefont
  {Cao}}, \bibinfo {author} {\bibfnamefont {R.-X.}\ \bibnamefont {Zhang}},
  \bibinfo {author} {\bibfnamefont {P.}~\bibnamefont {Tang}}, \bibinfo {author}
  {\bibfnamefont {G.}~\bibnamefont {Yang}}, \bibinfo {author} {\bibfnamefont
  {J.}~\bibnamefont {Sofo}}, \bibinfo {author} {\bibfnamefont {W.}~\bibnamefont
  {Duan}},\ and\ \bibinfo {author} {\bibfnamefont {C.-X.}\ \bibnamefont
  {Liu}},\ }\bibfield  {title} {\bibinfo {title} {Heavy {Dirac} fermions in a
  graphene/topological insulator hetero-junction},\ }\href@noop {} {\bibfield
  {journal} {\bibinfo  {journal} {2D Mater.}\ }\textbf {\bibinfo {volume}
  {3}},\ \bibinfo {pages} {034006} (\bibinfo {year} {2016})}\BibitemShut
  {NoStop}%
\bibitem [{\citenamefont {Steinberg}\ \emph {et~al.}(2015)\citenamefont
  {Steinberg}, \citenamefont {Orona}, \citenamefont {Fatemi}, \citenamefont
  {Sanchez-Yamagishi}, \citenamefont {Watanabe}, \citenamefont {Taniguchi},\
  and\ \citenamefont {Jarillo-Herrero}}]{steinberg2015tunneling}%
  \BibitemOpen
  \bibfield  {author} {\bibinfo {author} {\bibfnamefont {H.}~\bibnamefont
  {Steinberg}}, \bibinfo {author} {\bibfnamefont {L.~A.}\ \bibnamefont
  {Orona}}, \bibinfo {author} {\bibfnamefont {V.}~\bibnamefont {Fatemi}},
  \bibinfo {author} {\bibfnamefont {J.~D.}\ \bibnamefont {Sanchez-Yamagishi}},
  \bibinfo {author} {\bibfnamefont {K.}~\bibnamefont {Watanabe}}, \bibinfo
  {author} {\bibfnamefont {T.}~\bibnamefont {Taniguchi}},\ and\ \bibinfo
  {author} {\bibfnamefont {P.}~\bibnamefont {Jarillo-Herrero}},\ }\bibfield
  {title} {\bibinfo {title} {Tunneling in graphene--topological insulator
  hybrid devices},\ }\href@noop {} {\bibfield  {journal} {\bibinfo  {journal}
  {Phys. Rev. B}\ }\textbf {\bibinfo {volume} {92}},\ \bibinfo {pages} {241409}
  (\bibinfo {year} {2015})}\BibitemShut {NoStop}%
\bibitem [{\citenamefont {Bian}\ \emph {et~al.}(2016)\citenamefont {Bian},
  \citenamefont {Chung}, \citenamefont {Chen}, \citenamefont {Liu},
  \citenamefont {Chang}, \citenamefont {Wu}, \citenamefont {Belopolski},
  \citenamefont {Zheng}, \citenamefont {Xu}, \citenamefont {Sanchez},
  \citenamefont {Alidoust}, \citenamefont {Pierce}, \citenamefont {Quilliams},
  \citenamefont {Barletta}, \citenamefont {Lorcy}, \citenamefont {Avila},
  \citenamefont {Chang}, \citenamefont {Lin}, \citenamefont {Jeng},
  \citenamefont {Asensio}, \citenamefont {Chen},\ and\ \citenamefont
  {Hasan}}]{bian2016experimental}%
  \BibitemOpen
  \bibfield  {author} {\bibinfo {author} {\bibfnamefont {G.}~\bibnamefont
  {Bian}}, \bibinfo {author} {\bibfnamefont {T.-F.}\ \bibnamefont {Chung}},
  \bibinfo {author} {\bibfnamefont {C.}~\bibnamefont {Chen}}, \bibinfo {author}
  {\bibfnamefont {C.}~\bibnamefont {Liu}}, \bibinfo {author} {\bibfnamefont
  {T.-R.}\ \bibnamefont {Chang}}, \bibinfo {author} {\bibfnamefont
  {T.}~\bibnamefont {Wu}}, \bibinfo {author} {\bibfnamefont {I.}~\bibnamefont
  {Belopolski}}, \bibinfo {author} {\bibfnamefont {H.}~\bibnamefont {Zheng}},
  \bibinfo {author} {\bibfnamefont {S.-Y.}\ \bibnamefont {Xu}}, \bibinfo
  {author} {\bibfnamefont {D.~S.}\ \bibnamefont {Sanchez}}, \bibinfo {author}
  {\bibfnamefont {N.}~\bibnamefont {Alidoust}}, \bibinfo {author}
  {\bibfnamefont {J.}~\bibnamefont {Pierce}}, \bibinfo {author} {\bibfnamefont
  {B.}~\bibnamefont {Quilliams}}, \bibinfo {author} {\bibfnamefont {P.~P.}\
  \bibnamefont {Barletta}}, \bibinfo {author} {\bibfnamefont {S.}~\bibnamefont
  {Lorcy}}, \bibinfo {author} {\bibfnamefont {J.}~\bibnamefont {Avila}},
  \bibinfo {author} {\bibfnamefont {G.}~\bibnamefont {Chang}}, \bibinfo
  {author} {\bibfnamefont {H.}~\bibnamefont {Lin}}, \bibinfo {author}
  {\bibfnamefont {H.-T.}\ \bibnamefont {Jeng}}, \bibinfo {author}
  {\bibfnamefont {M.-C.}\ \bibnamefont {Asensio}}, \bibinfo {author}
  {\bibfnamefont {Y.~P.}\ \bibnamefont {Chen}},\ and\ \bibinfo {author}
  {\bibfnamefont {M.~Z.}\ \bibnamefont {Hasan}},\ }\bibfield  {title} {\bibinfo
  {title} {Experimental observation of two massless dirac-fermion gases in
  graphene-topological insulator heterostructure},\ }\href
  {https://doi.org/10.1088/2053-1583/3/2/021009} {\bibfield  {journal}
  {\bibinfo  {journal} {2D Mater.}\ }\textbf {\bibinfo {volume} {3}},\ \bibinfo
  {pages} {021009} (\bibinfo {year} {2016})}\BibitemShut {NoStop}%
\bibitem [{\citenamefont {Tian}\ \emph {et~al.}(2016)\citenamefont {Tian},
  \citenamefont {Chung}, \citenamefont {Miotkowski},\ and\ \citenamefont
  {Chen}}]{tian2016electrical}%
  \BibitemOpen
  \bibfield  {author} {\bibinfo {author} {\bibfnamefont {J.}~\bibnamefont
  {Tian}}, \bibinfo {author} {\bibfnamefont {T.-F.}\ \bibnamefont {Chung}},
  \bibinfo {author} {\bibfnamefont {I.}~\bibnamefont {Miotkowski}},\ and\
  \bibinfo {author} {\bibfnamefont {Y.~P.}\ \bibnamefont {Chen}},\ }\bibfield
  {title} {\bibinfo {title} {Electrical spin injection into graphene from a
  topological insulator in a van der {Waals} heterostructure},\ }\href@noop {}
  {\bibfield  {journal} {\bibinfo  {journal} {arXiv preprint arXiv:1607.02651}\
  } (\bibinfo {year} {2016})}\BibitemShut {NoStop}%
\bibitem [{\citenamefont {Chong}\ \emph {et~al.}(2018)\citenamefont {Chong},
  \citenamefont {Han}, \citenamefont {Nagaoka}, \citenamefont {Tsuchikawa},
  \citenamefont {Liu}, \citenamefont {Liu}, \citenamefont {Vardeny},
  \citenamefont {Pesin}, \citenamefont {Lee}, \citenamefont {Sparks},\ and\
  \citenamefont {Deshpande}}]{chong2018topological}%
  \BibitemOpen
  \bibfield  {author} {\bibinfo {author} {\bibfnamefont {S.~K.}\ \bibnamefont
  {Chong}}, \bibinfo {author} {\bibfnamefont {K.~B.}\ \bibnamefont {Han}},
  \bibinfo {author} {\bibfnamefont {A.}~\bibnamefont {Nagaoka}}, \bibinfo
  {author} {\bibfnamefont {R.}~\bibnamefont {Tsuchikawa}}, \bibinfo {author}
  {\bibfnamefont {R.}~\bibnamefont {Liu}}, \bibinfo {author} {\bibfnamefont
  {H.}~\bibnamefont {Liu}}, \bibinfo {author} {\bibfnamefont {Z.~V.}\
  \bibnamefont {Vardeny}}, \bibinfo {author} {\bibfnamefont {D.~A.}\
  \bibnamefont {Pesin}}, \bibinfo {author} {\bibfnamefont {C.}~\bibnamefont
  {Lee}}, \bibinfo {author} {\bibfnamefont {T.~D.}\ \bibnamefont {Sparks}},\
  and\ \bibinfo {author} {\bibfnamefont {V.~V.}\ \bibnamefont {Deshpande}},\
  }\bibfield  {title} {\bibinfo {title} {Topological insulator-based van der
  {Waals} heterostructures for effective control of massless and massive
  {Dirac} fermions},\ }\bibfield  {booktitle} {\emph {\bibinfo {booktitle}
  {Nano Letters}},\ }\href {https://doi.org/10.1021/acs.nanolett.8b04291}
  {\bibfield  {journal} {\bibinfo  {journal} {Nano Lett.}\ }\textbf {\bibinfo
  {volume} {18}},\ \bibinfo {pages} {8047} (\bibinfo {year}
  {2018})}\BibitemShut {NoStop}%
\bibitem [{\citenamefont {Jafarpisheh}\ \emph {et~al.}(2018)\citenamefont
  {Jafarpisheh}, \citenamefont {Cummings}, \citenamefont {Watanabe},
  \citenamefont {Taniguchi}, \citenamefont {Beschoten},\ and\ \citenamefont
  {Stampfer}}]{jafarpisheh2018proximity}%
  \BibitemOpen
  \bibfield  {author} {\bibinfo {author} {\bibfnamefont {S.}~\bibnamefont
  {Jafarpisheh}}, \bibinfo {author} {\bibfnamefont {A.~W.}\ \bibnamefont
  {Cummings}}, \bibinfo {author} {\bibfnamefont {K.}~\bibnamefont {Watanabe}},
  \bibinfo {author} {\bibfnamefont {T.}~\bibnamefont {Taniguchi}}, \bibinfo
  {author} {\bibfnamefont {B.}~\bibnamefont {Beschoten}},\ and\ \bibinfo
  {author} {\bibfnamefont {C.}~\bibnamefont {Stampfer}},\ }\bibfield  {title}
  {\bibinfo {title} {Proximity-induced spin-orbit coupling in
  graphene/{${\mathrm{Bi}}_{1.5}{\mathrm{Sb}}_{0.5}{\mathrm{Te}}_{1.7}{\mathrm{Se}}_{1.3}$}
  heterostructures},\ }\href {https://doi.org/10.1103/PhysRevB.98.241402}
  {\bibfield  {journal} {\bibinfo  {journal} {Phys. Rev. B}\ }\textbf {\bibinfo
  {volume} {98}},\ \bibinfo {pages} {241402} (\bibinfo {year}
  {2018})}\BibitemShut {NoStop}%
\bibitem [{\citenamefont {Khokhriakov}\ \emph {et~al.}(2018)\citenamefont
  {Khokhriakov}, \citenamefont {Cummings}, \citenamefont {Song}, \citenamefont
  {Vila}, \citenamefont {Karpiak}, \citenamefont {Dankert}, \citenamefont
  {Roche},\ and\ \citenamefont {Dash}}]{Khokhriakov2018}%
  \BibitemOpen
  \bibfield  {author} {\bibinfo {author} {\bibfnamefont {D.}~\bibnamefont
  {Khokhriakov}}, \bibinfo {author} {\bibfnamefont {A.~W.}\ \bibnamefont
  {Cummings}}, \bibinfo {author} {\bibfnamefont {K.}~\bibnamefont {Song}},
  \bibinfo {author} {\bibfnamefont {M.}~\bibnamefont {Vila}}, \bibinfo {author}
  {\bibfnamefont {B.}~\bibnamefont {Karpiak}}, \bibinfo {author} {\bibfnamefont
  {A.}~\bibnamefont {Dankert}}, \bibinfo {author} {\bibfnamefont
  {S.}~\bibnamefont {Roche}},\ and\ \bibinfo {author} {\bibfnamefont {S.~P.}\
  \bibnamefont {Dash}},\ }\bibfield  {title} {\bibinfo {title} {Tailoring
  emergent spin phenomena in {Dirac} material heterostructures},\ }\href
  {https://doi.org/10.1126/sciadv.aat9349} {\bibfield  {journal} {\bibinfo
  {journal} {Sci. Adv.}\ }\textbf {\bibinfo {volume} {4}},\ \bibinfo {pages}
  {eaat9349} (\bibinfo {year} {2018})}\BibitemShut {NoStop}%
\bibitem [{\citenamefont {Fu}(2009)}]{fu2009hexagonal}%
  \BibitemOpen
  \bibfield  {author} {\bibinfo {author} {\bibfnamefont {L.}~\bibnamefont
  {Fu}},\ }\bibfield  {title} {\bibinfo {title} {Hexagonal warping effects in
  the surface states of the topological insulator {Bi$_2$Te$_3$}},\ }\href
  {https://doi.org/10.1103/PhysRevLett.103.266801} {\bibfield  {journal}
  {\bibinfo  {journal} {Phys. Rev. Lett.}\ }\textbf {\bibinfo {volume} {103}},\
  \bibinfo {pages} {266801} (\bibinfo {year} {2009})}\BibitemShut {NoStop}%
\bibitem [{\citenamefont {Park}\ \emph {et~al.}(2008)\citenamefont {Park},
  \citenamefont {Yang}, \citenamefont {Son}, \citenamefont {Cohen},\ and\
  \citenamefont {Louie}}]{park2008anisotropic}%
  \BibitemOpen
  \bibfield  {author} {\bibinfo {author} {\bibfnamefont {C.-H.}\ \bibnamefont
  {Park}}, \bibinfo {author} {\bibfnamefont {L.}~\bibnamefont {Yang}}, \bibinfo
  {author} {\bibfnamefont {Y.-W.}\ \bibnamefont {Son}}, \bibinfo {author}
  {\bibfnamefont {M.~L.}\ \bibnamefont {Cohen}},\ and\ \bibinfo {author}
  {\bibfnamefont {S.~G.}\ \bibnamefont {Louie}},\ }\bibfield  {title} {\bibinfo
  {title} {Anisotropic behaviours of massless {Dirac} fermions in graphene
  under periodic potentials},\ }\href@noop {} {\bibfield  {journal} {\bibinfo
  {journal} {Nat. Phys.}\ }\textbf {\bibinfo {volume} {4}},\ \bibinfo {pages}
  {213} (\bibinfo {year} {2008})}\BibitemShut {NoStop}%
\bibitem [{\citenamefont {Yankowitz}\ \emph {et~al.}(2012)\citenamefont
  {Yankowitz}, \citenamefont {Xue}, \citenamefont {Cormode}, \citenamefont
  {Sanchez-Yamagishi}, \citenamefont {Watanabe}, \citenamefont {Taniguchi},
  \citenamefont {Jarillo-Herrero}, \citenamefont {Jacquod},\ and\ \citenamefont
  {LeRoy}}]{yankowitz2012emergence}%
  \BibitemOpen
  \bibfield  {author} {\bibinfo {author} {\bibfnamefont {M.}~\bibnamefont
  {Yankowitz}}, \bibinfo {author} {\bibfnamefont {J.}~\bibnamefont {Xue}},
  \bibinfo {author} {\bibfnamefont {D.}~\bibnamefont {Cormode}}, \bibinfo
  {author} {\bibfnamefont {J.~D.}\ \bibnamefont {Sanchez-Yamagishi}}, \bibinfo
  {author} {\bibfnamefont {K.}~\bibnamefont {Watanabe}}, \bibinfo {author}
  {\bibfnamefont {T.}~\bibnamefont {Taniguchi}}, \bibinfo {author}
  {\bibfnamefont {P.}~\bibnamefont {Jarillo-Herrero}}, \bibinfo {author}
  {\bibfnamefont {P.}~\bibnamefont {Jacquod}},\ and\ \bibinfo {author}
  {\bibfnamefont {B.~J.}\ \bibnamefont {LeRoy}},\ }\bibfield  {title} {\bibinfo
  {title} {Emergence of superlattice {Dirac} points in graphene on hexagonal
  boron nitride},\ }\href@noop {} {\bibfield  {journal} {\bibinfo  {journal}
  {Nat. Phys.}\ }\textbf {\bibinfo {volume} {8}},\ \bibinfo {pages} {382}
  (\bibinfo {year} {2012})}\BibitemShut {NoStop}%
\bibitem [{\citenamefont {Dean}\ \emph {et~al.}(2013)\citenamefont {Dean},
  \citenamefont {Wang}, \citenamefont {Maher}, \citenamefont {Forsythe},
  \citenamefont {Ghahari}, \citenamefont {Gao}, \citenamefont {Katoch},
  \citenamefont {Ishigami}, \citenamefont {Moon}, \citenamefont {Koshino},
  \citenamefont {Taniguchi}, \citenamefont {Watanabe}, \citenamefont {Shepard},
  \citenamefont {Hone},\ and\ \citenamefont {Kim}}]{dean2013hofstadter}%
  \BibitemOpen
  \bibfield  {author} {\bibinfo {author} {\bibfnamefont {C.~R.}\ \bibnamefont
  {Dean}}, \bibinfo {author} {\bibfnamefont {L.}~\bibnamefont {Wang}}, \bibinfo
  {author} {\bibfnamefont {P.}~\bibnamefont {Maher}}, \bibinfo {author}
  {\bibfnamefont {C.}~\bibnamefont {Forsythe}}, \bibinfo {author}
  {\bibfnamefont {F.}~\bibnamefont {Ghahari}}, \bibinfo {author} {\bibfnamefont
  {Y.}~\bibnamefont {Gao}}, \bibinfo {author} {\bibfnamefont {J.}~\bibnamefont
  {Katoch}}, \bibinfo {author} {\bibfnamefont {M.}~\bibnamefont {Ishigami}},
  \bibinfo {author} {\bibfnamefont {P.}~\bibnamefont {Moon}}, \bibinfo {author}
  {\bibfnamefont {M.}~\bibnamefont {Koshino}}, \bibinfo {author} {\bibfnamefont
  {T.}~\bibnamefont {Taniguchi}}, \bibinfo {author} {\bibfnamefont
  {K.}~\bibnamefont {Watanabe}}, \bibinfo {author} {\bibfnamefont
  {K.}~\bibnamefont {Shepard}}, \bibinfo {author} {\bibfnamefont
  {J.}~\bibnamefont {Hone}},\ and\ \bibinfo {author} {\bibfnamefont
  {P.}~\bibnamefont {Kim}},\ }\bibfield  {title} {\bibinfo {title}
  {{Hofstadter's} butterfly and the fractal quantum {Hall} effect in moir{\'e}
  superlattices},\ }\href@noop {} {\bibfield  {journal} {\bibinfo  {journal}
  {Nature}\ }\textbf {\bibinfo {volume} {497}},\ \bibinfo {pages} {598}
  (\bibinfo {year} {2013})}\BibitemShut {NoStop}%
\bibitem [{\citenamefont {Hunt}\ \emph {et~al.}(2013)\citenamefont {Hunt},
  \citenamefont {Sanchez-Yamagishi}, \citenamefont {Young}, \citenamefont
  {Yankowitz}, \citenamefont {LeRoy}, \citenamefont {Watanabe}, \citenamefont
  {Taniguchi}, \citenamefont {Moon}, \citenamefont {Koshino}, \citenamefont
  {Jarillo-Herrero},\ and\ \citenamefont {Ashoori}}]{hunt2013massive}%
  \BibitemOpen
  \bibfield  {author} {\bibinfo {author} {\bibfnamefont {B.}~\bibnamefont
  {Hunt}}, \bibinfo {author} {\bibfnamefont {J.~D.}\ \bibnamefont
  {Sanchez-Yamagishi}}, \bibinfo {author} {\bibfnamefont {A.~F.}\ \bibnamefont
  {Young}}, \bibinfo {author} {\bibfnamefont {M.}~\bibnamefont {Yankowitz}},
  \bibinfo {author} {\bibfnamefont {B.~J.}\ \bibnamefont {LeRoy}}, \bibinfo
  {author} {\bibfnamefont {K.}~\bibnamefont {Watanabe}}, \bibinfo {author}
  {\bibfnamefont {T.}~\bibnamefont {Taniguchi}}, \bibinfo {author}
  {\bibfnamefont {P.}~\bibnamefont {Moon}}, \bibinfo {author} {\bibfnamefont
  {M.}~\bibnamefont {Koshino}}, \bibinfo {author} {\bibfnamefont
  {P.}~\bibnamefont {Jarillo-Herrero}},\ and\ \bibinfo {author} {\bibfnamefont
  {R.~C.}\ \bibnamefont {Ashoori}},\ }\bibfield  {title} {\bibinfo {title}
  {Massive {Dirac} fermions and {Hofstadter} butterfly in a van der {Waals}
  heterostructure},\ }\href {https://doi.org/10.1126/science.1237240}
  {\bibfield  {journal} {\bibinfo  {journal} {Science}\ }\textbf {\bibinfo
  {volume} {340}},\ \bibinfo {pages} {1427} (\bibinfo {year}
  {2013})}\BibitemShut {NoStop}%
\bibitem [{\citenamefont {Ponomarenko}\ \emph {et~al.}(2013)\citenamefont
  {Ponomarenko}, \citenamefont {Gorbachev}, \citenamefont {Yu}, \citenamefont
  {Elias}, \citenamefont {Jalil}, \citenamefont {Patel}, \citenamefont
  {Mishchenko}, \citenamefont {Mayorov}, \citenamefont {Woods}, \citenamefont
  {Wallbank}, \citenamefont {Mucha-Kruczynski}, \citenamefont {Piot},
  \citenamefont {Grigorieva}, \citenamefont {Novoselov}, \citenamefont
  {Guinea}, \citenamefont {Fal'ko},\ and\ \citenamefont
  {Geim}}]{ponomarenko2013cloning}%
  \BibitemOpen
  \bibfield  {author} {\bibinfo {author} {\bibfnamefont {L.}~\bibnamefont
  {Ponomarenko}}, \bibinfo {author} {\bibfnamefont {R.}~\bibnamefont
  {Gorbachev}}, \bibinfo {author} {\bibfnamefont {G.}~\bibnamefont {Yu}},
  \bibinfo {author} {\bibfnamefont {D.}~\bibnamefont {Elias}}, \bibinfo
  {author} {\bibfnamefont {R.}~\bibnamefont {Jalil}}, \bibinfo {author}
  {\bibfnamefont {A.}~\bibnamefont {Patel}}, \bibinfo {author} {\bibfnamefont
  {A.}~\bibnamefont {Mishchenko}}, \bibinfo {author} {\bibfnamefont
  {A.}~\bibnamefont {Mayorov}}, \bibinfo {author} {\bibfnamefont
  {C.}~\bibnamefont {Woods}}, \bibinfo {author} {\bibfnamefont
  {J.}~\bibnamefont {Wallbank}}, \bibinfo {author} {\bibfnamefont
  {M.}~\bibnamefont {Mucha-Kruczynski}}, \bibinfo {author} {\bibfnamefont
  {B.}~\bibnamefont {Piot}}, \bibinfo {author} {\bibfnamefont {I.}~\bibnamefont
  {Grigorieva}}, \bibinfo {author} {\bibfnamefont {K.}~\bibnamefont
  {Novoselov}}, \bibinfo {author} {\bibfnamefont {F.}~\bibnamefont {Guinea}},
  \bibinfo {author} {\bibfnamefont {V.}~\bibnamefont {Fal'ko}},\ and\ \bibinfo
  {author} {\bibfnamefont {A.}~\bibnamefont {Geim}},\ }\bibfield  {title}
  {\bibinfo {title} {Cloning of {Dirac} fermions in graphene superlattices},\
  }\href@noop {} {\bibfield  {journal} {\bibinfo  {journal} {Nature}\ }\textbf
  {\bibinfo {volume} {497}},\ \bibinfo {pages} {594} (\bibinfo {year}
  {2013})}\BibitemShut {NoStop}%
\bibitem [{\citenamefont {Bernevig}\ \emph {et~al.}(2006)\citenamefont
  {Bernevig}, \citenamefont {Hughes},\ and\ \citenamefont
  {Zhang}}]{Bernevig2006quantum}%
  \BibitemOpen
  \bibfield  {author} {\bibinfo {author} {\bibfnamefont {B.~A.}\ \bibnamefont
  {Bernevig}}, \bibinfo {author} {\bibfnamefont {T.~L.}\ \bibnamefont
  {Hughes}},\ and\ \bibinfo {author} {\bibfnamefont {S.-C.}\ \bibnamefont
  {Zhang}},\ }\bibfield  {title} {\bibinfo {title} {Quantum spin hall effect
  and topological phase transition in {HgTe} quantum wells},\ }\href
  {https://doi.org/10.1126/science.1133734} {\bibfield  {journal} {\bibinfo
  {journal} {Science}\ }\textbf {\bibinfo {volume} {314}},\ \bibinfo {pages}
  {1757} (\bibinfo {year} {2006})}\BibitemShut {NoStop}%
\bibitem [{\citenamefont {Mong}\ and\ \citenamefont
  {Shivamoggi}(2011)}]{mong2011edge}%
  \BibitemOpen
  \bibfield  {author} {\bibinfo {author} {\bibfnamefont {R.~S.~K.}\
  \bibnamefont {Mong}}\ and\ \bibinfo {author} {\bibfnamefont {V.}~\bibnamefont
  {Shivamoggi}},\ }\bibfield  {title} {\bibinfo {title} {Edge states and the
  bulk-boundary correspondence in {Dirac} hamiltonians},\ }\href
  {https://doi.org/10.1103/PhysRevB.83.125109} {\bibfield  {journal} {\bibinfo
  {journal} {Phys. Rev. B}\ }\textbf {\bibinfo {volume} {83}},\ \bibinfo
  {pages} {125109} (\bibinfo {year} {2011})}\BibitemShut {NoStop}%
\bibitem [{\citenamefont {Wei\ss{}e}\ \emph {et~al.}(2006)\citenamefont
  {Wei\ss{}e}, \citenamefont {Wellein}, \citenamefont {Alvermann},\ and\
  \citenamefont {Fehske}}]{RevModPhys.78.275}%
  \BibitemOpen
  \bibfield  {author} {\bibinfo {author} {\bibfnamefont {A.}~\bibnamefont
  {Wei\ss{}e}}, \bibinfo {author} {\bibfnamefont {G.}~\bibnamefont {Wellein}},
  \bibinfo {author} {\bibfnamefont {A.}~\bibnamefont {Alvermann}},\ and\
  \bibinfo {author} {\bibfnamefont {H.}~\bibnamefont {Fehske}},\ }\bibfield
  {title} {\bibinfo {title} {The kernel polynomial method},\ }\href
  {https://doi.org/10.1103/RevModPhys.78.275} {\bibfield  {journal} {\bibinfo
  {journal} {Rev. Mod. Phys.}\ }\textbf {\bibinfo {volume} {78}},\ \bibinfo
  {pages} {275} (\bibinfo {year} {2006})}\BibitemShut {NoStop}%
\bibitem [{\citenamefont {Wilson}\ \emph {et~al.}(2018)\citenamefont {Wilson},
  \citenamefont {Pixley}, \citenamefont {Huse}, \citenamefont {Refael},\ and\
  \citenamefont {Das~Sarma}}]{Wilson-2018}%
  \BibitemOpen
  \bibfield  {author} {\bibinfo {author} {\bibfnamefont {J.~H.}\ \bibnamefont
  {Wilson}}, \bibinfo {author} {\bibfnamefont {J.~H.}\ \bibnamefont {Pixley}},
  \bibinfo {author} {\bibfnamefont {D.~A.}\ \bibnamefont {Huse}}, \bibinfo
  {author} {\bibfnamefont {G.}~\bibnamefont {Refael}},\ and\ \bibinfo {author}
  {\bibfnamefont {S.}~\bibnamefont {Das~Sarma}},\ }\bibfield  {title} {\bibinfo
  {title} {Do the surface fermi arcs in weyl semimetals survive disorder?},\
  }\href {https://doi.org/10.1103/PhysRevB.97.235108} {\bibfield  {journal}
  {\bibinfo  {journal} {Phys. Rev. B}\ }\textbf {\bibinfo {volume} {97}},\
  \bibinfo {pages} {235108} (\bibinfo {year} {2018})}\BibitemShut {NoStop}%
\bibitem [{\citenamefont {Kresse}\ and\ \citenamefont
  {Furthm\"uller}(1996)}]{vasp1}%
  \BibitemOpen
  \bibfield  {author} {\bibinfo {author} {\bibfnamefont {G.}~\bibnamefont
  {Kresse}}\ and\ \bibinfo {author} {\bibfnamefont {J.}~\bibnamefont
  {Furthm\"uller}},\ }\bibfield  {title} {\bibinfo {title} {Efficient iterative
  schemes for ab initio total-energy calculations using a plane-wave basis
  set},\ }\href {https://doi.org/10.1103/PhysRevB.54.11169} {\bibfield
  {journal} {\bibinfo  {journal} {Phys. Rev. B}\ }\textbf {\bibinfo {volume}
  {54}},\ \bibinfo {pages} {11169} (\bibinfo {year} {1996})}\BibitemShut
  {NoStop}%
\bibitem [{\citenamefont {Kresse}\ and\ \citenamefont
  {Furthm{\~A}{\OE}ller}(1996)}]{vasp2}%
  \BibitemOpen
  \bibfield  {author} {\bibinfo {author} {\bibfnamefont {G.}~\bibnamefont
  {Kresse}}\ and\ \bibinfo {author} {\bibfnamefont {J.}~\bibnamefont
  {Furthm{\~A}{\OE}ller}},\ }\bibfield  {title} {\bibinfo {title} {Efficiency
  of ab-initio total energy calculations for metals and semiconductors using a
  plane-wave basis set},\ }\href
  {https://doi.org/https://doi.org/10.1016/0927-0256(96)00008-0} {\bibfield
  {journal} {\bibinfo  {journal} {Comput. Mater. Sci.}\ }\textbf {\bibinfo
  {volume} {6}},\ \bibinfo {pages} {15 } (\bibinfo {year} {1996})}\BibitemShut
  {NoStop}%
\bibitem [{\citenamefont {Bl\"ochl}(1994)}]{PAW}%
  \BibitemOpen
  \bibfield  {author} {\bibinfo {author} {\bibfnamefont {P.~E.}\ \bibnamefont
  {Bl\"ochl}},\ }\bibfield  {title} {\bibinfo {title} {Projector augmented-wave
  method},\ }\href {https://doi.org/10.1103/PhysRevB.50.17953} {\bibfield
  {journal} {\bibinfo  {journal} {Phys. Rev. B}\ }\textbf {\bibinfo {volume}
  {50}},\ \bibinfo {pages} {17953} (\bibinfo {year} {1994})}\BibitemShut
  {NoStop}%
\bibitem [{\citenamefont {Perdew}\ \emph {et~al.}(1996)\citenamefont {Perdew},
  \citenamefont {Burke},\ and\ \citenamefont {Ernzerhof}}]{VASP_pbe}%
  \BibitemOpen
  \bibfield  {author} {\bibinfo {author} {\bibfnamefont {J.~P.}\ \bibnamefont
  {Perdew}}, \bibinfo {author} {\bibfnamefont {K.}~\bibnamefont {Burke}},\ and\
  \bibinfo {author} {\bibfnamefont {M.}~\bibnamefont {Ernzerhof}},\ }\bibfield
  {title} {\bibinfo {title} {Generalized gradient approximation made simple},\
  }\href {https://doi.org/10.1103/PhysRevLett.77.3865} {\bibfield  {journal}
  {\bibinfo  {journal} {Phys. Rev. Lett.}\ }\textbf {\bibinfo {volume} {77}},\
  \bibinfo {pages} {3865} (\bibinfo {year} {1996})}\BibitemShut {NoStop}%
\bibitem [{\citenamefont {Monkhorst}\ and\ \citenamefont
  {Pack}(1976)}]{MP_grid}%
  \BibitemOpen
  \bibfield  {author} {\bibinfo {author} {\bibfnamefont {H.~J.}\ \bibnamefont
  {Monkhorst}}\ and\ \bibinfo {author} {\bibfnamefont {J.~D.}\ \bibnamefont
  {Pack}},\ }\bibfield  {title} {\bibinfo {title} {Special points for
  {Brillouin}-zone integrations},\ }\href
  {https://doi.org/10.1103/PhysRevB.13.5188} {\bibfield  {journal} {\bibinfo
  {journal} {Phys. Rev. B}\ }\textbf {\bibinfo {volume} {13}},\ \bibinfo
  {pages} {5188} (\bibinfo {year} {1976})}\BibitemShut {NoStop}%
\bibitem [{\citenamefont {Marzari}\ \emph {et~al.}(2012)\citenamefont
  {Marzari}, \citenamefont {Mostofi}, \citenamefont {Yates}, \citenamefont
  {Souza},\ and\ \citenamefont {Vanderbilt}}]{wannier_review}%
  \BibitemOpen
  \bibfield  {author} {\bibinfo {author} {\bibfnamefont {N.}~\bibnamefont
  {Marzari}}, \bibinfo {author} {\bibfnamefont {A.~A.}\ \bibnamefont
  {Mostofi}}, \bibinfo {author} {\bibfnamefont {J.~R.}\ \bibnamefont {Yates}},
  \bibinfo {author} {\bibfnamefont {I.}~\bibnamefont {Souza}},\ and\ \bibinfo
  {author} {\bibfnamefont {D.}~\bibnamefont {Vanderbilt}},\ }\bibfield  {title}
  {\bibinfo {title} {Maximally localized {Wannier} functions: Theory and
  applications},\ }\href {https://doi.org/10.1103/RevModPhys.84.1419}
  {\bibfield  {journal} {\bibinfo  {journal} {Rev. Mod. Phys.}\ }\textbf
  {\bibinfo {volume} {84}},\ \bibinfo {pages} {1419} (\bibinfo {year}
  {2012})}\BibitemShut {NoStop}%
\bibitem [{\citenamefont {Mostofi}\ \emph {et~al.}(2008)\citenamefont
  {Mostofi}, \citenamefont {Yates}, \citenamefont {Lee}, \citenamefont {Souza},
  \citenamefont {Vanderbilt},\ and\ \citenamefont {Marzari}}]{mlwf}%
  \BibitemOpen
  \bibfield  {author} {\bibinfo {author} {\bibfnamefont {A.~A.}\ \bibnamefont
  {Mostofi}}, \bibinfo {author} {\bibfnamefont {J.~R.}\ \bibnamefont {Yates}},
  \bibinfo {author} {\bibfnamefont {Y.-S.}\ \bibnamefont {Lee}}, \bibinfo
  {author} {\bibfnamefont {I.}~\bibnamefont {Souza}}, \bibinfo {author}
  {\bibfnamefont {D.}~\bibnamefont {Vanderbilt}},\ and\ \bibinfo {author}
  {\bibfnamefont {N.}~\bibnamefont {Marzari}},\ }\bibfield  {title} {\bibinfo
  {title} {wannier90: A tool for obtaining maximally-localised {Wannier}
  functions},\ }\href
  {https://doi.org/http://dx.doi.org/10.1016/j.cpc.2007.11.016} {\bibfield
  {journal} {\bibinfo  {journal} {Comput. Phys. Commun.}\ }\textbf {\bibinfo
  {volume} {178}},\ \bibinfo {pages} {685 } (\bibinfo {year}
  {2008})}\BibitemShut {NoStop}%
\bibitem [{\citenamefont {Mostofi}\ \emph {et~al.}(2014)\citenamefont
  {Mostofi}, \citenamefont {Yates}, \citenamefont {Pizzi}, \citenamefont {Lee},
  \citenamefont {Souza}, \citenamefont {Vanderbilt},\ and\ \citenamefont
  {Marzari}}]{mlwf_new}%
  \BibitemOpen
  \bibfield  {author} {\bibinfo {author} {\bibfnamefont {A.~A.}\ \bibnamefont
  {Mostofi}}, \bibinfo {author} {\bibfnamefont {J.~R.}\ \bibnamefont {Yates}},
  \bibinfo {author} {\bibfnamefont {G.}~\bibnamefont {Pizzi}}, \bibinfo
  {author} {\bibfnamefont {Y.-S.}\ \bibnamefont {Lee}}, \bibinfo {author}
  {\bibfnamefont {I.}~\bibnamefont {Souza}}, \bibinfo {author} {\bibfnamefont
  {D.}~\bibnamefont {Vanderbilt}},\ and\ \bibinfo {author} {\bibfnamefont
  {N.}~\bibnamefont {Marzari}},\ }\bibfield  {title} {\bibinfo {title} {An
  updated version of wannier90: A tool for obtaining maximally-localised
  {Wannier} functions},\ }\href
  {https://doi.org/https://doi.org/10.1016/j.cpc.2014.05.003} {\bibfield
  {journal} {\bibinfo  {journal} {Comput. Phys. Commun.}\ }\textbf {\bibinfo
  {volume} {185}},\ \bibinfo {pages} {2309 } (\bibinfo {year}
  {2014})}\BibitemShut {NoStop}%
\bibitem [{\citenamefont {Jung}\ \emph {et~al.}(2015)\citenamefont {Jung},
  \citenamefont {DaSilva}, \citenamefont {MacDonald},\ and\ \citenamefont
  {Adam}}]{GhBN_gap}%
  \BibitemOpen
  \bibfield  {author} {\bibinfo {author} {\bibfnamefont {J.}~\bibnamefont
  {Jung}}, \bibinfo {author} {\bibfnamefont {A.~M.}\ \bibnamefont {DaSilva}},
  \bibinfo {author} {\bibfnamefont {A.~H.}\ \bibnamefont {MacDonald}},\ and\
  \bibinfo {author} {\bibfnamefont {S.}~\bibnamefont {Adam}},\ }\bibfield
  {title} {\bibinfo {title} {Origin of band gaps in graphene on hexagonal boron
  nitride},\ }\href {https://doi.org/10.1038/ncomms7308} {\bibfield  {journal}
  {\bibinfo  {journal} {Nat. Commun.}\ }\textbf {\bibinfo {volume} {6}},\
  \bibinfo {pages} {6308} (\bibinfo {year} {2015})}\BibitemShut {NoStop}%
\bibitem [{\citenamefont {Wallbank}\ \emph {et~al.}(2013)\citenamefont
  {Wallbank}, \citenamefont {Patel}, \citenamefont
  {Mucha-Kruczy\ifmmode~\acute{n}\else \'{n}\fi{}ski}, \citenamefont {Geim},\
  and\ \citenamefont {Fal'ko}}]{GhBN_miniband}%
  \BibitemOpen
  \bibfield  {author} {\bibinfo {author} {\bibfnamefont {J.~R.}\ \bibnamefont
  {Wallbank}}, \bibinfo {author} {\bibfnamefont {A.~A.}\ \bibnamefont {Patel}},
  \bibinfo {author} {\bibfnamefont {M.}~\bibnamefont
  {Mucha-Kruczy\ifmmode~\acute{n}\else \'{n}\fi{}ski}}, \bibinfo {author}
  {\bibfnamefont {A.~K.}\ \bibnamefont {Geim}},\ and\ \bibinfo {author}
  {\bibfnamefont {V.~I.}\ \bibnamefont {Fal'ko}},\ }\bibfield  {title}
  {\bibinfo {title} {Generic miniband structure of graphene on a hexagonal
  substrate},\ }\href {https://doi.org/10.1103/PhysRevB.87.245408} {\bibfield
  {journal} {\bibinfo  {journal} {Phys. Rev. B}\ }\textbf {\bibinfo {volume}
  {87}},\ \bibinfo {pages} {245408} (\bibinfo {year} {2013})}\BibitemShut
  {NoStop}%
\bibitem [{\citenamefont {Jung}\ \emph {et~al.}(2014)\citenamefont {Jung},
  \citenamefont {Raoux}, \citenamefont {Qiao},\ and\ \citenamefont
  {MacDonald}}]{Jung2014}%
  \BibitemOpen
  \bibfield  {author} {\bibinfo {author} {\bibfnamefont {J.}~\bibnamefont
  {Jung}}, \bibinfo {author} {\bibfnamefont {A.}~\bibnamefont {Raoux}},
  \bibinfo {author} {\bibfnamefont {Z.}~\bibnamefont {Qiao}},\ and\ \bibinfo
  {author} {\bibfnamefont {A.~H.}\ \bibnamefont {MacDonald}},\ }\bibfield
  {title} {\bibinfo {title} {Ab initio theory of moir\'e superlattice bands in
  layered two-dimensional materials},\ }\href
  {https://doi.org/10.1103/PhysRevB.89.205414} {\bibfield  {journal} {\bibinfo
  {journal} {Phys. Rev. B}\ }\textbf {\bibinfo {volume} {89}},\ \bibinfo
  {pages} {205414} (\bibinfo {year} {2014})}\BibitemShut {NoStop}%
\bibitem [{\citenamefont {Li}\ and\ \citenamefont
  {Koshino}(2019)}]{Li2019SOCproximity}%
  \BibitemOpen
  \bibfield  {author} {\bibinfo {author} {\bibfnamefont {Y.}~\bibnamefont
  {Li}}\ and\ \bibinfo {author} {\bibfnamefont {M.}~\bibnamefont {Koshino}},\
  }\bibfield  {title} {\bibinfo {title} {Twist-angle dependence of the
  proximity spin-orbit coupling in graphene on transition-metal
  dichalcogenides},\ }\href {https://doi.org/10.1103/PhysRevB.99.075438}
  {\bibfield  {journal} {\bibinfo  {journal} {Phys. Rev. B}\ }\textbf {\bibinfo
  {volume} {99}},\ \bibinfo {pages} {075438} (\bibinfo {year}
  {2019})}\BibitemShut {NoStop}%
\bibitem [{\citenamefont {Das~Sarma}\ and\ \citenamefont
  {Li}(2013)}]{PhysRevB.88.081404}%
  \BibitemOpen
  \bibfield  {author} {\bibinfo {author} {\bibfnamefont {S.}~\bibnamefont
  {Das~Sarma}}\ and\ \bibinfo {author} {\bibfnamefont {Q.}~\bibnamefont {Li}},\
  }\bibfield  {title} {\bibinfo {title} {Many-body effects and possible
  superconductivity in the two-dimensional metallic surface states of
  three-dimensional topological insulators},\ }\href
  {https://doi.org/10.1103/PhysRevB.88.081404} {\bibfield  {journal} {\bibinfo
  {journal} {Phys. Rev. B}\ }\textbf {\bibinfo {volume} {88}},\ \bibinfo
  {pages} {081404} (\bibinfo {year} {2013})}\BibitemShut {NoStop}%
\bibitem [{\citenamefont {Schouteden}\ \emph {et~al.}(2016)\citenamefont
  {Schouteden}, \citenamefont {Li}, \citenamefont {Chen}, \citenamefont {Song},
  \citenamefont {Partoens}, \citenamefont {Van~Haesendonck},\ and\
  \citenamefont {Park}}]{Schouteden2016}%
  \BibitemOpen
  \bibfield  {author} {\bibinfo {author} {\bibfnamefont {K.}~\bibnamefont
  {Schouteden}}, \bibinfo {author} {\bibfnamefont {Z.}~\bibnamefont {Li}},
  \bibinfo {author} {\bibfnamefont {T.}~\bibnamefont {Chen}}, \bibinfo {author}
  {\bibfnamefont {F.}~\bibnamefont {Song}}, \bibinfo {author} {\bibfnamefont
  {B.}~\bibnamefont {Partoens}}, \bibinfo {author} {\bibfnamefont
  {C.}~\bibnamefont {Van~Haesendonck}},\ and\ \bibinfo {author} {\bibfnamefont
  {K.}~\bibnamefont {Park}},\ }\bibfield  {title} {\bibinfo {title} {Moir{\'e}
  superlattices at the topological insulator {Bi$_2$Te$_3$}},\ }\href
  {https://doi.org/10.1038/srep20278} {\bibfield  {journal} {\bibinfo
  {journal} {Sci. Rep.-UK}\ }\textbf {\bibinfo {volume} {6}},\ \bibinfo {pages}
  {20278} (\bibinfo {year} {2016})}\BibitemShut {NoStop}%
\bibitem [{\citenamefont {Wang}\ \emph
  {et~al.}(2020{\natexlab{b}})\citenamefont {Wang}, \citenamefont {Yuan},\ and\
  \citenamefont {Fu}}]{LiangFupaper}%
  \BibitemOpen
  \bibfield  {author} {\bibinfo {author} {\bibfnamefont {T.}~\bibnamefont
  {Wang}}, \bibinfo {author} {\bibfnamefont {N.~F.~Q.}\ \bibnamefont {Yuan}},\
  and\ \bibinfo {author} {\bibfnamefont {L.}~\bibnamefont {Fu}},\ }\bibfield
  {title} {\bibinfo {title} {Enhanced superconductivity at high-order van
  {Hove} singularity}} (\bibinfo {year} {2020}{\natexlab{b}})\BibitemShut
  {NoStop}%
\end{thebibliography}%

\end{document}